\documentclass[10pt,twocolumn,a4paper]{article}
\usepackage{preprint}
\usepackage{commands}

\definecolor{mycolor}{RGB}{233,211,63}
\newcolumntype{C}[1]{>{\centering\arraybackslash}m{#1}}
\newcolumntype{L}[1]{>{\raggedright\arraybackslash}m{#1}}
\title{An \fes\ model for shear-deformable beams considering periodic lattice-like truss mesostructures}
\author{Julian Ochs\textsuperscript{*} \quad Jens Wackerfu{\ss}}

\date{October 1, 2026}
\renewcommand{\preprintheaderauthors}{Julian Ochs, Jens Wackerfu{\ss}}
\renewcommand{\preprintversion}{2026-10-01}
\hypersetup{pdftitle={An FE2 model for shear-deformable beams considering periodic lattice-like truss mesostructures}}
\begin{document}
\abstract{
The increasing use of lattice-like architectures in load-bearing members, facilitated by advances in additive manufacturing, calls for efficient multiscale approaches to their structural analysis.
For slender lattice-like structures, beam models provide a computationally efficient macroscopic representation while retaining the ability to account for the underlying lattice architecture through an homogenization approach.
Although \textit{finite element squared} (\fes) approaches combining beam models at the macroscopic scale with continuum models at the microscopic scale have been established, an \fes\ framework combining a beam-based macroscopic model with a truss-based microscopic model of a lattice structure has not yet been reported.
To address this gap, an \fes\ framework is developed for three-dimensional shear-deformable beams with periodically repeated lattice-like microstructures and periodic displacement boundary conditions.

The proposed \fes\ framework introduces constraint equations that (a) prevent rigid-body translations and rotations of the representative volume element (RVE) without imposing additional kinematic restrictions and (b) ensure that the homogenized stress resultants and material matrix are independent of the selected RVE length.
To mitigate artificial boundary effects arising at the boundaries of the lattice-like RVE, local correction factors are introduced.
The framework is assessed through numerical studies of two- and three-dimensional lattice-like structures with different geometries, considering both geometric and material linearity and nonlinearity.
The results demonstrate the independence of the homogenized stress resultants and material matrix from the selected RVE length and show good agreement with the corresponding reference models in all investigated cases.
The proposed \fes\ framework therefore enables accurate and computationally efficient modeling of slender lattice-like structures.
}

\keywords{\fes, trusses-to-beam homogenization, lattice-like mesostructure, multiscale, Timoshenko beam}
\maketitle
\section{Introduction\label{sec:intro}}
Recent advances in additive manufacturing enable the highly precise fabrication of lattice-like structures with complex geometries and topologies \cite{additiveManufacturing2011}.
Lattice-like microstructures offer exceptional mechanical properties with low material consumption \cite{properties1}, making them promising for aerospace, biomedical, and energy absorption applications \cite{application1, application2, application3}.

The targeted design of these microstructures requires structural models that reliably predict their mechanical properties.
For use in engineering structures, direct modeling is often infeasible or prohibitively expensive because of the geometric and topological complexity involved.
Alternative approaches are therefore needed to describe the microstructural properties in a homogenized form, replacing the complex lattice geometry with an equivalent homogeneous medium.
Numerous homogenization approaches have been proposed for lattice structures \cite{numericalMethods1}, including
{beam theory approaches} \cite{numericalMethodsBeamTheory1},
{strain energy equivalence approaches} \cite{numericalMethodsStrainEnergyEquivalence1},
{micropolar theory} \cite{numericalMethodsMicroPolarTheory1},
{asymptotic homogenization} \cite{numericalMethodsAsymptoticHomogenisation1},
{multiscale homogenization} \cite{Vigliotti.2014}
and {machine learning-based approaches} \cite{numericalMethodsMachineLearning1}.
This work focuses on {multiscale homogenization}, also referred to as the \fes\ (\textit{finite element squared}) method.

The \fes\ method is classified below according to the coarse-scale model and the fine-scale architecture.
At the coarse scale, a distinction is made between continuum and structural models; at the fine scale, between continuum and lattice-like microstructures.
This yields four categories of \fes\ models, summarized in Table~\ref{tab:fesquaremethodscomparison}.
\begin{table*}
    \centering
    {\normalsize
        \begin{tabular}{ll|L{6cm}L{6cm}}

            & & \multicolumn{2}{c}{coarse scale} \\[1em]

            & &
            \multicolumn{1}{c}{continuum model} &
            \multicolumn{1}{c}{structural model}
            \\\hline

            \multirow{14}{*}{
                \rotatebox[origin=c]{90}{fine scale}
            }
            &
            \multirow{2}{*}{
                \rotatebox[origin=c]{90}{continuous}
            }
            &
            \multicolumn{1}{L{6cm}}{
                \raggedright\vspace{3ex}
                I: (e.g. \cite{Feyel.1999},
                \cite{wriggers2004},
                \cite{ramm2010},
                \cite{geers2010a})
                \begin{minipage}{5.8cm}

                    \centering
                    \includegraphics{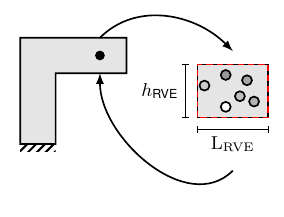}
                \end{minipage}
            }
            &
            \multicolumn{1}{L{6cm}}{
                \raggedright\vspace{3ex}
                III: (e.g. \cite{geers2007},
                \cite{gruttmann2013},
                \cite{klarmann2020},
                \cite{gruttmann2024})
                \begin{minipage}{5.8cm}

                    \centering
                    \includegraphics{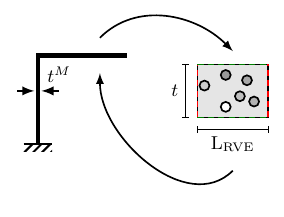}
                \end{minipage}
            }
            \\[4em]

            &
            \multirow{2}{*}{
                \rotatebox[origin=c]{90}{lattice-like}
            }
            &
            \multicolumn{1}{L{6cm}}{
                \raggedright\vspace{3ex}
                II: (e.g. \cite{Vigliotti.2012},
                \cite{Glaesener.2019},
                \cite{Weeger.2021},
                \cite{Danesh.2025})
                \begin{minipage}{5.8cm}
                    \centering
                    \includegraphics{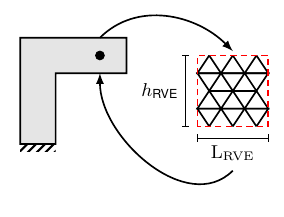}
                \end{minipage}
            }
            &
            \multicolumn{1}{L{6cm}}{
                \raggedright\vspace{3ex}
                IV: \cite{RoyChowdhury.2019}
                \begin{minipage}{5.8cm}
                    \centering
                    \includegraphics{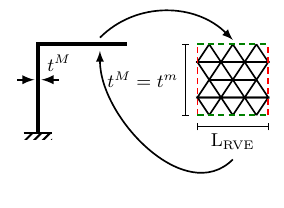}
                \end{minipage}
            }
            \\
        \end{tabular}
    }
    \caption{Categorization of \fes-Methods (The references listed for Categories I, II, and III are incomplete.)}
    \label{tab:fesquaremethodscomparison}
\end{table*}
These categories are briefly described below in the context of existing research.

I) The starting point is a continuum model at the coarse (macroscopic) scale coupled to a continuum at the fine (microscopic) scale.
This configuration is shown at the upper left of Table~\ref{tab:fesquaremethodscomparison}.
The term \fes\ was first introduced in this context in \cite{Feyel.1999}.
The theoretical foundations have since been studied extensively \cite{Feyel.2000, Feyel.2003, Miehe.1999, Miehe.2002, Miehe.2002b, miehe2002}.
These foundations have supported substantial further development; see, for example, \cite{wriggers2004, ramm2010, geers2010a, Ricken.2022}.
In general, the \fes\ method couples two finite element analyses at hierarchically arranged macroscopic and microscopic scales.
The microscopic boundary value problem is solved at every integration point of the discretized macroscopic boundary value problem.
The resulting computational cost is therefore governed primarily by the discretization of both scales.
In the classical \fes\ method, both scales are modeled as continua.

II) Several approaches have also been investigated for a macroscopic continuum coupled to a lattice-like microstructure; see the lower left of Table~\ref{tab:fesquaremethodscomparison}.
In \cite{Vigliotti.2012,Vigliotti.2012b,Vigliotti.2014}, effective material properties were determined for materially linear and nonlinear lattice structures.
The lattices were modeled using shear-rigid beams.
Building on this approach, \cite{Danesh.2025} investigated a lattice of truss members with materially nonlinear behavior.
That study focused on the plastic regime.
In \cite{Weeger.2021}, a second-order linear elastic homogenization theory was developed to capture size and localization effects.
The lattice was modeled using shear-deformable beams.
A second-order homogenization strategy was also adopted in \cite{Nguyen.2014}.
A {Mindlin strain-gradient} continuum was used at the macroscopic scale.
The higher-order homogenization approach was intended to investigate microscopic stability.
In \cite{Glaesener.2019}, a multiscale model for two-dimensional linear elastic lattices was presented, combining a generalized macroscopic continuum with second-order homogenization.
The microstructure was modeled using shear-rigid beams.
The model was extended to three-dimensional microstructures in \cite{Glaesener.2020}.
Both studies showed that higher-order homogenization improves the description of localization effects while having only a minor influence on the global response.
In \cite{Glaesener.2021}, the approach was extended to material nonlinearity, particularly viscoelasticity.
In \cite{Radi.2024}, microstructures composed of truss elements were investigated for impact mitigation and energy absorption.

III) A variant of the \fes\ method arises when a structural model is used at the macroscopic scale; see the upper right of Table~\ref{tab:fesquaremethodscomparison}.
The RVE dimensions are then directly linked to those of the underlying structural model.
For a shell, the RVE thickness equals the shell thickness; for a beam, the RVE cross section matches the shape and dimensions of the macroscopic beam cross section.
The fine scale is therefore referred to as the \textit{meso}scopic rather than the \textit{micro}scopic scale.
For structural elements, scale separation consequently applies only along the reference surface of a shell or the reference axis of a beam.
A framework for the numerical homogenization of shells was first presented in \cite{geers2007}.
Numerous extensions and variants have since been developed, primarily addressing the underlying shell kinematics and the boundary conditions imposed on the RVE.
The different approaches are briefly reviewed below.

A homogenization strategy for thin, shear-rigid shells was first proposed in \cite{geers2010b}.
It was subsequently applied to masonry \cite{Mercatoris.2009}, sandwich plates \cite{Gigliotti.2015}, and microelectromechanical systems \cite{Lucas.2017}.
A nonlinear multiscale model for Kirchhoff--Love shells was also developed in \cite{papadopoulos2022}.
An appropriate strain projection removes large rotations from the kinematic constraints imposed on the RVE.

The authors of \cite{luscher2012} developed general orthogonality conditions for second-order multiscale homogenization of statistical volume elements.
To account for thickness changes, a numerical multiscale formulation for Reissner--Mindlin plates with seven degrees of freedom was developed in \cite{diebels2013, Helfen.2014}.
Reference \cite{larsson2013} presented a shell theory based on two-scale homogenization.
The authors used a first-order Taylor expansion for the in-plane displacements and a second-order expansion for the normal components in the thickness direction.

Layered shells were investigated in \cite{gruttmann2013}.
That study was restricted to linear displacement boundary conditions.
In \cite{gruttmann2016}, the model was applied to sandwich plates with cores consisting of periodically arranged cell walls.
For geometrically and materially linear microscopic problems, homogenization can be performed in advance for a single RVE.
According to \cite{cong2015}, this is also possible for geometrically nonlinear shell problems at the macroscopic scale.

In \cite{fraemby2016}, the shear stiffness and shear stresses obtained with existing models were found to depend on the RVE length.
The origin of this effect was identified in \cite{klarmann2020}.
Imposing shear strains on the RVE also generates a parasitic moment.
Its magnitude depends on the RVE length.
Constraints at the center of the RVE were introduced to eliminate this effect.

Building on \cite{luscher2012}, additional constraints were also introduced in \cite{hii2022} to eliminate the length dependence of the shear stiffness.
A predictor--corrector scheme for the shear strain is required to obtain the correct values.

In \cite{klarmann2022}, a homogenization procedure based on Irving--Kirkwood theory was presented.
Its implementation is straightforward because existing homogenization algorithms and boundary conditions can be reused.

In addition to the classical hierarchically nested \fes\ method, monolithic approaches couple the macroscopic and microscopic or mesoscopic problems within a common system of equations or Newton iteration.
An early monolithic formulation for a two-scale shell model was presented in \cite{gruttmann2013}.
The Direct-\fes\ method of \cite{Tan.2020} also follows a monolithic approach, combining the macroscopic and microscopic scales into a single finite element system through kinematic constraints, which can facilitate implementation in commercial finite element software.
Monolithic approaches directly couple the two scales without nested iterations, although their computational and memory requirements can increase substantially when the microscopic model has many degrees of freedom.
An alternative monolithic solution approach was presented in \cite{Lange.2021}.
The Direct-\fes\ approach has also been extended to various structural mechanics problems.
In \cite{zhi2023}, it was combined with a Reissner--Mindlin shell to model thin-walled fiber-reinforced composite laminates, whereas \cite{Yeoh.2022} extended it to shear-deformable beams based on Timoshenko--Ehrenfest beam theory.

A multiscale method for the mechanical analysis of planar laminates with discontinuities and mixed boundary conditions was described in \cite{turon2023}.
Reference \cite{boerjesson2023} presented a variationally consistent method for developing a homogenization framework for linear Reissner--Mindlin plates.
In \cite{klarmann2023}, three types of boundary conditions satisfying the Hill--Mandel condition \cite{hill1952} were introduced to couple the two scales.

IV) The combination of a macroscopic structural element and a lattice-like mesostructure, shown at the lower right of Table~\ref{tab:fesquaremethodscomparison}, has so far been investigated in only one published study.
That study introduced a geometrically exact formulation for micropolar beams and applied it within a two-scale model \cite{RoyChowdhury.2019}.
At the macroscopic scale, a new two-dimensional micropolar Timoshenko beam with five beam strains was considered: one axial strain, symmetric and antisymmetric shear strains, and three curvatures, two describing the macroscopic beam curvature and one the mesostructural curvature.
At the mesoscopic scale, a standard two-dimensional Timoshenko beam was used.
The macroscopic strains were imposed on the RVE through linear displacement boundary conditions.
Homogenization was based on the Hill--Mandel condition \cite{hill1952}.
The model in \cite{RoyChowdhury.2019} can be regarded as the first category IV \fes\ model, although several limitations remain.
First, the model is restricted to two dimensions and therefore cannot explicitly resolve the cross-sectional geometry of members with C-, I-, or U-shaped sections.
Second, linear displacement boundary conditions can introduce kinematic restrictions and boundary effects, potentially causing the homogenized properties to depend on the selected RVE length.
Such a dependence is also apparent in the results presented in \cite{RoyChowdhury.2019}.
However, the influence of RVE length on homogenization results, particularly on the homogenized material tangent, was not investigated systematically.
Third, the influence of geometric and material nonlinearity on the homogenized properties was likewise not examined systematically.
Fourth, the influence of geometric eccentricities within the mesostructure on the homogenized response was not considered.

In summary, it can be concluded that \fes\ models in categories I, II, and III are highly advanced and already well-established, whereas there is still a great need for research on \fes\ models in category IV.
This work aims to present a robust category IV \fes\ model that addresses the gaps identified above.
The principal features and contributions of the proposed \fes\ method are summarized as follows:

\begin{enumerate}
    \item A homogenization strategy is developed for beams with Timoshenko kinematics and periodic lattice-like mesostructures composed of truss members.
    Various two- and three-dimensional lattices are investigated.
    \item Appropriate constraints eliminate the parasitic rigid-body rotations of the RVE arising under periodic boundary conditions.
    \item Specific correction factors eliminate parasitic boundary effects arising at the model boundaries of lattice-like microstructures.
    These ensure that the homogenized quantities are independent of the selected RVE length.
    \item The constraints are incorporated into the system of equations using master--slave elimination.
    \item The method is verified through numerous numerical examples and validated against reference models, including cases with eccentricities and geometric and material nonlinearity.
    New benchmark tests are developed for this purpose.
    Beams with both solid and open cross-sectional shapes were examined.
\end{enumerate}
\begin{figure*}
    \centering
    \includegraphics{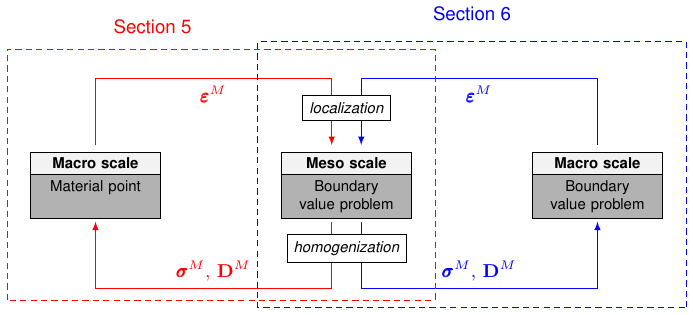}
    \caption{Test environments for the proposed \fes-method, used in Sections 5 and 6}
    \label{fig:numericalExample_twoStrategies}
\end{figure*}
The remainder of this article is organized as follows.
Section \ref{sec:latticeStructure} describes the lattice structures and introduces a consistent notation.
Section \ref{sec:fe2} presents the multiscale model.
The macroscopic model is introduced first, followed by the mesoscopic model and the coupling strategy.
Section \ref{sec:constraints} presents the additional constraints imposed at the mesoscopic scale.
Sections \ref{sec:numericalExamples_I} and \ref{sec:numericalExamples_II} demonstrate the reliability of the proposed \fes\ method through numerical examples.
Section \ref{sec:numericalExamples_I} focuses on tests with a single material point at the macroscopic scale.
Section \ref{sec:numericalExamples_II} considers more complex examples that also require solving a macroscopic boundary value problem; Figure~\ref{fig:numericalExample_twoStrategies} illustrates the two cases schematically.
The paper concludes with Section \ref{sec:conclusion}.

\section{Lattice-like microstructures \label{sec:latticeStructure}}
\begin{figure*}
    \begin{subfigure}{\textwidth}
        \centering
        \includegraphics{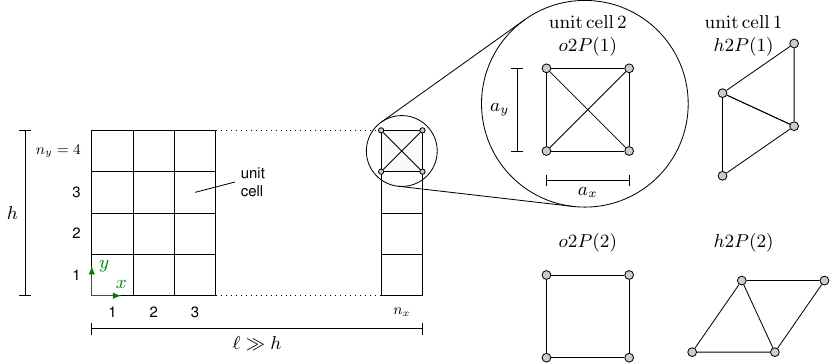}
        \caption{2-dimensional case}
        \label{fig:microstructure2dOverview}
    \end{subfigure}

    \vspace{1ex}
    \begin{subfigure}{\textwidth}
        \centering
        \includegraphics{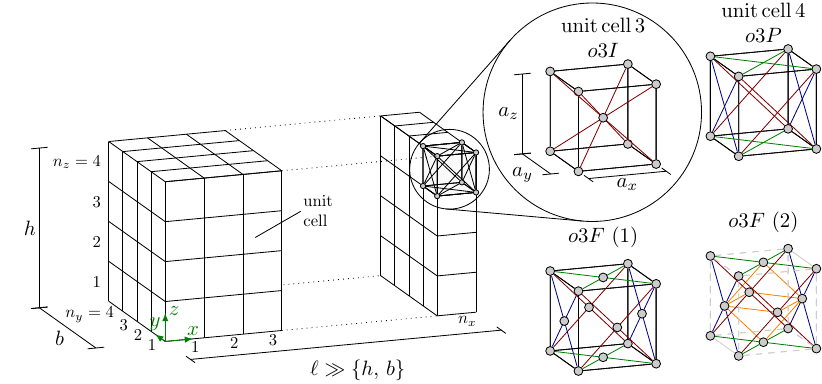}
        \caption{3-dimensional case}
        \label{fig:microstructure3dOverview}
    \end{subfigure}

    \caption{Construction of a slender, lattice-like microstructure composed of orthorhombic unit cells.
     The dashed lines shown there serve only as visual guides and do not represent actual connections.}
    \label{fig:microstructureOverview}
\end{figure*}

This work considers two- and three-dimensional lattice-like mesostructures.
In general, a lattice structure consists of individual members connected at lattice points.
The members are assumed to be straight.
Only periodic lattice structures are considered, i.e., structures composed of repeating unit cells, as shown in Figure \ref{fig:microstructureOverview}.
The number of unit cells in each spatial direction is denoted by \mbox{$n_i$}, \mbox{where $i\in\{x,y,z\}$}.
A mathematical description of a unit cell requires geometric, topological, and material data.
The following geometric description follows \cite{tadmor2011}.
The lattice vectors span a {unit cell}, also referred to as a \textit{primitive unit cell}.
Repeating the {unit cell} throughout space reproduces the original lattice.
In addition to the lattice vectors, the lattice spacing $a_i$, with $i\in\{x,y,z\}$, is introduced below.
The lattice spacing describes the distance between two lattice points in the $x$-, $y$-, and $z$-directions.
This definition is unambiguous only if at least one of the {primitive lattice vectors} is aligned with the orthogonal basis.
There are 14 distinct lattices in three dimensions and five in two dimensions, referred to as \textsc{Bravais} lattices \cite{tadmor2011}.

Four arrangements are distinguished for three-dimensional orthorhombic lattices: primitive ($P$), body-centered ($I$), face-centered ($F$), and base-centered ($S$).
This work is restricted to orthorhombic ($o2P$) and hexagonal ($h2P$) lattices in two dimensions and orthorhombic ($o3I$, $o3P$) lattices in three dimensions; see the right-hand side of Figure \ref{fig:microstructureOverview}.
The lattices shown can also be rotated arbitrarily in space.

The multiscale method presented in this work considers slender lattice structures.
Thus, the number of unit cells in the $x$-direction is much larger than the number in the $y$- and $z$-directions.
Accordingly, the length \mbox{$\ell=n_xa_x$} is much larger than the width \mbox{$b=n_ya_y$} and height \mbox{$h=n_za_z$}:
\begin{equation}
    \ell\gg \{b,\,h\} \; .
\end{equation}
Slender mesostructures with smooth boundaries at both ends are considered, i.e., all lattice points at each end lie in the same plane.
Mesostructures composed of unit cell h2P(2), shown in Figure \ref{fig:microstructure2dOverview}, do not satisfy this condition and are therefore excluded from the following analysis.

\paragraph{Modeling the mesostructure}
The cross-sectional geometry and material properties of each mesostructural member are assumed to remain constant along its axis.
The members are assumed to be connected by ideal hinges (pinned connection).
The mesostructure must therefore be designed to exclude mechanisms within a geometrically linear analysis, e.g. \cite{Wackerfuss2020}.
Examples of mesostructures exhibiting mechanisms include those composed of unit cells $o2P(2)$, $o3F(2)$, and $o3F(1)$; see Figure \ref{fig:microstructure3dOverview}.
The first two give rise to global mechanisms, whereas the last also exhibits local mechanisms on each outer face of the unit cell.
The following analysis considers only geometrically stable mesostructures, namely $h2P(1)$, $o2P(1)$, $o3I$, and $o3P$, referred to for simplicity as unit cells '1', '2', '3', and '4', respectively.

\section{\fes \, model\label{sec:fe2}}
\begin{figure*}
    \centering
    \includegraphics[width=\textwidth]{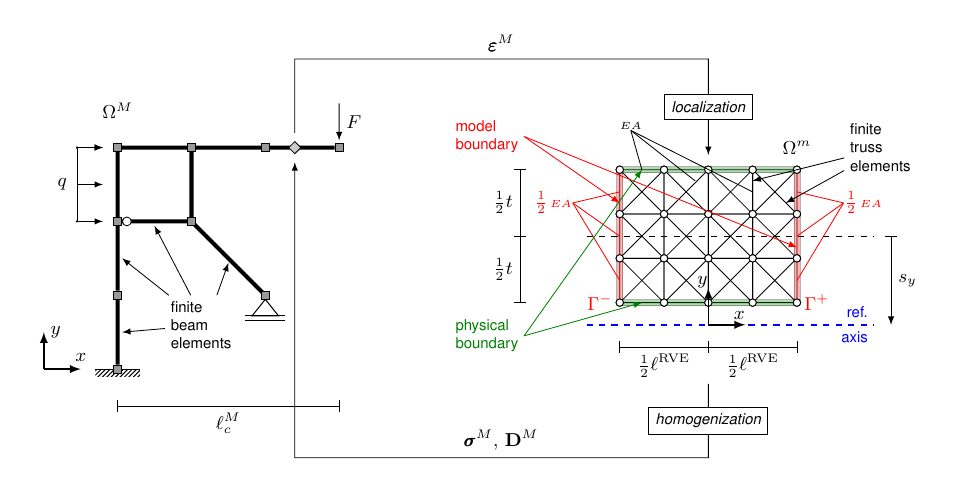}
    \caption{Terminology, symbols, and design of the proposed FE Method (2-dimensional case)}
    \label{fig:fesScheme}
\end{figure*}

Figure \ref{fig:fesScheme} schematically illustrates the \fes\ model considered in this work.
The macroscopic and mesoscopic models and their coupling through localization and homogenization are described below.

\subsection{Macro level \label{macroLevel}}
The macroscopic scale is represented by a spatial frame structure $\Omega^M$.
Each macroscopic beam is spatially discretized using finite beam elements with (Timoshenko kinematics, finite rotations).
The beam stiffnesses, i.e., the cross-sectional and material properties, are determined by the mesostructure described at the mesoscopic scale.
The generalized displacements $\mathbf{u}\macro$, strains $\pmb\varepsilon\macro$, and stress resultants $\pmb\sigma\macro$ of the beam model are
\begin{equation} \label{eq:macro_01}
	\mathbf{u}\macro =
	\begin{bmatrix}
		u_x \\
		u_y    \\
		u_z    \\
		\varphi_x    \\
		\varphi_y    \\
		\varphi_z
	\end{bmatrix}
;\;\,
	\pmb\varepsilon\macro =
\begin{bmatrix}
	\varepsilon \\
	\gamma_y    \\
	\gamma_z    \\
	\vartheta    \\
	\kappa_{y}    \\
	\kappa_{z}
\end{bmatrix}
;\;\,
\pmb\sigma\macro =
\begin{bmatrix}
	N    \\
	Q_y  \\
	Q _z \\
	M_T  \\
	M_y  \\
	M_z
\end{bmatrix} \;,
\end{equation}
where $u_x$, $u_y$, and $u_z$ denote the displacements and $\varphi_x$, $\varphi_y$, and $\varphi_z$ denote the rotations.
The strain components are the axial strain $\varepsilon$, shear strains $\gamma_y$ and $\gamma_z$, twist $\vartheta$, and curvatures $\kappa_{y}$ and $\kappa_{z}$.
The work-conjugate beam stress resultants are the axial force $N$, shear forces $Q_y$ and $Q_z$, torsional moment $M_T$, and bending moments $M_y$ and $M_z$.
The physical units of the first three generalized displacements, beam strains, and stress resultants are [length], [1], and [force], respectively; those of the remaining components are [1], [1/length], and [force length], respectively.
All components refer to the reference axis $x$ selected for each beam, which need not coincide with the centroidal axis; thus, eccentricities $s_y$ and $s_z$ can be taken into account.
The cross-sectional axes $y$ and $z$ are perpendicular to the $x$-axis and need not be principal axes.

The strains $\pmb\varepsilon\macro$ are derived using the kinematic assumptions expressed by the following product ansatz
\begin{equation} \label{eq:macro_02}
	\begin{split}
		\begin{bmatrix}
			u_x \\
			u_y \\
			u_z
		\end{bmatrix}
		&=
		\begin{bmatrix}
			1 & 0 & 0 & 0  & z & -y \\
			0 & 1 & 0 & -z & 0 & 0 \\
			0 & 0 & 1 &  y & 0 & 0 \\
		\end{bmatrix}
		\mathbf{u}^M \\
		\mathbf{u}(x,y,z)&=\mathbf{A}(y,z)\,\mathbf{u}^M(x)
	\end{split} \;,
\end{equation}
where $\mathbf{u}$ denotes the displacements at a material point within the beam.
The product ansatz \eqref{eq:macro_02} is valid only for moderate rotations $\varphi_x$, $\varphi_y$, and $\varphi_z$, i.e., when
\begin{equation} \label{eq:macro_03}
	\{ \varphi_x, \varphi_y, \varphi_z \} < \varphi^\text{mod}
\end{equation}
holds.
For a threshold value of $\varphi^\text{mod}=7^\circ$, the small-angle approximations underlying Eq. \eqref{eq:macro_03}, $\sin(\varphi)\approx\varphi$ and $\cos(\varphi)\approx 1$, yield acceptable relative errors of 0.25\% and 0.75\%, respectively.

The strains $\pmb\varepsilon\macro$ are obtained from the generalized displacements $\mathbf{u}\macro$ of the macroscopic beam model through the kinematic equations, whereas the corresponding stress resultants $\pmb\sigma\macro$ and beam stiffnesses $\vector{D}\macro$ are determined by homogenization at the mesoscopic scale.

The beam is assumed to be straight and consists of a prismatic lattice-like mesostructure with an arbitrary cross-sectional shape and dimensions, e.g., a rectangle of height $t^M$ and width $b^M$.

\paragraph{State of equilibrium}
The weak form of the macroscopic boundary value problem is given by
\begin{equation} \label{eq:macro_04}
\begin{split}
	g_\text{int}^M(\mathbf{u},\delta\mathbf{u}) + g_\text{ext}^M(\delta\mathbf{u}) = \mathbf{0}&  \qquad \text{in $\Omega^M$} \\
	\mathbf{u} - \bar{\mathbf{u}} = \mathbf{0}& \qquad \text{on $\Gamma^M_u$} \\
\end{split}
\end{equation}
with the internal contribution
\begin{equation} \label{eq:macro_05}
		g_\text{int}^M(\mathbf{u},\delta\mathbf{u}) = \int_{\Omega^M}  [\delta\pmb\varepsilon^M]^T \pmb \sigma^M  \, \text{d}x
\end{equation}
and the external contribution
\begin{equation} \label{eq:macro_06}
		g_\text{ext}^M(\delta\mathbf{u}) = - \int_{\Omega^M} \delta \mathbf{u}^T \bar{\mathbf{b}}  \, \text{d}x + [\delta \mathbf{u}^T \bar{\mathbf{t}}]|_{\Gamma_\sigma^M}
\end{equation}
where $\bar{\mathbf{b}}$ and $\bar{\mathbf{t}}$ denote conservative distributed loads (e.g., line loads) and concentrated loads (Neumann boundary conditions), respectively.
The Dirichlet boundary conditions \eqref{eq:macro_04}$_2$, with prescribed displacements $\bar{\mathbf{u}}$, are defined on the displacement boundary $\Gamma^M_u$.

\subsection{Meso-level \label{MesoLevel}}
At the mesoscopic scale $\Omega^m$, the lattice-like mesostructure of the beam is resolved explicitly.
It consists of 'numel' pin-jointed truss members with length $\ell$ and cross-sectional area $A$.
Each member is spatially discretized using a single finite truss element.
The strain $\varepsilon$ in a member follows from the kinematic relation $\varepsilon=u'$ for geometric linearity or $\varepsilon=u'+\frac{1}{2}(u')^2$ for geometric nonlinearity, where $u$ is the displacement along the member axis.
The strain is assumed to be small.
The axial force $N$ follows from the constitutive relation $N=CA\varepsilon$, where $CA$ denotes the axial stiffness.
The numerical studies presented in Sections \ref{sec:numericalExamples_I} and \ref{sec:numericalExamples_II} consider both elastic behavior, with $C=E$ (Young's modulus $E$), and elastoplastic behavior ($J_2$ plasticity) with linear isotropic hardening, with $C=C_T^\text{ep}$ (tangent modulus $C_T^\text{ep}$, yield stress $\sigma_Y$, and hardening modulus $H$).

\paragraph{RVE}
The prismatic mesostructure consists of unit cells and is periodic in the $x$-direction; see Figure \ref{fig:microstructureOverview}.
Any segment whose extent in the $x$-direction corresponds to a complete period can therefore serve as an RVE of the mesostructure.
The period length is an integer multiple of the lattice spacing $a_x$.
The RVE is aligned with the beam axes $x$, $y$, and $z$.
Without loss of generality, the RVE is positioned in the $x$-direction such that the plane at $x=0$ is a plane of symmetry.
The origin can be selected freely within this plane, allowing eccentricities relative to the $y$- and $z$-axes to be taken into account.
The RVE boundaries at $x=\pm\frac{1}{2}\ell\rve$ are perpendicular to the $x$-axis.
These are referred to as the \textit{model boundaries} $\Gamma^\pm$, whereas all remaining boundaries are referred to as \textit{physical boundaries}; see Figure \ref{fig:fesScheme}.

\begin{figure}
    \centering
    \begin{subfigure}{\columnwidth}
        \centering
        \includegraphics{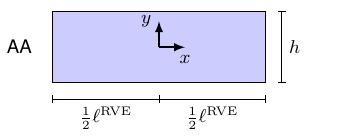}
        \caption{homogeneous composition}
    \end{subfigure}
    \hfill
    \begin{subfigure}{\columnwidth}
        \centering
        \vspace{2ex}
        \includegraphics{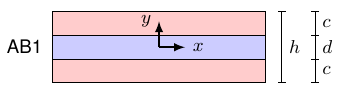}\\
        \vspace{2ex}
        \includegraphics{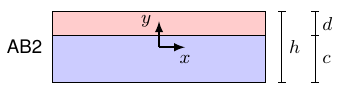}\\
        \vspace{2ex}
        \includegraphics{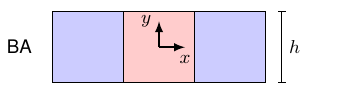}\\
        \vspace{2ex}
        \includegraphics{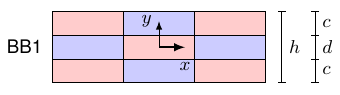}\\
        \vspace{2ex}
        \includegraphics{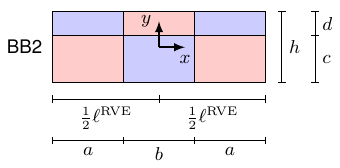}
        \caption{heterogeneous composition}
    \end{subfigure}
    \caption{Different types of RVE (two-dimensional case). Each individual color represents a lattice-like microstructure with specific geometric, topological, and material properties}
    \label{fig:RVE_classification}
\end{figure}

The RVE width $b=n_ya_y$ and height $h=n_za_z$ are prescribed by the mesostructure under consideration; their cross-sectional dimensions are therefore identical.
In contrast, the RVE length $\ell\rve=n_x a_x$ is a model parameter that must be selected appropriately, subject to the requirement of \textit{scale separation}
\begin{equation} \label{eq:meso_01}
    \ell_c\macro \gg \ell\rve \;.
\end{equation}
Here, $\ell_c\macro$ is a characteristic length at the macroscopic scale.
To minimize the computational cost of the \fes\ simulation, $\ell\rve$ should be as small as possible; its minimum value depends on the presence and extent of heterogeneity in the $x$-direction.
Figure \ref{fig:RVE_classification} schematically illustrates all types of heterogeneity possible in periodic mesostructures
using a simple two-phase, two-dimensional RVE, with colors representing material properties.
For RVEs with a homogeneous composition in the $x$-direction (C1, C3a, and C3b), the minimum length $\ell^\text{RVE}=a_x$ is sufficient.

To ensure the periodicity of the lattice-like RVE, all members lying along the model boundaries $\Gamma^+$ and $\Gamma^-$ must be modified.
The axial stiffness $EA$ of these members must be halved; see Figure \ref{fig:fesScheme}.
Thus, for member $i$,
\begin{equation}
    EA_i =
    \begin{cases}
        \frac{1}{2}EA & \quad \text{for}\quad i \in \Gamma^\pm \\
        EA & \quad \text{else} \;.
    \end{cases}
\end{equation}
Without this modification, artificial boundary effects arise, particularly for small values of $\ell^\text{RVE}$.
This boundary effect is examined numerically in Section \ref{sec:Rve_length}; see also Figure \ref{fig:numericalResultsI_2d_centric_o2P-1_h2P-1_DiiOverLen}.

\paragraph{State of equilibrium}
The equilibrium configuration of the RVE is determined by solving the following weak form of the boundary value problem
\begin{equation} \label{eq:meso_02}
	\begin{split}
		g_\text{int}^m(\mathbf{u},\delta\mathbf{u}) + g_\text{ext}^m(\delta\mathbf{u}) = \mathbf{0}&  \qquad \text{in $\Omega^m$} \\
		\mathbf{c}_1(\mathbf{u}) = \mathbf{0}& \qquad \text{on $\Gamma^\pm$} \\
		\mathbf{c}_2(\mathbf{u}) = \mathbf{0}& \qquad \text{in $\Omega^m$} \;,
	\end{split}
\end{equation}
with
\begin{equation} \label{eq:meso_03}
\begin{split}
	g_\text{int}^m(\mathbf{u},\delta\mathbf{u}) &= \int_{\Omega^m} \delta \varepsilon \, N \, \text{d}x \\
	g_\text{ext}^m(\delta\mathbf{u}) &= 0 \;,
\end{split}
\end{equation}
where no external loads are applied.
Equations \eqref{eq:meso_02}$_2$ and \eqref{eq:meso_02}$_3$ can be interpreted as constraints and are satisfied in strong form; they are specified in Sections \ref{sec:localization} and \ref{sec:constraints}.
They depend linearly on the unknown displacements $\mathbf{u}$.
The Dirichlet boundary conditions are considerd as constraints in \eqref{eq:meso_02}$_3$.

\paragraph{FE discretization}
Equation \eqref{eq:meso_02}$_1$ is discretized using the Bubnov--Galerkin method.
The RVE is spatially discretized using two-node finite truss elements.
The numbers of elements and nodes are denoted by 'numel' and N, respectively.
Equation \eqref{eq:meso_02}$_1$ then yields a system of $3N$ nonlinear equations
\begin{equation} \label{eq:meso_04}
	\mathbf{F}_\text{int}(\mathbf{V}) = \mathbf{0} \;,
\end{equation}
where $\mathbf{F}_\text{int}\in\mathbb{R}^{3N}$ is the internal nodal force vector, whose components depend on the unknown nodal displacements $\mathbf{V}\in\mathbb{R}^{3N}$.
Within the finite element formulation, the linear constraints \eqref{eq:meso_02}$_2$ and \eqref{eq:meso_02}$_3$ become multipoint constraints
\begin{equation} \label{eq:meso_05}
\begin{split}
	\mathbf{c}_1(\mathbf{V}) &= \mathbf{0} \\
	\mathbf{c}_2(\mathbf{V}) &= \mathbf{0} \;,
\end{split}
\end{equation}
with $\mathbf{c}_1\in\mathbb{R}^{n_1}$ and $\mathbf{c}_2\in\mathbb{R}^{n_2}$, where the dimensions $n_1$ and $n_2$ are specified below.

The nonlinear equations \eqref{eq:meso_04}, subject to the linear constraints \eqref{eq:meso_05}, are solved using master--slave elimination \cite{boungard2024}\footnote{Other approaches, such as the method of Lagrange multipliers, can also be used.}.
Without loss of generality, the degrees of freedom on the model boundaries $\Gamma^+$ and $\Gamma^-$ are selected as master and slave degrees of freedom, respectively.
This yields the reduced system
\begin{equation} \label{eq:meso_06}
	\matrix{K}\micro_\text{red} \, \Delta\vector{V}\micro_\text{red} = - 	\vector{R}\micro_\text{red} \;,
\end{equation}
which is solved for the unknown increments $\Delta\vector{V}\micro_\text{red}\in\mathbb{R}^{(3N-n_1-n_2)}$ in each iteration of the underlying Newton method.
Here, $\matrix{K}\micro_\text{red}$ is the symmetric tangent stiffness matrix of the reduced system.

\subsection{Hill-Mandel condition}
The macroscopic and mesoscopic scales are coupled through the Hill--Mandel condition \cite{hill1952}.
It requires the virtual internal energy at a material point of the macroscopic domain $\Omega^M$ to equal the averaged virtual internal energy of the associated mesoscopic domain $\Omega^m$, i.e.
\begin{equation}\label{FE2_HillMandel_01}
	\delta W\internal\macro = \langle \delta W\internal\micro \rangle = \frac{1}{\ell\rve} \delta W\internal\micro\;,
\end{equation}
where, in the present setting, the volume average $\langle \bullet \rangle$ reduces to an average over the RVE length $\ell\rve$.

\subsection{Localization} \label{sec:localization}
During localization, the macroscopic strains $\pmb\varepsilon\macro$ are imposed on the RVE.
For small strains, it can be shown (e.g., \cite{gross2016}) that the periodic boundary conditions
\begin{equation}\label{FE2_localization_01}
    \vector{x}^+ - \vector{x}^- = \vector{F}^M (\vector{X}^+ - \vector{X}^-)
    \;,\qquad
    \vector{t}^+ = -\vector{t}^-
\end{equation}
satisfy the Hill--Mandel condition \eqref{FE2_HillMandel_01}.
In \eqref{FE2_localization_01}$_1$, $\mathbf{x}^+$ and $\mathbf{x}^-$ denote the current positions of two opposing material points on the model boundaries $\Gamma^+$ and $\Gamma^-$, respectively, and $\mathbf{X}^+$ and $\mathbf{X}^-$ denote their reference positions\footnote{This uses the geometric identity of the model boundaries $\Gamma^+$ and $\Gamma^-$, which is satisfied a priori for the periodic mesostructures considered here.}.
Equation \eqref{FE2_localization_01}$_1$ can be interpreted as a linear mapping valid only for small strains.
In \eqref{FE2_localization_01}$_1$, $\mathbf{F}^M$ is the deformation gradient, which, under the small-strain assumption, follows from the linearized Green--Lagrange strain tensor $\mathbf{E}^M_\text{lin}$ and the identity tensor $\mathbf{1}$
\begin{equation}\label{FE2_localization_01x}
    \mathbf{F}^M-\mathbf{1}=\mathbf{E}^M_\text{lin} \;.
\end{equation}

For a Timoshenko beam that is straight and untwisted in the reference configuration, the kinematic assumptions \eqref{eq:macro_02} yield only three nonzero components of $\mathbf{E}^M_\text{lin}$ (cf. \cite{klarmann2020})
\begin{equation}\label{FE2_localization_01a}
\mathbf{E}^M_\text{lin} =
\begin{bmatrix}
	E_{xx} & 0 & 0 \\
	E_{{yx}} & 0 & 0 \\
	E_{{zx}} & 0 & 0
\end{bmatrix}^M \;,
\end{equation}
which are obtained from the beam strain vector $\pmb\varepsilon^M$ introduced in Eq. \eqref{eq:macro_01} and the matrix $\mathbf{A}$ introduced in \eqref{eq:macro_02} as
\begin{equation}\label{FE2_localization_01b}
\mathbf{E}^M_\text{lin} \, \mathbf{e}_x  = \mathbf{A}\,\pmb\varepsilon^M \; .
\end{equation}
Component-wise, \eqref{FE2_localization_01b} becomes
\begin{equation}\label{FE2_localization_01b1}
        \begin{bmatrix}
            E_{xx} \\
            E_{{yx}} \\
            E_{{zx}}
        \end{bmatrix}^M =
        \begin{bmatrix}
            \varepsilon^M + z\, \kappa_{y}^M - y\, \kappa_{z}^M \\
            \gamma_y^M - z \, \vartheta \\
            \gamma_z^M + y \, \vartheta \\
        \end{bmatrix} \;.
\end{equation}

Using Eqs. \eqref{FE2_localization_01x} and \eqref{FE2_localization_01b}, together with $\mathbf{u}^\pm:=\mathbf{x}^\pm-\mathbf{X}^\pm$ and $\vector{X}^+ - \vector{X}^-=\ell^\text{RVE}\mathbf{e}_x$, the periodic displacement boundary condition \eqref{FE2_localization_01}$_1$ can be expressed as a relation between the microscopic displacements $\mathbf{u}^\pm$ and the macroscopic beam strains $\pmb\varepsilon^M$
\begin{equation}\label{FE2_localization_03}
    \begin{split}
        \vector{u}^+ - \vector{u}^-
        &= (\vector{F}^M - \mathbf{1}) \, (\vector{X}^+ - \vector{X}^-) \\
        &= \mathbf{E}^M_\text{lin} \, (\vector{X}^+ - \vector{X}^-) \\
        &= \ell^\text{RVE} \mathbf{E}^M_\text{lin} \mathbf{e}_x  \\
        &= \ell^\text{RVE} \mathbf{A}\,\pmb\varepsilon^M  \;.
    \end{split}
\end{equation}

\begin{figure}
    \includegraphics{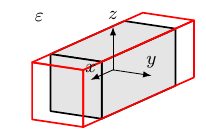}
    \includegraphics{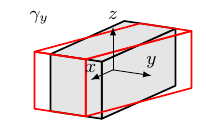}

    \includegraphics{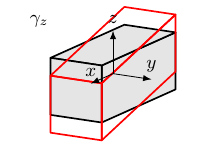}
    \includegraphics{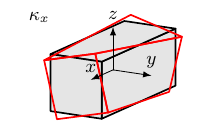}

    \includegraphics{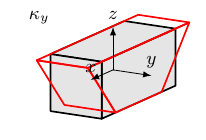}
    \includegraphics{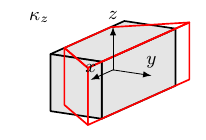}
\caption{Illustration of the displacement fields $\vector{u}^\pm=\vector{u}(\pm\frac{1}{2}\ell^\text{RVE},y,z)$ at the model boundaries $\Gamma^\pm$ of the RVE generated by the individual components of the macro-strain vector $\pmb\varepsilon^M$, as described by Eqn. \eqref{FE2_localization_03}. Note that the modes shown do not represent equilibrium states}
\label{fig:modesMatrixA}
\end{figure}

Evaluating the boundary conditions \eqref{FE2_localization_03} separately for each component of $\pmb\varepsilon^M$ yields the six RVE loading modes illustrated in Figure \ref{fig:modesMatrixA}.

In \eqref{FE2_localization_01}$_2$, $\vector{t}^+$ and $\vector{t}^-$ denote the forces acting on the model boundaries $\Gamma^+$ and $\Gamma^-$.

\paragraph{Lattice-like RVE}
The periodic boundary conditions formulated above for continuum RVEs are now transferred to the lattice-like RVEs.
The number of nodes on each model boundary, $\Gamma^+$ and $\Gamma^-$, is denoted by $n_\Gamma$.
Equations \eqref{FE2_localization_03} and \eqref{FE2_localization_01}$_2$ must therefore be satisfied for $n_\Gamma$ node pairs, with one constraint for each of the three spatial directions.
This yields a total of $n_1=3n_\Gamma$ constraints of the form \eqref{eq:meso_05}$_1$.
For node pair $i$, Eqs. \eqref{FE2_localization_03} and \eqref{FE2_localization_01}$_2$ become
\begin{equation}\label{FE2_localization_05}
	\begin{split}
		\vector{u}_i^+(y_i,z_i) - \vector{u}_i^-(y_i,z_i) &= \ell^\text{RVE} \mathbf{A}_i(y_i,z_i) \, \pmb \varepsilon^M \\
		\vector{t}_i^+(y_i,z_i) + \vector{t}_i^-(y_i,z_i) &= \mathbf{0} \; ,
	\end{split}
\end{equation}
where $y_i$ and $z_i$ are the coordinates of node $i$ on the model boundary $\Gamma^+$ or $\Gamma^-$.
In the continuum setting, $\vector{t}^\pm$ denote traction vectors with physical units of [force/length$^2$].
In the discrete lattice setting, $\vector{t}_i^\pm$ denote nodal boundary forces with physical units of [force].
Collecting \eqref{FE2_localization_05} for all node pairs one obtains
\begin{equation} \label{FE2_localization_06}
\begin{split}
	\begin{bmatrix}
		\mathbf{u}_1^+ \\
		\mathbf{u}_2^+ \\
		\vdots\\
		\mathbf{u}_{n_\Gamma}^+ \\
	\end{bmatrix}
	-
	\begin{bmatrix}
		\mathbf{u}_1^- \\
		\mathbf{u}_2^- \\
		\vdots\\
		\mathbf{u}_{n_\Gamma}^- \\
	\end{bmatrix}
	&= \ell^\text{RVE}
	\begin{bmatrix}
		\mathbf{A}_1 \\
		\mathbf{A}_2 \\
		\vdots\\
		\mathbf{A}_{n_\Gamma} \\
	\end{bmatrix}
	\pmb\varepsilon^M \\
	\begin{bmatrix}
	\mathbf{t}_1^+ \\
	\mathbf{t}_2^+ \\
	\vdots\\
	\mathbf{t}_{n_\Gamma}^+ \\
\end{bmatrix}
+
\begin{bmatrix}
	\mathbf{t}_1^- \\
	\mathbf{t}_2^- \\
	\vdots\\
	\mathbf{t}_{n_\Gamma}^- \\
\end{bmatrix}
&=
\begin{bmatrix}
	\mathbf{0} \\
	\mathbf{0} \\
	\vdots\\
	\mathbf{0}
\end{bmatrix} \;,
\end{split}
\end{equation}
or, in compact notation,
\begin{equation} \label{FE2_localization_07}
\begin{split}
	\mathbf{V}^+ - \mathbf{V}^- &= \ell^\text{RVE} \mathbb{A} \, \pmb\varepsilon^M \\
	\mathbf{T}^+ + \mathbf{T}^- &= \mathbf{0} \;,
\end{split}
\end{equation}
with the vectors $\mathbf{V}^\pm\in\mathbb{R}^{n_1}$, $\mathbf{T}^\pm\in\mathbb{R}^{n_1}$, and the matrix $\mathbb{A}\in\mathbb{R}^{(n_1 \times 6)}$:
\begin{equation} \label{FE2_localization_08}
\mathbb{A} =
\begin{bmatrix}
	\mathbf{A}_1 \\
	\mathbf{A}_2 \\
	\vdots\\
	\mathbf{A}_{n_\Gamma}
\end{bmatrix}; \quad
\mathbf{A}_i =
\begin{bmatrix}
	1 & 0 & 0 & 0  & z_i & -y_i \\
	0 & 1 & 0 & -z_i & 0 & 0 \\
	0 & 0 & 1 &  y_i & 0 & 0 \\
\end{bmatrix} .
\end{equation}
The matrices $\mathbf{A}_i$ follow from \eqref{eq:macro_02}.
The $n_1$ periodic displacement boundary conditions \eqref{FE2_localization_07}$_1$ can be interpreted as constraint equations.

The vectors $\mathbf{V}^+$ and $\mathbf{V}^-$ can formally be extracted from $\mathbf{V}\in\mathbb{R}^{3N}$ using a permutation matrix $\mathbf{P}\in\mathbb{R}^{(3N\times 3N)}$
\begin{equation} \label{FE2_localization_09}
	\begin{bmatrix}
		\mathbf{V}^+ \\
		\mathbf{V}^- \\
		\mathbf{V}^{\Omega^*} \\
	\end{bmatrix}
	=
	\mathbf{P} \, \mathbf{V}
	=
	\begin{bmatrix}
		\mathbf{P}^+ \\
		\mathbf{P}^- \\
		\mathbf{P}^{\Omega^*} \\
	\end{bmatrix}
	\mathbf{V} \;.
\end{equation}
Here, $\mathbf{V}^+$ and $\mathbf{V}^-$ contain the nodal degrees of freedom on the model boundaries $\Gamma^+$ and $\Gamma^-$, respectively, and $\mathbf{V}^{\Omega^*}$ contains those of the remaining nodes in the RVE interior, i.e., $\Omega^*=\{\Omega^m\setminus\Gamma^\pm\}$.
Similarly, $\mathbf{T}^+$ and $\mathbf{T}^-$ can be extracted from $\mathbf{T}$ using the permutation matrix $\mathbf{P}$ introduced in \eqref{FE2_localization_09}.

\subsection{Limitations\label{sec:localication_limits}}
The periodic boundary condition \eqref{FE2_localization_07}$_1$ contains the matrix $\mathbf{A}_i$ given in \eqref{FE2_localization_08}$_2$.
In general, the use of this matrix assumes a geometrically linear analysis at the mesoscopic scale.

When this matrix is used in a geometrically nonlinear analysis, parasitic effects may arise in the RVE subjected to macroscopic strains.
These effects result from the geometric linearity assumed in defining $\mathbf{A}_i$.
According to \eqref{eq:macro_03}, this matrix is valid only for moderate rotations $\varphi_x$, $\varphi_y$, and $\varphi_z$.
For the RVE, this means that the model boundaries $\Gamma^\pm$ may undergo only moderate rotations when the periodic displacement boundary conditions \eqref{FE2_localization_07}$_1$ are imposed.
According to \eqref{FE2_localization_07}$_1$ and \eqref{FE2_localization_08}, rotations are induced only by the macroscopic strains $\vartheta^M$, $\kappa_{y}^M$, and $\kappa_{z}^M$, since only these components are multiplied by the cross-sectional coordinate $y$ or $z$.
The rotations at the model boundaries $\Gamma^\pm$ are
\begin{equation} \label{FE2_localization_limit_01}
\begin{bmatrix}
    \varphi_x^\pm \\
    \varphi_y^\pm \\
    \varphi_z^\pm
\end{bmatrix}
= \frac{1}{2} \ell\rve
\begin{bmatrix}
    \vartheta^M \\
    \kappa_y^M \\
    \kappa_z^M
\end{bmatrix} \;.
\end{equation}
Combining \eqref{eq:macro_03} with \eqref{FE2_localization_limit_01} provides an upper bound on the admissible RVE length $\ell\rve$
\begin{equation} \label{FE2_localization_limit_02}
	\ell\rve < 2 \, \varphi^\text{mod} \min \left \{ \frac{1}{\vartheta^M},\frac{1}{\kappa_{y}^M},\frac{1}{\kappa_{z}^M} \right \} \; .
\end{equation}
The previously discussed limit of $7^\circ$ ($\approx 0.122$ \, rad) for moderate rotations can be adopted for $\varphi^\text{mod}$.

The parasitic effects are detailed below, distinguishing three mechanisms.
First,
the upper bound in \eqref{FE2_localization_limit_02} depends on the magnitude of the imposed macroscopic strain. In a geometrically nonlinear analysis, this introduces an artificial dependence of the homogenized quantities on the selected RVE length $\ell\rve$.
This dependence is examined in the numerical study in Section \ref{sec:GeometricNonlinearity} and is visible in the curves in Figures \ref{fig:numericalResultsI_geometricNonlinear_simulation3} and \ref{fig:numericalResultsI_geometricNonlinear_simulation4}.
Second,
under geometric nonlinearity, imposing the periodic boundary conditions \eqref{FE2_localization_07}$_1$ introduces an artificial axial kinematic constraint for the loading cases $\vartheta^M$, $\kappa_y^M$, and $\kappa_z^M$.
This becomes apparent when \eqref{FE2_localization_07}$_1$ is evaluated at the model boundaries $\Gamma^\pm$, with $x^\pm=\pm\tfrac{1}{2}\ell\rve$, at the reference axis, i.e., $y=z=0$
\begin{equation} \label{FE2_localization_limit_03}
\begin{bmatrix}
    u_x(x^+,0,0) \\
    u_y(x^+,0,0) \\
    u_z(x^+,0,0) \\
\end{bmatrix}
-
\begin{bmatrix}
    u_x(x^-,0,0) \\
    u_y(x^-,0,0) \\
    u_z(x^-,0,0) \\
\end{bmatrix}
=
\begin{bmatrix}
    0 \\
    0 \\
    0
\end{bmatrix} \;.
\end{equation}
Equation \eqref{FE2_localization_limit_03} enforces a constant distance between the model boundaries at this location, independent of the loading magnitude, which is exact under geometric linearity.
Under geometric nonlinearity, however, this introduces an axial kinematic constraint that induces a parasitic axial force $N$ during homogenization.
This phenomenon is also examined in Section \ref{sec:GeometricNonlinearity}.
Figures \ref{fig:numericalResultsI_geometricNonlinear_simulation3} and \ref{fig:numericalResultsI_geometricNonlinear_simulation4} show that the macroscopic strains $\vartheta$ and $\kappa_y$ induce not only the respective stress resultants $M_T$ and $M_y$, but also a parasitic axial force $N$.
Third,
the artificial axial constraint also occurs for $\gamma_y$ and $\gamma_z$, as can be seen by evaluating \eqref{FE2_localization_07}$_1$ at the model boundaries $\Gamma^\pm$
\begin{equation} \label{FE2_localization_limit_04}
    \begin{bmatrix}
        u_x(x^+,y,z) \\
        u_y(x^+,y,z) \\
        u_z(x^+,y,z) \\
    \end{bmatrix}
    -
    \begin{bmatrix}
        u_x(x^-,y,z) \\
        u_y(x^-,y,z) \\
        u_z(x^-,y,z) \\
    \end{bmatrix}
    = \ell\rve
    \begin{bmatrix}
        0 \\
        \gamma_y \\
        \gamma_z
    \end{bmatrix} .
\end{equation}
Equation \eqref{FE2_localization_limit_04} enforces the model boundaries $\Gamma^\pm$ to translate parallel to one another in the transverse direction, in proportion to $\gamma_y$ or $\gamma_z$, while maintaining their horizontal distance.
Under geometric nonlinearity, this induces a parasitic homogenized axial force $N$; see, for example, Figure \ref{fig:numericalResultsI_geometricNonlinear_simulation2}.

The significance of these parasitic effects is discussed below.
They arise not only for macroscopic beam models but also for plate and shell models.
They also occur independent of whether the mesoscopic model has a continuous or a lattice-like structure.
For the lattice-like microstructures considered here, it must additionally be ensured that the loaded RVE remains structurally stable at all times, e.g., that individual truss members do not buckle.
Extensive numerical experiments on geometrically nonlinear lattice-like RVEs showed that the parasitic effects described above appeared only after local instabilities (buckling) had already developed within the RVE.
The regions in which structural instabilities occurred are shaded gray in Figure \ref{fig:numericalResultsI_geometricNonlinear}.
Both the artificial length dependence and the parasitic axial force become significant only after the first instability occurs.
This observation cannot be generalized, since the onset of structural instability depends on the geometry and topology of the selected unit cell and on the magnitude of the tensile and bending stiffnesses of the truss members.
Nevertheless, the examples incorporating geometric nonlinearity at the mesoscopic scale (Sections \ref{sec:GeometricNonlinearity}, \ref{sec:SimplySupportedBeam}, and \ref{sec:SpatialFrameStructure}) also yielded length-independent results in very good agreement with independent reference solutions.

Finally, the parasitic effects discussed above vanish as the RVE length decreases.
If geometric linearity is assumed at the mesoscopic scale, these effects are absent a priori, even for arbitrarily long RVEs.

\subsection{Homogenization} \label{sec:homogenization}
The RVE must be in equilibrium before the homogenized stress resultants $\vectorSym{\sigma}\macro$ and material matrix $\matrix{D}\macro$ are evaluated.
This requires solving the generally nonlinear equations \eqref{eq:meso_04} subject to the constraints \eqref{FE2_localization_07}$_1$ and \eqref{eq:constraintSummary_02}.
Using Newton's method with master--slave elimination yields the system \eqref{eq:meso_06} to be solved at each iteration, which is partitioned as follows:
\begin{equation} \label{homogenization_sig_and_D_01}
\begin{bmatrix}
	\matrix{K}\micro_\text{red,aa} & \matrix{K}\micro_\text{red,ab} \\
	\matrix{K}\micro_\text{red,ba} & \matrix{K}\micro_\text{red,bb}
\end{bmatrix}
\begin{bmatrix}
	\Delta\vector{V}\micro_\text{red,a} \\
	\Delta\vector{V}\micro_\text{red,b}
\end{bmatrix}
= -
\begin{bmatrix}
	\vector{R}\micro_\text{red,a} \\
	\vector{R}\micro_\text{red,b}
\end{bmatrix} .
\end{equation}
In \eqref{homogenization_sig_and_D_01}, $\Delta\vector{V}\micro_\text{red,b}\in\mathbb{R}^{n_1}$ contains the increments associated with the nodes on the model boundary $\Gamma^+$, and $\Delta\vector{V}\micro_\text{red,a}\in\mathbb{R}^{(3N-2n_1-n_2)}$ contains all remaining increments.
At equilibrium, $\vector{R}\micro_\text{red,a}=\vector{0}$.
The increments $\Delta\vector{V}\micro_\text{red,a}$ are eliminated from \eqref{homogenization_sig_and_D_01} by static condensation.
This yields
\begin{equation} \label{homogenization_sig_and_D_02}
	\matrix{K}^m_\text{sc} \, \Delta\vector{V}\micro_\text{red,b} = - \matrix{R}^m_\text{sc} \;,
\end{equation}
with $\matrix{R}^m_\text{sc}\in \mathbb{R}^{n_1}$ and $\matrix{K}^m_\text{sc}\in \mathbb{R}^{(n_1 \times n_1)}$ defined by
\begin{equation} \label{homogenization_sig_and_D_03}
	\begin{split}
		\matrix{R}^m_\text{sc} &:= \vector{R}\micro_\text{red,b} - \matrix{K}\micro_\text{red,ba} [\matrix{K}\micro_\text{red,aa}]^{-1} \matrix{R}\micro_\text{red,a} \\
		\matrix{K}^m_\text{sc} &:= \vector{K}\micro_\text{red,bb} - \matrix{K}\micro_\text{red,ba} [\matrix{K}\micro_\text{red,aa}]^{-1} \matrix{K}\micro_\text{red,ab} \;.
	\end{split}
\end{equation}
Using the matrix $\mathbb{A}\in\mathbb{R}^{(n_1\times 6)}$ from \eqref{FE2_localization_08}, the quantities associated with the nodes on the model boundary $\Gamma^+$ are projected onto the reference axis
\begin{equation} \label{homogenization_sig_and_D_04}
\begin{split}
	\vectorSym{\sigma}\macro &= \mathbb{A}\transpose \matrix{R}^m_\text{sc} \\
	\matrix{D}\macro &= \mathbb{A}\transpose \matrix{K}^m_\text{sc} \mathbb{A} \;.
\end{split}
\end{equation}
This gives the homogenized stress resultants $\vectorSym{\sigma}\macro\in\mathbb{R}^6$ and the symmetric homogenized material matrix $\mathbf{D}^M\in\mathbb{R}^{6\times 6}$.

\paragraph{Proof}
The consistency of the proposed stress homogenization is verified below.
The virtual internal energy at a material point on the macroscopic scale is
\begin{equation}\label{homogenization_01}
    \delta W\internal\macro
    = [\vectorSym{\sigma}\macro]\transpose\,\delta \vectorSym{\varepsilon}\macro
    =
    \left(\begin{bmatrix}
        N    \\
        Q_y  \\
        Q _z \\
        M_T  \\
        M_y  \\
        M_z
    \end{bmatrix}\macro\right)^T
    \begin{bmatrix}
        \delta \varepsilon \\
        \delta \gamma_y    \\
        \delta \gamma_z    \\
        \delta \vartheta    \\
        \delta \kappa_{y}    \\
        \delta \kappa_{z}
    \end{bmatrix}\macro .
\end{equation}
The virtual internal energy of the lattice-like mesoscopic model is the sum of the contributions of its '\numberOfElements' truss members:
\begin{equation}\label{homogenization_02}
\delta W\internal\micro
    =  \sum_{i=1}^\numberOfElements \int_{\ell_i} N_i \, \delta \varepsilon_i \, \d s_i
    =  \sum_{i=1}^\numberOfElements N_i \, \ell_i \, \delta \varepsilon_i \;.
\end{equation}
Here, $N_i$, $\varepsilon_i$, $\ell_i$, and $s_i$ denote the axial force, strain, length, and coordinate of truss member $i$, respectively.
Within the finite element formulation, \eqref{homogenization_02} can be rewritten as
\begin{equation}\label{homogenization_03}
    \delta W\internal\micro    = [\vector{F}\internal\micro]\transpose \delta \vector{V}\micro \;,
\end{equation}
where $\delta\mathbf{V}^m\in\mathbb{R}^{3N}$ and $\delta\mathbf{F}^m_\text{int}\in\mathbb{R}^{3N}$ contain the virtual displacements and work-conjugate internal nodal forces of the RVE, respectively.
Partitioning $\vector{F}\internal\micro$ and $\delta \vector{V}\micro$ as in \eqref{FE2_localization_09} decomposes \eqref{homogenization_02} into three contributions
\begin{equation}\label{homogenization_04}
    \delta W\internal\micro
    = [\vector{F}^+_{\mathrm{int}}]\transpose \delta\vector{V}^+
    + [\vector{F}^-_{\mathrm{int}}]\transpose \delta\vector{V}^-
    + [\vector{F}^{\Omega^*}_{\mathrm{int}}]\transpose \delta\vector{V}^{\Omega^*}.
\end{equation}
The forces $\vector{F}^\pm_{\mathrm{int}}$ are identical to the boundary forces $\vector{T}^\pm$ in \eqref{FE2_localization_07}, i.e.,
\begin{equation}\label{homogenization_05}
    \vector{F}^+_{\mathrm{int}} = \vector{T}^+
\;;\qquad
    \vector{F}^-_{\mathrm{int}} = \vector{T}^- \;.
\end{equation}
Homogenization can be performed only once the RVE is in equilibrium.
In this state, the internal nodal forces cancel at all nodes outside the model boundaries $\Gamma^\pm$, i.e.,
\begin{equation}\label{homogenization_06}
   \vector{F}^{\Omega^*}_{\mathrm{int}} = \mathbf{0} \;.
\end{equation}
Using \eqref{FE2_localization_07}, \eqref{homogenization_05}, and \eqref{homogenization_06}, Eq. \eqref{homogenization_04} reduces to
\begin{equation}\label{homogenization_07}
    \begin{split}
        \delta W\internal\micro
        &= [\vector{F}^+_{\mathrm{int}}]^T \left( \delta\vector{V}^+ - \delta\vector{V}^- \right) \\
        &= [\vector{F}^+_{\mathrm{int}}]^T \ell^\text{RVE} \mathbb{A} \, \delta \pmb\varepsilon^M \;.
    \end{split}
\end{equation}
Substituting \eqref{homogenization_01} and \eqref{homogenization_07} into \eqref{FE2_HillMandel_01} gives
\begin{equation}\label{homogenization_08}
    [\vectorSym{\sigma}\macro]\transpose\,\delta \vectorSym{\varepsilon}\macro =
    \cancel{\frac{\ell^\text{RVE}}{\ell\rve}}\underbrace{[\vector{F}^+_{\mathrm{int}}]^T \mathbb{A}}_{=:[\vectorSym{\sigma}\macro]\transpose} \, \delta \pmb\varepsilon^M \; ,
\end{equation}
and thus defines the homogenized beam stress resultants $\pmb\sigma^M$.
This definition is independent of the RVE length $\ell^\text{RVE}$.
The stress resultants can also be expressed as a sum over the $n_\Gamma$ nodes on the model boundary $\Gamma^+$:
\begin{equation}\label{homogenization_09}
    \vectorSym{\sigma}\macro
    := \sum_{i=1}^{n_\Gamma} \vectorSym{\sigma}\macro_i
    \;;\qquad
    \vectorSym{\sigma}\macro_i
    = \mathbf{A}_i^T \vector{F}^+_{\mathrm{int},i} \;.
\end{equation}
Substituting \eqref{FE2_localization_08}$_2$ into \eqref{homogenization_09}$_2$ yields
\begin{equation}\label{homogenization_10}
    \begin{bmatrix}
        N    \\
        Q_y  \\
        Q _z \\
        M_T  \\
        M_y  \\
        M_z
    \end{bmatrix}\macro
    =
    \sum_{i=1}^{n_\Gamma}
    \begin{bmatrix}
        F^+_{\mathrm{int},i,x} \\
        F^+_{\mathrm{int},i,y} \\
        F^+_{\mathrm{int},i,z} \\
        F^+_{\mathrm{int},i,z} \,y_i - F^+_{\mathrm{int},i,y} \,z_i\\
        F^+_{\mathrm{int},i,x} \,z_i\\
        -F^+_{\mathrm{int},i,x} \,y_i
    \end{bmatrix} \; .
\end{equation}
According to \eqref{homogenization_10}, the beam stress resultants can be computed from the nodal forces on the RVE model boundary $\Gamma^+$.
For the periodic boundary conditions considered here, $N$, $Q_y$, and $Q_z$ follow from three force equilibrium equations, and $M_T$, $M_y$, and $M_z$ follow from three moment equilibrium equations.
The homogenization procedure therefore yields stress resultants compatible with the macroscopic beam model; cf. Eq. \eqref{eq:macro_01}.

\subsection{Entries of matrix $\mathbf{D}^M$}
General properties of the entries of the symmetric material matrix $\mathbf{D}^M$ are discussed below\footnote{The physical units of entries $D_{ij}$ follow once the units of the components of $\pmb\sigma^M$ and $\pmb\varepsilon^M$ are specified; see Section \ref{macroLevel}.}.
The microstructures are still assumed to be periodic and composed exclusively of truss members.
The assumptions regarding the RVE architecture stated in Section \ref{MesoLevel} also remain applicable.

\paragraph{Geometric linearity}
For a geometrically linear RVE model, the resulting material matrix $\mathbf{D}^M$ contains 12 independent nonzero entries arranged as follows:
\begin{equation} \label{homogenization_D_structure_01}
\mathbf{D}^M =
    \begin{bmatrix}
        D_{11}  & 0       & 0      & 0      & D_{15} & D_{16} \\
        & D_{22} & D_{23} & D_{24} & 0      & 0        \\
        &        & D_{33} & D_{34} & 0      & 0        \\
        &        &        & D_{44} & 0      & 0        \\
        & \mathrm{sym.} & &        & D_{55} & D_{56} \\
        &        &        &       &         & D_{66}
    \end{bmatrix}
\end{equation}
This matrix structure is independent of both the lattice-like mesostructure, i.e., the underlying unit cell type, and the material properties of the individual truss members.

Comparing \eqref{homogenization_D_structure_01} with the material matrix of a classical Timoshenko beam with a continuum cross section and linear elastic behavior (cf. \cite{wackerfuss2011, Schramm.1997}) shows that the two matrices have the same structure.
Comparing their individual components allows the entries $D_{ij}$ to be identified with beam stiffness parameters.
For the axial and bending contributions
\begin{equation} \label{homogenization_D_structure_02}
    \begin{aligned}
        D_{11} &= EA        & \quad D_{55} &= EI_y \\
        D_{15} &= EA \, s_z & \quad D_{56} &= -EI_{yz} \\
        D_{16} &= -EA \, s_y& \quad D_{66} &= EI_z \; ,
    \end{aligned}
\end{equation}
where $EA$ is the axial stiffness, $EI_y$ and $EI_z$ are the bending stiffnesses, $EI_{yz}$ is the bending coupling stiffness, and $s_y$ and $s_z$ are the centroid coordinates.
The shear contributions are
\begin{equation} \label{homogenization_D_structure_03}
    \begin{split}
        D_{22} &= GA \, \shearFac{y} \\
        D_{23} &= GA \, \shearFac{yz}  \\
        D_{24} &=-GA \, \shearFac{y} \, m_z \\
        D_{33} &= GA \, \shearFac{z} \\
        D_{34} &= GA \, \shearFac{z} \, m_y \;,
        \end{split}
\end{equation}
where $GA$ is the shear stiffness, $\shearFac{y}$, $\shearFac{z}$, and $\shearFac{yz}$ are dimensionless shear correction factors, and $m_y$ and $m_z$ are the shear center coordinates.
The torsional contribution is
\begin{equation} \label{homogenization_D_structure_04}
    \begin{split}
        D_{44} &= GI_T^*= G(I_T + A\,\shearFac{y}\,m_z^2 + A\,\shearFac{z}\,m_y^2)
    \end{split}
\end{equation}
where $I_T$ is the Saint-Venant torsion constant.
This gives the following 12 independent stiffness and geometric parameters:
\begin{equation} \label{homogenization_D_structure_05}
\begin{split}
    &\{EA, EI_y, EI_z, EI_{yz}, s_y, s_z,\\
    &\;\, GA\,\shearFac{y}, GA\,\shearFac{z}, GA\,\shearFac{yz}, GI_T, m_y, m_z \}
\end{split}
\end{equation}
These can be uniquely identified from the 12 nonzero components of $\mathbf{D}^M$
\begin{equation} \label{homogenization_D_structure_06}
	\begin{split}
		s_y &= -\frac{D_{16}}{D_{11}}\;;\qquad
		s_z = \frac{D_{15}}{D_{11}} \\
		m_y &= \frac{D_{34}}{D_{33}}\;;\qquad
        m_z = -\frac{D_{24}}{D_{22}} \\
        GI_T &= D_{44} - \frac{D_{24}^2}{D_{22}} - \frac{D_{34}^2}{D_{33}} \;.
    \end{split}
\end{equation}
In Sections \ref{sec:numericalExamples_I} and \ref{sec:numericalExamples_II}, the entries of $\mathbf{D}^M$ are described using the notation in \eqref{homogenization_D_structure_05}.

If the RVE cross section has at least one axis of symmetry, has no eccentricity, and the $y$- and $z$-axes are principal axes, i.e., $s_y=s_z=m_y=m_z=0$, $I_{yz}=0$, and $k_{yz}=0$, the matrix $\mathbf{D}^M$ is diagonal.
All of these properties were verified in the numerical studies reported in Section \ref{sec:numericalExamples_I}.

\begin{remark}
If both geometric and material linearity are assumed, the stiffnesses recovered using \eqref{homogenization_D_structure_02}--\eqref{homogenization_D_structure_04} can also serve as input to a single-scale finite element simulation in which the slender lattice-like mesostructure is modeled a priori as a continuum Timoshenko beam.
\end{remark}

\paragraph{Geometric nonlinearity}
Including geometric nonlinearity in the RVE model changes the entries of the material matrix $\mathbf{D}^M$.
These changes arise from the geometric contribution to the tangent stiffness matrix $\mathbf{K}_\text{red}^m$ in \eqref{eq:meso_06} and therefore depend directly on the loading magnitude\footnote{In contrast, the material contribution to matrix $\mathbf{K}_\text{red}^m$ is represented exclusively by the components in \eqref{homogenization_D_structure_01}, which are independent of loading and therefore constant for linear elasticity.}.

Compared with geometric linearity, additional changes occur in the first row and column of the matrix, except for $D_{11}$.
First, the off-diagonal entries $D_{12},D_{13}$, and $D_{14}$ become nonzero, giving the following matrix structure
\begin{equation} \label{homogenization_D_structure_nili_01}
    \mathbf{D}^M =
    \begin{bmatrix}
        D_{11}   & D_{12} & D_{13} & D_{14} & D_{15} & D_{16} \\
        & D_{22} & D_{23} & D_{24} & 0      & 0        \\
        &        & D_{33} & D_{34} & 0      & 0        \\
        &        &        & D_{44} & 0      & 0        \\
        & \mathrm{sym.} & &        & D_{55} & D_{56} \\
        &        &        &       &         & D_{66}
    \end{bmatrix} \; .
\end{equation}
Second, the entries $D_{15}$ and $D_{16}$ are also nonzero, even for a centric RVE.
These changes are caused by the parasitic axial forces discussed in Section \ref{sec:localication_limits}.
They occur when the RVE is subjected to the macroscopic strains $\gamma_y^M$, $\gamma_z^M$, $\vartheta^M$, $\kappa_y^M$, and $\kappa_z^M$.

All of these changes depend on the magnitude of the imposed macroscopic strains; thus, \eqref{homogenization_D_structure_nili_01} approaches \eqref{homogenization_D_structure_01} as the loading decreases.

As in the geometrically linear case, the matrix \eqref{homogenization_D_structure_nili_01} is symmetric, and its structure is independent of the mesostructural architecture and the material behavior of the individual truss members.
This is verified numerically in Section \ref{sec:GeometricNonlinearity}.

\subsection{FE$^2$ formulation}
In the FE$^2$ method, the macroscopic and mesoscopic scales are coupled at every integration point of a finite beam element (Figure \ref{fig:fesScheme}), i.e., at every integration point in the macroscopic finite element mesh where the constitutive relation is evaluated.
During localization, the macroscopic beam strains $\pmb\varepsilon^M$ at that point are imposed on the mesoscopic RVE through the periodic displacement boundary conditions \eqref{FE2_localization_07}$_1$.
These conditions are applied only to the model boundaries $\Gamma^\pm$, not to the physical boundaries of the RVE.
Additional constraints are required to suppress rigid-body motions; these are specified in Section \ref{sec:constraints}.
The equilibrium configuration of the RVE can then be determined, using the Newton--Raphson method for a nonlinear mesostructural model.
In the final homogenization step (Section \ref{sec:homogenization}), the homogenized beam stress resultants $\pmb\sigma^M$ and the associated material matrix $\mathbf{D}^M$ are computed and returned to the macroscopic scale.

\section{Additional constraints\label{sec:constraints}}
This section specifies the constraints $\mathbf{c}_2(\mathbf{V})=\mathbf{0}$ introduced in Eq. \eqref{eq:meso_05}$_2$.
These constraints eliminate possible rigid-body translations and rotations of the RVE without introducing artificial kinematic restrictions.

\subsection{Rigid body rotations about $y$- and $z$-axis ('discrete constant moment constraint')\label{sec:constraints_shear}}
For both continuum and lattice-like RVEs, imposing the shear strains $\gamma_y^M$ or $\gamma_z^M$ produces a pure rigid-body rotation about the $z$- or $y$-axis, respectively, causing the homogenized shear forces $Q_y$ or $Q_z$ and shear stiffnesses $GA_{sy}$ or $GA_{sz}$ to vanish.
This phenomenon was first described in \cite{geers2007} for a macroscopic shell coupled to a mesoscopic continuum model.
A constraint was introduced to suppress this parasitic behavior and was subsequently extended to layered shells in \cite{gruttmann2024}; see also Appendix \ref{app:constraints_shear_conti}.

A new constraint required for lattice-like mesostructures is introduced below.
This '\textit{discrete constant moment constraint}' builds on the '\textit{constant moment constraint}' for continuum mesostructures, specifically on its simplified form \eqref{eq:appendix_cmc_09} derived in Appendix \ref{app:constraints_shear_conti}.

\begin{figure}
    \centering
    \includegraphics[width=0.4 \textwidth,clip]{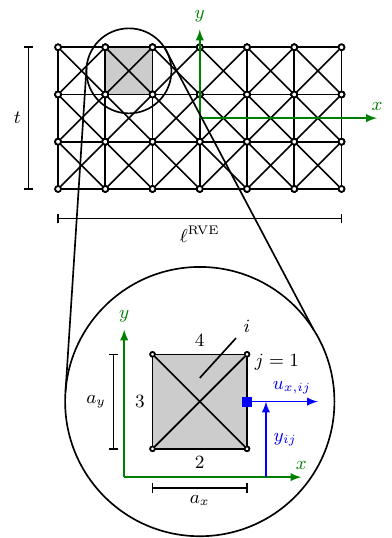}
    \caption{Illustration of the abbreviations defined in Eq. \eqref{eq:constraintShear_02} using the example of a 2-dimensional RVE}
    \label{fig:constraints_shear}
\end{figure}

The derivation uses the fact that the lattice-like RVE is composed of unit cells.
The integral in \eqref{eq:appendix_cmc_09} is expressed as a sum over the $n$ unit cells:
\begin{equation} \label{eq:constraintShear_01}
    \sum_{i=1}^n f_i \, V_i \overset{!}{=} 0 \; ,
\end{equation}
with the contribution $f_i$ of unit cell $i$ given by
\begin{equation} \label{eq:constraintShear_02}
    f_i := \frac{1}{m_i} \sum_{j=1}^{m_i} D_{ij} \, u_{x,ij} \, y_{ij} \;,
\end{equation}
and weighted by its volume $V_i$; see also Figure \ref{fig:constraints_shear}.
The contribution $f_i$ is obtained by averaging the contributions of the $m_i$ members of unit cell $i$.
In \eqref{eq:constraintShear_02}, $y_{ij}$ denotes the distance from the midpoint of member $j$ in unit cell $i$ to the $x$-axis.
The displacement of this midpoint in the $x$-direction is denoted by $u_{x,ij}$.
A weighted axial stiffness $D_{ij}$ is introduced as
\begin{equation} \label{eq:constraintShear_03}
    D_{ij} := \frac{C_{\text{tg},ij}A_{ij}}{[\ell_{ij}]^3} \;,
\end{equation}
where $C_{\text{tg},ij}$, $A_{ij}$, and $\ell_{ij}$ denote the tangent modulus, cross-sectional area, and length of member $j$, respectively.
For linear elasticity with Young's modulus $E$, $C_{\text{tg},ij}= E$.

Substituting \eqref{eq:constraintShear_02} into \eqref{eq:constraintShear_01} and assuming that all unit cells have the same type and size, i.e.,
\begin{equation} \label{eq:constraintShear_04}
\begin{split}
    m_1=m_2=\ldots=m_m =: m \\
    V_1=V_2=\ldots=V_m =: V \\
\end{split}
\end{equation}
gives
\begin{equation} \label{eq:constraintShear_05}
    \sum_{i=1}^n \sum_{j=1}^{m} D_{ij} \, u_{x,ij} \, y_{ij} \, r_{ij} \overset{!}{=} 0 \;,
\end{equation}
where the factor $r_{ij}$ prevents multiple counting of members shared by neighboring unit cells.

For computational implementation, it is convenient to reformulate \eqref{eq:constraintShear_05} as a sum over members rather than unit cells, thereby excluding multiple counting a priori.
The 'discrete constant moment constraint' then becomes
\begin{equation} \label{eq:constraintShear_06}
    \sum_{e=1}^\text{numel}  D^{(e)} \, u_x^{(e)} \, y^{(e)} \overset{!}{=} 0 \;,
\end{equation}
with
\begin{equation} \label{eq:constraintShear_07}
\begin{split}
    D^{(e)} &= \frac{C_\text{tg}^{(e)}A^{(e)}}{[\ell^{(e)]^3}} \\
    u_x^{(e)} &= \frac{1}{2} (u_{x1}^{(e)} + u_{x2}^{(e)}) \\
    y^{(e)} &= \frac{1}{2} (y_1^{(e)} + y_2^{(e)}) \;,
\end{split}
\end{equation}
where '1' and '2' denote the start and end nodes of member $e$, respectively, and numel is the total number of members.

\paragraph{Eccentric RVE.}
For eccentric RVEs, i.e., $s_y\ne 0$ or $s_z\ne 0$, an additional constraint is required to ensure that the homogenized shear forces and shear stiffnesses are independent of the RVE length $\ell^\text{RVE}$ \cite{klarmann2020}.
Starting from the reduced '\textit{constant normal force constraint}' for continuum RVEs in the appendix, Eq. \eqref{eq:appendix_cnc_03}, and following the same reasoning as in Eqs. \eqref{eq:constraintShear_01}--\eqref{eq:constraintShear_07}, the '\textit{discrete constant normal force constraint}' for lattice-like RVEs is obtained as
\begin{equation} \label{eq:constraintShear_08}
    \sum_{e=1}^\text{numel} D^{(e)} \, u_x^{(e)} \overset{!}{=} 0 \;.
\end{equation}

\subsection{Remaining rigid body motions\label{sec:constraints_rigidBodyMotions}}
The macroscopic strains $\pmb\varepsilon^M$ are imposed on the RVE through \eqref{FE2_localization_07}.
Supports (Dirichlet boundary conditions) must suppress all six possible rigid-body motions: three translations and three rotations.
These supports can alternatively be formulated as constraints on the nodal degrees of freedom $\mathbf{V}$.
Following the form of \eqref{eq:constraintShear_06} and \eqref{eq:constraintShear_08} yields the following three constraints suppressing rigid-body translations:
\begin{equation} \label{eq:constraint_rbm_01}
    \sum_{e=1}^\text{numel} u_x^{(e)} \overset{!}{=} 0
    \;;\quad
    \sum_{e=1}^\text{numel} u_y^{(e)} \overset{!}{=} 0
    \;;\quad
    \sum_{e=1}^\text{numel} u_z^{(e)} \overset{!}{=} 0
\end{equation}
The rigid-body rotations about the $y$- and $z$-axes are already eliminated by the 'discrete constant moment constraints'; cf. \eqref{eq:constraintShear_06}.
The remaining rigid-body rotation about the $x$-axis is eliminated using
\begin{equation} \label{eq:constraint_rbm_02}
        \sum_{e=1}^\text{numel} u_z^{(e)} \, y^{(e)} - u_y^{(e)} \, z^{(e)} \overset{!}{=} 0
\end{equation}

\subsection{Summary and implementation aspects\label{sec:constraints_summary}}
Summarizing the constraints \eqref{eq:constraintShear_06}, \eqref{eq:constraintShear_08}, \eqref{eq:constraint_rbm_01}, and \eqref{eq:constraint_rbm_02} gives
\begin{equation} \label{eq:constraintSummary_01}
	\begin{bmatrix}
		c_1 \\
		c_2 \\
		c_3 \\
		c_4 \\
		c_5 \\
		c_6 \\
	\end{bmatrix}
	=
	\sum_{e=1}^\text{numel}
	\begin{bmatrix}
		D^{(e)} y^{(e)} & 0 & 0 \\[1mm]
		D^{(e)} z^{(e)} & 0 & 0 \\[1mm]
		D^{(e)}         & 0 & 0 \\[1mm]
		1 & 0 & 0 \\[1mm]
		0 & 1 & 0 \\[1mm]
		0 & 0 & 1 \\[1mm]
		0 & - z^{(e)} & y^{(e)}
	\end{bmatrix}
	\begin{bmatrix}
	u_x^{(e)} \\
	u_y^{(e)} \\
	u_z^{(e)} \\
\end{bmatrix}
	\overset{!}{=} 0
\end{equation}
or, in compact notation,
\begin{equation} \label{eq:constraintSummary_02}
	\mathbf{c}_2(\mathbf{V}) = \sum_{e=1}^\text{numel} \mathbf{B}^{(e)} \mathbf{u}^{(e)} = \mathbf{0} \; .
\end{equation}
Together with the constraints imposing the macroscopic strains $\pmb\varepsilon^M$ through \eqref{FE2_localization_07},
this yields a total of $6+3n_\Gamma$ constraints that depend linearly on the RVE nodal degrees of freedom $\mathbf{V}$
\begin{equation} \label{eq:constraintSummary_03}
\begin{bmatrix}
	\mathbf{c}_1(\mathbf{V}) \\
	\mathbf{c}_2(\mathbf{V}) \\
\end{bmatrix}
=
\begin{bmatrix}
	\mathbf{V}^+ - \mathbf{V}^- - \ell^\text{RVE} \mathbb{A} \, \pmb\varepsilon^M \\
	\sum\limits_{e=1}^\text{numel} \mathbf{B}^{(e)} \mathbf{u}^{(e)} \\
\end{bmatrix}
\overset{!}{=}
\begin{bmatrix}
	\mathbf{0} \\
	\mathbf{0} \\
\end{bmatrix} .
\end{equation}
Equation \eqref{eq:constraintSummary_03} and the nonlinear equations \eqref{eq:meso_04} fully define the RVE boundary value problem.

\subsection{Symmetry of RVE state variables}
When the constraints in Eq. \eqref{eq:constraintSummary_03} are imposed on the RVE, its state variables $s(x,y,z)$ have symmetric or antisymmetric distributions at equilibrium.
Symmetry and antisymmetry with respect to the coordinate directions are defined as follows:
\begin{equation}
    s(x,y,z) =
    \begin{cases}
        s(-x,y,z)   & \; \text{sym. w.r.t. $x$-axis} \\
        s(x,-y,z)   & \; \text{sym. w.r.t. $y$-axis} \\
        s(x,y,-z)   & \; \text{sym. w.r.t. $z$-axis} \\
        -s(-x,y,z)  & \; \text{anti. w.r.t. $x$-axis} \\
        -s(x,-y,z)  & \; \text{anti. w.r.t. $y$-axis} \\
        -s(x,y,-z)  & \; \text{anti. w.r.t. $z$-axis} \;.
    \end{cases}
\end{equation}
For clarity, the symmetry properties are presented only for two-dimensional RVEs.
Table \ref{tab:symmetry} considers the six RVE types classified in Figure \ref{fig:RVE_classification}, including both homogeneous and heterogeneous compositions.
The strain $\varepsilon(x,y)$ in a lattice member is selected as the state variable $s(x,y)$, with the three macroscopic strains $\varepsilon^M$, $\gamma^M$, and $\kappa^M$ considered separately.
\begin{table}
    \begin{subtable}{\linewidth}
        \centering
        \begin{subtable}{\linewidth}
            \centering
            \begin{tabular}{ccccc}
                \toprule
                         & RVE & $\varepsilon^M$ & $\gamma^M$  & $\kappa^M$  \\ \hline
                $\varepsilon(x,y)$ & AA  &   \texttt{SS}   & \texttt{AA} & \texttt{SA} \\
                         & AB1 &   \texttt{SS}   & \texttt{AA} & \texttt{SA} \\
                         & AB2 &   \texttt{SA}   & \texttt{AA} & \texttt{SA} \\
                         & BA  &   \texttt{SS}   & \texttt{AA} & \texttt{SA} \\
                         & BB1 &   \texttt{SS}   & \texttt{AA} & \texttt{SA} \\
                         & BB2 &   \texttt{SA}   & \texttt{AA} & \texttt{SA} \\ \bottomrule
            \end{tabular}
            \caption{lattice-like RVE}
            \label{tab:symmetry_3D}
        \end{subtable}
        \begin{tabular}{ccccc}
            \toprule
                                    & RVE & $\varepsilon^M$ & $\gamma^M$  & $\kappa^M$  \\ \hline
            $\varepsilon_{xx}(x,y)$ & AA  &   \texttt{SS}   & \texttt{0}  & \texttt{SA} \\
                                    & AB1 &   \texttt{SS}   & \texttt{0}  & \texttt{SA} \\
                                    & AB2 &   \texttt{SS}   & \texttt{0}  & \texttt{SA} \\
                                    & BA  &   \texttt{SS}   & \texttt{AA} & \texttt{SA} \\
                                    & BB1 &   \texttt{SS}   & \texttt{AA} & \texttt{SA} \\
                                    & BB2 &   \texttt{S-}   & \texttt{A-} & \texttt{S-} \\ \hline
            $\varepsilon_{yy}(x,y)$ & AA  &   \texttt{SS}   & \texttt{0}  & \texttt{SA} \\
                                    & AB1 &   \texttt{SS}   & \texttt{0}  & \texttt{SA} \\
                                    & AB2 &   \texttt{SS}   & \texttt{0}  & \texttt{SA} \\
                                    & BA  &   \texttt{SS}   & \texttt{AA} & \texttt{SA} \\
                                    & BB1 &   \texttt{SS}   & \texttt{AA} & \texttt{SA} \\
                                    & BB2 &   \texttt{S-}   & \texttt{A-} & \texttt{S-} \\ \hline
            $\varepsilon_{xy}(x,y)$ & AA  &   \texttt{0}    & \texttt{SS} & \texttt{AS} \\
                                    & AB1 &   \texttt{0}    & \texttt{SS} & \texttt{AS} \\
                                    & AB2 &   \texttt{0}    & \texttt{S-} & \texttt{AS} \\
                                    & BA  &   \texttt{AA}   & \texttt{SS} & \texttt{AS} \\
                                    & BB1 &   \texttt{AA}   & \texttt{SS} & \texttt{AS} \\
                                    & BB2 &   \texttt{AA}   & \texttt{S-} & \texttt{AA} \\ \bottomrule
        \end{tabular}
        \caption{continuous RVE}
        \label{tab:symmetry_2D}
    \end{subtable}
    \caption{Symmetry properties of the state variables for all two-dimensional RVE types shown in Figure \ref{fig:RVE_classification}. Properties are indicated separately for the $x$- and $y$-directions: 'S' denotes symmetry, 'A' antisymmetry, '0' a value that vanishes throughout the domain, and '-' the absence of a general symmetry statement}
    \label{tab:symmetry}
\end{table}
Table \ref{tab:symmetry} also lists the symmetry properties of continuum RVEs, using the three strain components as state variables.
Table \eqref{tab:symmetry} can also be extended to three-dimensional RVEs.
The tabulated properties assume geometric and material linearity.

\section{Numerical examples I (benchmark problems) \label{sec:numericalExamples_I}}
The proposed multiscale model can be straightforwardly implemented in finite element codes providing the standard interfaces required for FE$^2$ analyses. Here, the in-house research code dockSIM is used.

This section verifies the implementation of the proposed \fes\ model and validates the derived equations.
To examine localization and homogenization in isolation, only a single material point is considered at the macroscopic scale, without solving a separate macroscopic boundary value problem.
The method can thus be studied without discretization errors caused by the macro model.
In contrast, Section \ref{sec:numericalExamples_II} also considers macroscopic boundary value problems; Figure \ref{fig:numericalExample_twoStrategies} illustrates both cases.

In both cases, discretization errors cases on the meso scale are absent a priori because each mesostructural member (rectangular cross section of height $h$ and width $b$) can be represented exactly by a single finite truss element.

Dedicated benchmark tests were designed to verify and validate the method systematically.
The focus is on the effective material properties, i.e., the homogenized stress resultants $\vectorSym{\sigma}\macro$ and material tangent $\matrix{D}\macro$, for two- and three-dimensional lattice-like mesostructures.
Table \ref{tab:numericalExamples_parameters} specifies the geometric, topological, and material data for all mesostructures investigated in Sections \ref{sec:numericalExamples_I} and \ref{sec:numericalExamples_II}.
\begin{table*}
    \begin{subtable}{\textwidth}
        \centering
        \begin{tabular}{ccccccccccc}
            \toprule
                                                                                              \multicolumn{2}{c}{type}                                                                                    & \multicolumn{2}{c}{parameters} &                                                \multicolumn{5}{c}{section \ref{sec:numericalExamples_I}}                                                 &   \multicolumn{2}{c}{section \ref{sec:numericalExamples_II}}    \\ \cmidrule
                         (lr){1-2}  \cmidrule(lr){3-4}\cmidrule(lr){5-9}\cmidrule(lr){10-11}               &                                                                                              &               &                &  \ref{sec:Rve_length}  & \ref{sec:heterogeneousRve} & \ref{sec:GeometricNonlinearity} & \ref{sec:MaterialNonlinearity} & \ref{sec:c_shaped_crossSection} & \ref{sec:SimplySupportedBeam} & \ref{sec:SpatialFrameStructure} \\ \midrule
            \multirow{5}{*}{\parbox[c][1.75cm][c]{1em}{\centering\rotatebox[origin=c]{90}{unit cell '1'}}} & \multirow{5}{*}{\includegraphics{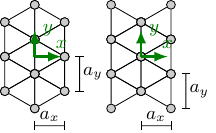}} &     $a_x$     &  \sUnitLength  & $\tfrac{1}{2}\sqrt{3}$ &                            &                                 &                                &                                 &    $\tfrac{1}{4}\sqrt{3}$     &                                 \\
                                                                                                           &                                                                                              &     $a_y$     &  \sUnitLength  &       \num{1.0}        &                            &                                 &                                &                                 &           \num{0.5}           &                                 \\
                                                                                                           &                                                                                              &     $s_y$     &  \sUnitLength  &       \num{2.0}        &                            &                                 &                                &                                 &           \num{0.0}           &                                 \\
                                                                                                           &                                                                                              &     $n_x$     &   \sUnitNone   &          var.          &                            &                                 &                                &                                 &            \num{2}            &                                 \\
                                                                                                           &                                                                                              &     $n_y$     &   \sUnitNone   &        \num{5}         &                            &                                 &                                &                                 &           \num{26}            &                                 \\ \midrule
            \multirow{4}{*}{\parbox[c][1.50cm][c]{1em}{\centering\rotatebox[origin=c]{90}{unit cell '2'}}} & \multirow{4}{*}{\includegraphics{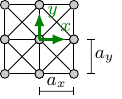}} &   $a_x,a_y$   &  \sUnitLength  &       \num{1.0}        &         \num{16.0}         &                                 &                                &                                 &           \num{0.5}           &                                 \\
                                                                                                           &                                                                                              &     $s_y$     &  \sUnitLength  &       \num{0.0}        &         \num{0.0}          &                                 &                                &                                 &           \num{0.0}           &                                 \\
                                                                                                           &                                                                                              &     $n_x$     &   \sUnitNone   &          var.          &            var.            &                                 &                                &                                 &            \num{2}            &                                 \\
                                                                                                           &                                                                                              &     $n_y$     &   \sUnitNone   &        \num{4}         &          \num{1}           &                                 &                                &                                 &           \num{16}            &                                 \\ \midrule
            \multirow{4}{*}{\parbox[c][1.50cm][c]{1em}{\centering\rotatebox[origin=c]{90}{unit cell '3'}}} & \multirow{4}{*}{\includegraphics{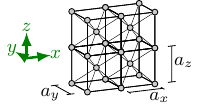}} & $a_x,a_y,a_z$ &  \sUnitLength  &       \num{1.0}        &                            &                                 &                                &                                 &                               &            \num{0.5}            \\
                                                                                                           &                                                                                              &   $s_y,s_z$   &  \sUnitLength  &       \num{0.0}        &                            &                                 &                                &                                 &                               &            \num{0.0}            \\
                                                                                                           &                                                                                              &     $n_x$     &   \sUnitNone   &          var.          &                            &                                 &                                &                                 &                               &             \num{1}             \\
                                                                                                           &                                                                                              &   $n_y,n_z$   &   \sUnitNone   &        \num{4}         &                            &                                 &                                &                                 &                               &            \num{12}             \\ \midrule
            \multirow{4}{*}{\parbox[c][1.50cm][c]{1em}{\centering\rotatebox[origin=c]{90}{unit cell '4'}}} & \multirow{4}{*}{\includegraphics{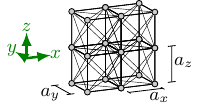}} & $a_x,a_y,a_z$ &  \sUnitLength  &       \num{1.0}        &                            &            \num{1.0}            &           \num{1.0}            &            \num{1.0}            &                               &            \num{0.5}            \\
                                                                                                           &                                                                                              &     $s_y$     &  \sUnitLength  &       \num{2.0}        &                            &            \num{0.0}            &           \num{0.0}            &          \num{23.271}           &                               &            \num{0.0}            \\
                                                                                                           &                                                                                              &     $s_z$     &  \sUnitLength  &       \num{2.0}        &                            &            \num{0.0}            &           \num{0.0}            &           \num{36.0}            &                               &            \num{0.0}            \\
                                                                                                           &                                                                                              &     $n_x$     &   \sUnitNone   &          var.          &                            &              var.               &              var.              &              var.               &                               &             \num{1}             \\
                                                                                                           &                                                                                              &   $n_y,n_z$   &   \sUnitNone   &        \num{4}         &                            &             \num{4}             &            \num{4}             &              var.               &                               &            \num{12}             \\ \bottomrule
        \end{tabular}
        \caption{Geometry parameter of the lattice structure (with $n_i = \{n_x,\,n_y,\,n_z\}$)}
        \label{tab:numericalResultsI_geometryParameters}
    \end{subtable}

    \vspace{3ex}

    \begin{subtable}{\textwidth}
        \centering
        \begin{tabular}{clcccc}
            \toprule
            \multicolumn{2}{c}{parameters}      &                                              \multicolumn{2}{c}{section \ref{sec:numericalExamples_I}}                                              &   \multicolumn{2}{c}{section \ref{sec:numericalExamples_II}}    \\ \cmidrule(lr){1-2} \cmidrule(lr){3-4} \cmidrule(lr){5-6}
            &                            & \ref{sec:Rve_length}, \ref{sec:heterogeneousRve}, \ref{sec:GeometricNonlinearity}, \ref{sec:c_shaped_crossSection} & \ref{sec:MaterialNonlinearity} & \ref{sec:SimplySupportedBeam} & \ref{sec:SpatialFrameStructure} \\ \midrule
            $E$     & \sUnitForcePerLengthLength &                                                    \num{21000}                                                     &          \num{21000}           &          \num{21000}          &           \num{21000}           \\
            $H$     & \sUnitForcePerLengthLength &                                                                                                                    &           \num{500}            &          \num{10500}          &                                 \\
            $\sigma_y$ & \sUnitForcePerLengthLength &                                                                                                                    &            \num{30}            &           \num{10}            &                                 \\
            $b$     & \sUnitLength               &                                                     \num{0.1}                                                      &           \num{0.1}            &          \num{0.15}           &           \num{0.15}            \\
            $h$     & \sUnitLength               &                                                     \num{0.1}                                                      &           \num{0.1}            &          \num{0.15}           &           \num{0.15}            \\ \bottomrule
        \end{tabular}
        \caption{Material and geometry parameters of the trusses (rectangular cross-section)}
        \label{tab:numericalResultsI_materialParameters}
    \end{subtable}
    \caption{Material and geometric parameters for all numerical examples presented in Sections \ref{sec:numericalExamples_I} and \ref{sec:numericalExamples_II} }
    \label{tab:numericalExamples_parameters}
\end{table*}
\paragraph{Reference solution ('full-scale model')}
Since analytical solutions are unavailable for the boundary value problems considered, additional analyses of fully resolved mesostructures, referred to as 'full-scale models' (or 'Ref'), were performed for each benchmark described below and used as reference solutions.
The relative error is defined as $e:=(\bullet^\text{FE$^2$}-\bullet^\text{Ref})/\bullet^\text{Ref}$.
Accurately representing the different boundary conditions of the full-scale models is challenging, particularly when determining the shear and torsional stiffnesses of a lattice-like mesostructure.
Although these reference solutions involve minor uncertainties for some mesostructures, they provide an important basis for validating the proposed method.
A detailed description of the reference models is beyond the scope of this article and is intended for a separate publication.

\subsection{Homogeneous RVE} \label{sec:Rve_length}
This example examines whether the homogenized beam stiffnesses obtained with the proposed \fes\ method, i.e., the entries of $\matrix{D}\macro$, are independent of the selected RVE length $\ell\rve$.
It also demonstrates that omitting the axial stiffness modification introduced in Section \ref{MesoLevel} for truss members on the model boundaries $\Gamma^\pm$ produces artificial boundary effects and incorrect, length-dependent homogenized beam stiffnesses.

\begin{figure}
    \centering
    \begin{subfigure}{0.4\columnwidth}
        \centering
        \includegraphics{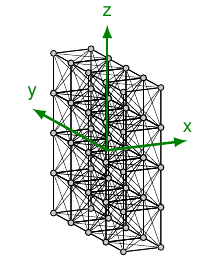}
        \caption{$\ncx=1$}
    \end{subfigure}
    \hfill
    \begin{subfigure}{0.575\columnwidth}
        \centering
        \includegraphics{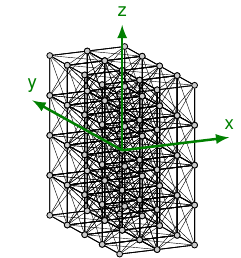}
        \caption{$\ncx=2$}
    \end{subfigure}

    \begin{subfigure}{\columnwidth}
        \centering
        \includegraphics{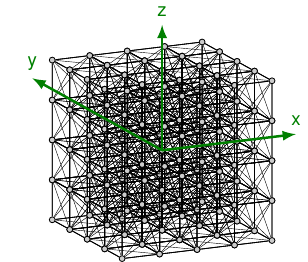}
        \caption{$\ncx=4$}
    \end{subfigure}
    \caption{Homogeneous RVE: Illustration of different RVEs constructed with unit cells of type ‘4’. These differ in their lengths $n_x$ (with $\ell\rve=n_x\, a_x$), while the cross-sectional dimensions $\ncy=\ncz=4$ remain constant}
    \label{fig:numericalExamples:I_lengthIndependency_rve}
\end{figure}

Two two-dimensional and two three-dimensional mesostructures are investigated, with one eccentric case in each dimension; see Table \ref{tab:numericalExamples_parameters}.
Each member of each RVE is discretized using a geometrically linear finite truss element.
Linear elastic behavior is assumed.
The two-dimensional RVEs contain five type '1' or four type '2' unit cells in the $y$-direction, i.e., $n_y=5$ or 4, respectively.
Both three-dimensional RVEs contain four unit cells in each transverse direction, i.e., $n_y=n_z=4$.
The number $n_x$ of unit cells in the $x$-direction is varied for all RVEs.
Figure \ref{fig:numericalExamples:I_lengthIndependency_rve} shows three examples of RVEs consisting of type '4' unit cells.

\begin{figure}
    \centering
    \begin{subfigure}[t]{0.49\columnwidth}
        \centering
        \includegraphics{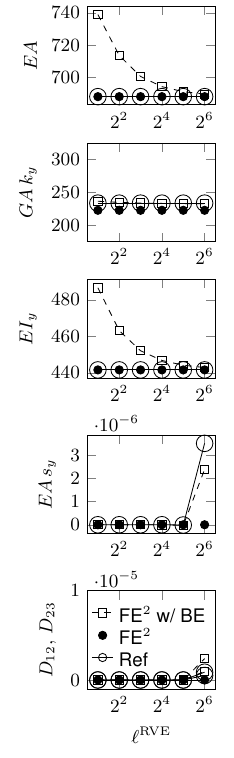}
        \caption{unit cell of type '1'}
        \label{fig:numericalResultsI_2d_centric_h2P-1_DiiOverLen}
    \end{subfigure}
    \hfill
    \begin{subfigure}[t]{0.49\columnwidth}
        \centering
        \includegraphics{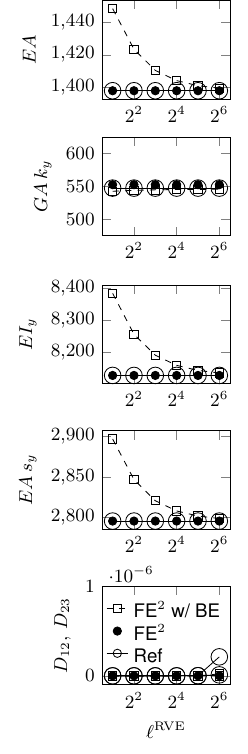}
        \caption{unit cell of type '2'}
        \label{fig:numericalResultsI_2d_centric_o2P-1_DiiOverLen}
    \end{subfigure}
    \caption{Homogeneous RVE (2-dimensional lattices): Independence of the homogenized beam stiffnesses from the length $\ell\rve$ of the RVE}
    \label{fig:numericalResultsI_2d_centric_o2P-1_h2P-1_DiiOverLen}
\end{figure}

\begin{figure*}
    \centering
    \begin{subfigure}{0.49\textwidth}
        \centering
        \includegraphics{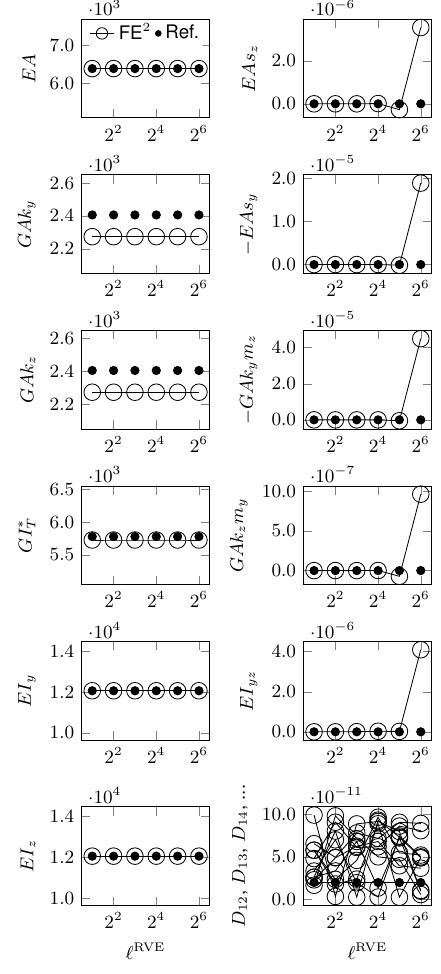}
        \caption{unit cell of type '3'}
        \label{fig:numericalResultsI_3d_o3I-1_DiiOverLen}
    \end{subfigure}
    \hfill
    \begin{subfigure}{0.49\textwidth}
        \centering
        \includegraphics{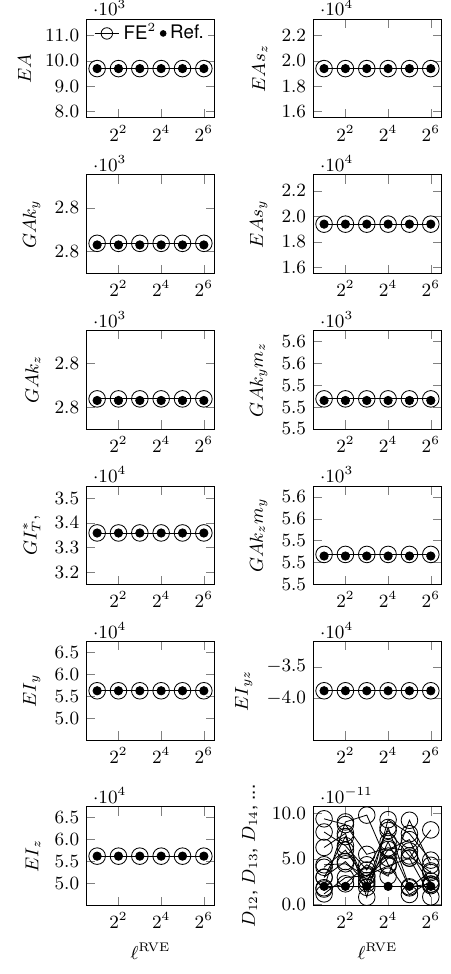}
        \caption{unit cell of type '4'}
        \label{fig:numericalResultsI_3d_o3I-P_DiiOverLen}
    \end{subfigure}
    \caption{Homogeneous RVE (3-dimensional lattices): Independence of the homogenized beam stiffnesses from the length $\ell\rve$ of the selected RVE}
    \label{fig:numericalResultsI_3d_o3P-1_o3I-1_DiiOverLen}
\end{figure*}

The numerical results are presented in Figures \ref{fig:numericalResultsI_2d_centric_o2P-1_h2P-1_DiiOverLen} and \ref{fig:numericalResultsI_3d_o3P-1_o3I-1_DiiOverLen} and Table \ref{tab:numericalResultsI_error}.
\begin{table*}
    \centering
    \begin{subtable}{\linewidth}
            \centering
            \begin{tabular}{p{0.4cm}ccccccc}
                \toprule
                         &             &        \multicolumn{3}{c}{unit cell type '1'}        &        \multicolumn{3}{c}{unit cell type '2'}        \\ \cmidrule(lr){3-5} \cmidrule(lr){6-8}
                         &             &   \fes    &   Ref.    & error $e$ &   \fes    &   Ref.    & error $e$ \\ \midrule
                $D_{11}$ &    $EA$     & 6.883e+02 & 6.883e+02 &    1.000e-08 \%    & 1.398e+03 & 1.398e+03 &    1.000e-08 \%    \\
                $D_{13}$ &  $EA\,s_y$  & 3.281e-13 & 1.000e-12 &     -6.719e-13     & 2.796e+03 & 2.796e+03 &    1.000e-08 \%    \\
                $D_{33}$ &  $EI_{z}$   & 4.417e+02 & 4.417e+02 &    1.000e-08 \%    & 8.127e+03 & 8.127e+03 &    1.000e-08 \%    \\
                $D_{22}$ & $GA\,k_{y}$ & 2.334e+02 & 2.347e+02 &    5.715e-01 \%    & 5.472e+02 & 5.525e+02 &    9.686e-01 \%    \\ \midrule
                         &    $s_y$    & 4.766e-16 & 0.000e+00 &         -          & 2.000e+00 & 2.000e+00 &         -          \\ \bottomrule
            \end{tabular}
            \caption{2-dimensional lattices}
            \label{tab:numericalResultsI_2d_h2P-1_o2P-1_error}
    \end{subtable}

    \vspace{3ex}

    \begin{subtable}{\linewidth}
            \centering
            \begin{tabular}{p{0.4cm}ccccccc}
                \toprule
                         &                 &        \multicolumn{3}{c}{unit cell type '3'}        &         \multicolumn{3}{c}{unit cell type '4'}         \\ \cmidrule(lr){3-5} \cmidrule(lr){6-8}
                         &                 &   \fes    &   Ref.    & error $e$ &    \fes    &    Ref.    & error $e$ \\ \midrule
                $D_{11}$ &      $EA$       & 6.391e+03 & 6.391e+03 &   1.000e-08  \%    & 9.695e+03  & 9.695e+03  &    1.000e-08 \%    \\
                $D_{15}$ &    $EA\,s_z$    & 4.914e-11 & 1.000e-12 &   1.000e-08  \%    & 1.939e+04  & 1.939e+04  &    1.000e-08 \%    \\
                $D_{16}$ &   $-EA\,s_y$    & 2.899e-11 & 1.000e-12 &   1.000e-08  \%    & 1.939e+04  & 1.939e+04  &    1.000e-08 \%    \\
                $D_{55}$ &    $EI_{y}$     & 1.207e+04 & 1.207e+04 &   1.000e-08  \%    & 5.622e+04  & 5.622e+04  &    1.000e-08 \%    \\
                $D_{66}$ &    $EI_{z}$     & 1.207e+04 & 1.207e+04 &   1.000e-08  \%    & 5.622e+04  & 5.622e+04  &    1.000e-08 \%    \\
                $D_{56}$ &   $-EI_{yz}$    & 3.268e-12 & 1.000e-12 &    1.000e-08 \%    & -3.878e+04 & -3.878e+04 &    1.000e-08 \%    \\
                $D_{22}$ &   $GA\,k_{y}$   & 2.274e+03 & 2.406e+03 &    5.490e-00 \%    & 2.764e+03  & 2.763e+03  &    4.101e-03 \%    \\
                $D_{33}$ &   $GA\,k_{z}$   & 2.274e+03 & 2.406e+03 &    5.490e-00 \%    & 2.764e+03  & 2.763e+03  &    4.101e-03 \%    \\
                $D_{23}$ &  $GA\,k_{yz}$   & 5.782e-07 & 1.000e-12 &     5.782e-07      & 2.124e-06  & 0.000e+00  &    1.000e-08 \%    \\
                $D_{24}$ & $-GA\,k_y\,m_z$ & 3.126e-10 & 1.000e-12 &     3.116e-10      & 5.528e+03  & 5.526e+03  &    4.101e-03 \%    \\
                $D_{34}$ & $GA\,k_z\,m_y$  & 5.263e-11 & 1.000e-12 &     5.163e-11      & 5.528e+03  & 5.526e+03  &    4.101e-03 \%    \\
                $D_{44}$ &   $GI_{T}^*$    & 5.727e+03 & 5.779e+04 &    8.920e-01 \%    & 3.359e+04  & 3.359e+04  &    5.955e-04 \%    \\ \midrule
                         &      $s_y$      & 7.688e-15 & 1.000e-12 &         -          & 2.000e+00  & 2.000e+00  &         -          \\
                         &      $s_z$      & 4.536e-15 & 1.000e-12 &         -          & 2.000e+00  & 2.000e+00  &         -          \\
                         &      $m_y$      & 1.374e-13 & 1.000e-12 &         -          & 2.000e+00  & 2.000e+00  &         -          \\
                         &      $m_z$      & 2.314e-14 & 1.000e-12 &         -          & 2.000e+00  & 2.000e+00  &         -          \\ \bottomrule
            \end{tabular}
            \caption{3-dimensional lattices}
            \label{tab:numericalResultsI_3d_o3I-1_o3P-1_error}
    \end{subtable}

    \caption{Homogeneous RVE: Verification of homogenized beam stiffnesses by comparison with reference and analytical solutions (in case the reference value is zero, the absolute error is reported instead of the relative error)}
    \label{tab:numericalResultsI_error}
\end{table*}
The figures show all nine, respectively 36, components of $\matrix{D}\macro$ as functions of $\ell\rve$, with components having a numerical value of zero combined into a single diagram.
The component names follow the quantities introduced in Eqs. \eqref{homogenization_10} and \eqref{homogenization_D_structure_01}.
For all four lattice-like mesostructures, the components are independent of $\ell^\text{RVE}$.
Figure \ref{fig:numericalResultsI_2d_centric_o2P-1_h2P-1_DiiOverLen} also shows a strong length dependence when the modification to the axial stiffness introduced in Section \ref{MesoLevel} is omitted, with relative errors of up to 7\% (for $EA$) occurring at small $\ell^\text{RVE}$.

\subsection{Heterogeneous RVE} \label{sec:heterogeneousRve}
\begin{figure}
    \centering
    \includegraphics{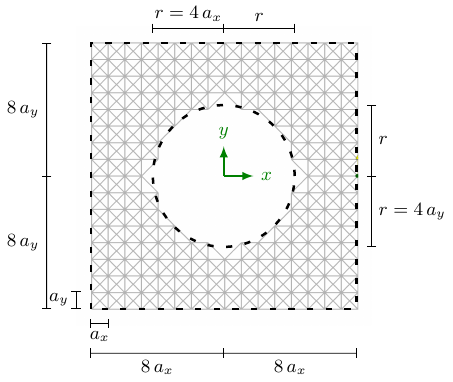}
    \caption{Heterogeneous RVE: Illustration of the 'super unit cell' of the microstructure}
    \label{fig:numericalResultsI_heterogen_rve}
\end{figure}

This example demonstrates that the constraints introduced in Section \ref{sec:constraints_shear} yield correct values for the homogenized quantities also for heterogeneous mesostructures.
A 'super unit cell' composed of $16\times 16$ type '2' unit cells, with dimensions $16 \, a_x \times 16 \, a_x$, is considered.
Its architecture is interrupted by a circular hole of radius $r=4 \, a_x$; see Figure \ref{fig:numericalResultsI_heterogen_rve}.
The two-dimensional RVE is constructed from this super unit cell, varying the number $n_x$ of super unit cells in the $x$-direction while setting $n_y=1$.
This allows the independence of the homogenized beam stiffnesses from the selected RVE length $\ell\rve=16\, a_x\,n_x$ to be examined.
Each RVE member is discretized using a geometrically linear finite truss element.
Linear elastic behavior is assumed.
The model parameters are listed in Table \ref{tab:numericalExamples_parameters}.

\begin{figure}
    \centering
    \includegraphics{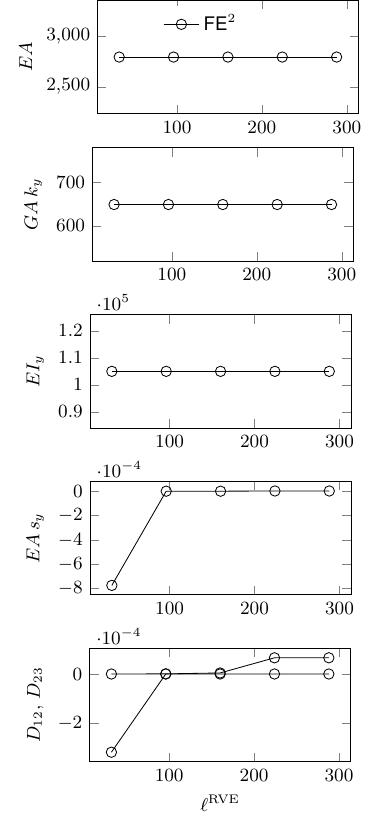}
    \caption{Heterogeneous RVE: Independence of the homogenized beam stiffnesses from the length $\ell\rve$ of the selected RVE}
    \label{fig:numericalResultsI_2d_heterogen_o2P-1_DiiOverLen}
\end{figure}

\begin{figure}
    \centering
    \begin{subfigure}{\columnwidth}
        \centering
        \includegraphics[width=\columnwidth]{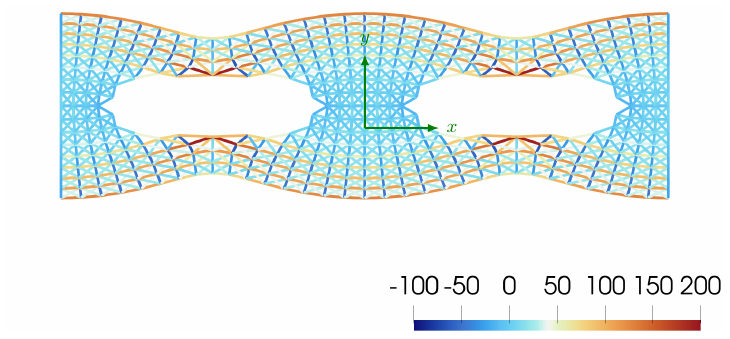}
        \caption{load case $\varepsilon$}
        \label{fig:numericalResultsI_heterogen_deformedMesh_eps}
    \end{subfigure}

    \begin{subfigure}{\columnwidth}
        \centering
        \includegraphics[width=\columnwidth]{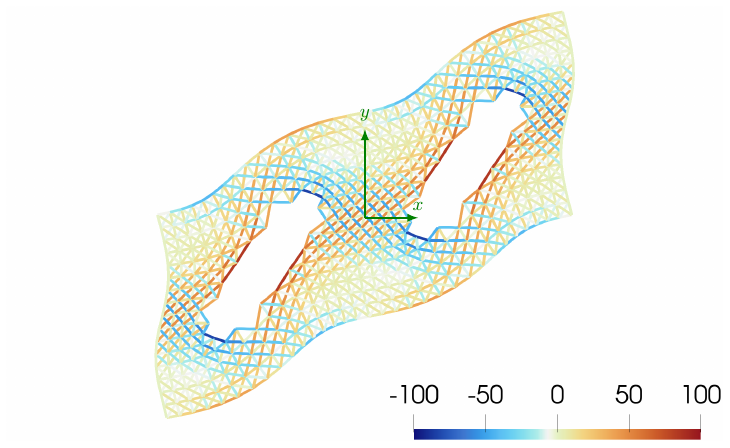}
        \caption{load case $\gamma$}
        \label{fig:numericalResultsI_heterogen_deformedMesh_gam}
    \end{subfigure}

    \begin{subfigure}{\columnwidth}
        \centering
        \includegraphics[width=\columnwidth]{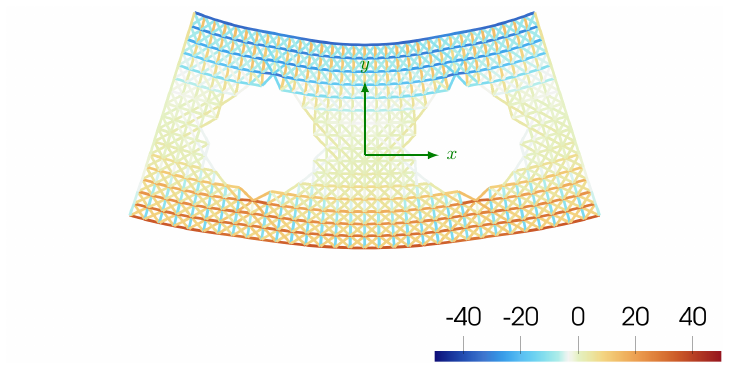}
        \caption{load case $\kappa$}
        \label{fig:numericalResultsI_heterogen_deformedMesh_kap}
    \end{subfigure}

    \caption{Heterogeneous RVE: Deformed RVE (magnified by a factor of 10) and (in color) normal force $N$ for each individual truss member}
    \label{fig:numericalResultsI_heterogen_deformedMesh}
\end{figure}

The numerical results are shown in Figures \ref{fig:numericalResultsI_2d_heterogen_o2P-1_DiiOverLen} and \ref{fig:numericalResultsI_heterogen_deformedMesh}.
The plots show all nine components of $\matrix{D}\macro$ as functions of $\ell\rve$, with components having a numerical value of zero combined into a single diagram.
The names and units of the components refer to the quantities listed in \eqref{homogenization_D_structure_05}.
For this heterogeneous mesostructure, the components are independent of $\ell\rve$.
Figure \ref{fig:numericalResultsI_heterogen_deformedMesh} also shows the deformed RVE for $n_x=2$ under the three load cases listed in Table \ref{tab:numericalResultsI_heterogen_loadCase}.
\begin{table}[]
    \begin{tabular}{ccccc}
        \toprule
                                        &                &    \multicolumn{3}{c}{load case}    \\ \cmidrule(lr){3-5}
        $\vectorSym{\varepsilon}\macro$ &      unit      & $\varepsilon$ & $\gamma$ & $\kappa$ \\ \midrule
                 $\varepsilon$          &   \unitNone    &     0.05      &    0     &    0     \\
                  $\gamma_y$            &   \unitNone    &       0       &   0.05   &    0     \\
                 $\kappa_{z}$           & \unitPerLength &       0       &    0     &  0.002   \\ \bottomrule
    \end{tabular}
    \caption{Heterogeneous RVE: Load cases considered}
    \label{tab:numericalResultsI_heterogen_loadCase}
\end{table}
The loading is applied in a single step.
For all load cases, the deformations and axial forces $N$ are symmetric or antisymmetric with respect to the $y$-axis.
Under axial strain $\varepsilon$ and simple shear $\gamma$, pronounced narrowing occurs around the holes.
The largest member axial forces also occur in these regions.
Under pure bending $\kappa$, the largest axial forces occur at the physical boundaries of the RVE.

\subsection{Geometric nonlinearity} \label{sec:GeometricNonlinearity}
This example examines the influence of geometric nonlinearity at the mesoscopic scale on the components of the homogenized material matrix $\matrix{D}\macro$.
It also considers the parasitic effects induced as the RVE length $\ell^\text{RVE}$ increases.
These effects were discussed in detail in Section \ref{sec:localication_limits}.

A centric lattice-like mesostructure composed of type '4' unit cells is considered.
The RVE contains four unit cells in each transverse direction, i.e., $n_y=n_z=4$.
The number $n_x$ in the $x$-direction is varied to study the dependence of the solution on $\ell^\text{RVE}$.
Each RVE member is discretized using a geometrically nonlinear finite truss element.
Linear elastic behavior is assumed.
The geometric and material parameters are listed in Table \ref{tab:numericalExamples_parameters}.

The load cases in Table \ref{tab:numericalResultsI_geometricNonlinear_loadCase} are considered.
Each loading is applied in ten steps.
Quadratic convergence was observed at every load step, requiring at most four iterations.
\begin{table}
    \centering
        \begin{tabular}{cccccc}
            \toprule
                                                                  &                &           \multicolumn{4}{c}{load case}           \\ \cmidrule
            (lr){3-6}
            $\vectorSym{\varepsilon}\macro$ &      unit      & $\varepsilon$ & $\gamma$ & $\vartheta$ & $\kappa$ \\ \midrule
                                $\varepsilon$                     &   \unitNone    &     0.02      &    0     &      0      &    0     \\
                                 $\gamma_y$                       &   \unitNone    &       0       &   0.02   &      0      &    0     \\
                                 $\gamma_z$                       &   \unitNone    &       0       &    0     &      0      &    0     \\
                                 $\vartheta$                      & \unitPerLength &       0       &    0     &    0.06     &    0     \\
                                $\kappa_{y}$                      & \unitPerLength &       0       &    0     &      0      &   0.02   \\
                                $\kappa_{z}$                      & \unitPerLength &       0       &    0     &      0      &    0     \\ \bottomrule
        \end{tabular}
        \caption{Geometric nonlinearity: Load cases considered}
        \label{tab:numericalResultsI_geometricNonlinear_loadCase}
\end{table}

\begin{figure*}
    \centering
    \begin{subfigure}{\textwidth}
        \centering
        \includegraphics{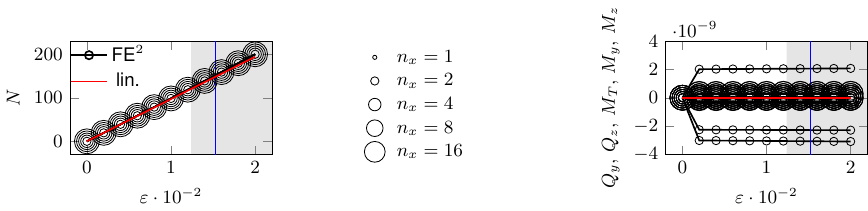}
        \caption{load case $\varepsilon$}
        \label{fig:numericalResultsI_geometricNonlinear_simulation1}
    \end{subfigure}
    \begin{subfigure}{\textwidth}
        \centering
        \includegraphics{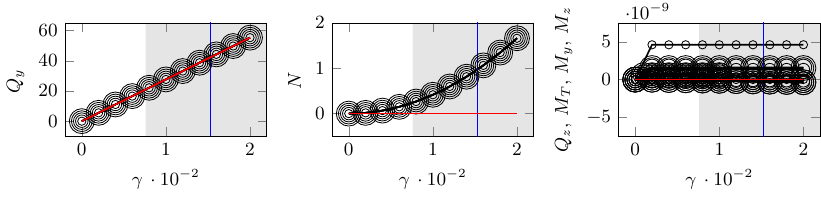}
        \caption{load case $\gamma$}
        \label{fig:numericalResultsI_geometricNonlinear_simulation2}
    \end{subfigure}
    \begin{subfigure}{\textwidth}
        \centering
        \includegraphics{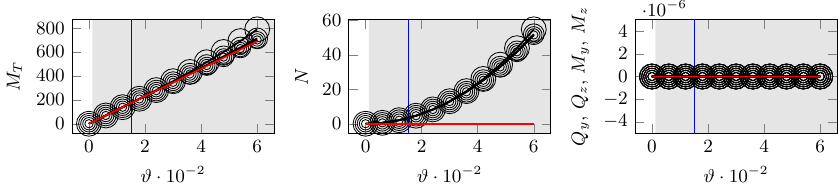}
        \caption{load case $\vartheta$}
        \label{fig:numericalResultsI_geometricNonlinear_simulation3}
    \end{subfigure}
    \begin{subfigure}{\textwidth}
        \centering
        \includegraphics{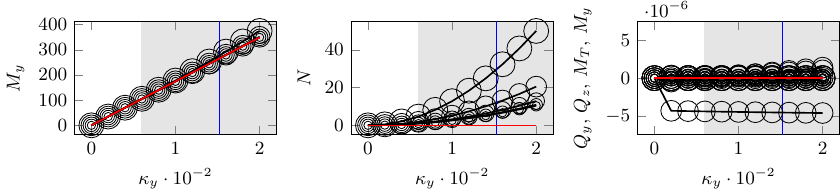}
        \caption{load case $\kappa$}
        \label{fig:numericalResultsI_geometricNonlinear_simulation4}
    \end{subfigure}
    \caption{Geometric nonlinearity: Dependence of the homogenized beam stress resultants on the applied beam strains}
    \label{fig:numericalResultsI_geometricNonlinear}
\end{figure*}

Figure \ref{fig:numericalResultsI_geometricNonlinear} presents the results as stress resultant--strain resultant curves, together with the geometrically linear solution for comparison.
The first column plots the work-conjugate stress resultant against the imposed beam strain.
The second column shows the parasitic axial force $N$.
Even for a centric RVE, i.e., $s_y=0$, this force introduces the additional coupling terms in $\matrix{D}\macro$ shown in \eqref{homogenization_D_structure_nili_01}.
For example, load case $\gamma$ produces coupling between the shear and axial forces.
Load case $\varepsilon$ is an exception, since no parasitic axial force is induced a priori.
The third column shows the remaining stress resultants, all of which vanish to numerical precision.

For load cases $\varepsilon$ and $\gamma$, the first column shows a linear relation between the imposed strain and its work-conjugate stress resultant, independent of the selected RVE length at every load step.
The behavior differs for load cases $\vartheta$ and $\kappa$.
In these cases, a linear response is observed only for short RVEs.
A weak nonlinearity develops as the RVE length increases.
This results in an artificial dependence on the RVE length.

The second column shows that the parasitic axial force $N$ increases exponentially with loading.
For load case $\gamma$, its magnitude is independent of the selected RVE length.
For the load cases $\vartheta$ and $\kappa$, a weak and strong dependence on length, respectively, can be observed.
For load case $\gamma$, the induced shear force $Q_y$ exceeds the parasitic axial force by more than one order of magnitude.

All plots in Figure \ref{fig:numericalResultsI_geometricNonlinear} contain a gray shaded region.
At strains within this region, the compressive axial force $N$ in at least one RVE member exceeds its theoretical Euler buckling load for a member pinned at both ends (case II).

A vertical blue line is also shown for the longest RVE, i.e., $n_x=16$\footnote{For clarity, the lines for the other RVE lengths are omitted.}.
It indicates the strain at which $\mathbf{A}_i$ ceases to be valid according to \eqref{FE2_localization_limit_02}.

For all load cases investigated, structural instability is the governing limitation.
These observations confirm the discussion in Section \ref{sec:localication_limits}.

\subsection{Material nonlinearity} \label{sec:MaterialNonlinearity}
This example demonstrates that the homogenized stress resultants $\vectorSym{\sigma}\macro$ remain independent of the RVE length $\ell\rve$ throughout the entire deformation process, even when material nonlinearities are considered.
It also shows that if the RVE is loaded only by a single component of the vector $\vectorSym{\varepsilon}\macro$, then only the work-conjugate component of the vector $\vectorSym{\sigma}\macro$ is induced there.

A centric lattice-like mesostructure, i.e., $s_y=s_z=0$, composed of type '4' unit cells is considered.
The RVE contains four unit cells in the $y$- and $z$-directions.
The number $n_x$ in the $x$-direction is varied.
Each RVE member is discretized using a geometrically linear finite truss element.
Elastoplastic material behavior with linear isotropic hardening is assumed.
The geometric and material parameters are listed in Table~\ref{tab:numericalExamples_parameters}.
The four load cases in Table \ref{tab:numericalResultsI_physicalNonlinear_loadCase} are considered.
The loading is increased in six equally spaced steps and subsequently reduced to zero.
Quadratic convergence was observed at every load step, requiring at most five iterations.
\begin{table}[]
    \begin{tabular}{cccccc}
        \toprule
                                        &                &           \multicolumn{4}{c}{load case}           \\ \cmidrule(lr){3-6}
        $\vectorSym{\varepsilon}\macro$ &      unit      & $\varepsilon$ & $\gamma$ & $\vartheta$ & $\kappa$ \\ \midrule
                 $\varepsilon$          &   \unitNone    &     0.005     &    0     &      0      &    0     \\
                  $\gamma_y$            &   \unitNone    &       0       &  0.005   &      0      &    0     \\
                  $\gamma_z$            &   \unitNone    &       0       &    0     &      0      &    0     \\
                  $\vartheta$           & \unitPerLength &       0       &    0     &    0.005    &    0     \\
                 $\kappa_{y}$           & \unitPerLength &       0       &    0     &      0      &  0.005   \\
                 $\kappa_{z}$           & \unitPerLength &       0       &    0     &      0      &    0     \\ \bottomrule
    \end{tabular}

    \caption{Material nonlinearity: load cases considered}
    \label{tab:numericalResultsI_physicalNonlinear_loadCase}
\end{table}

\begin{figure}

    \begin{subfigure}{\columnwidth}
        \centering
        \includegraphics{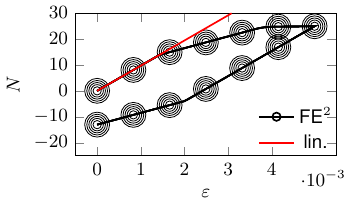}
        \caption{Load case $\varepsilon$}
        \label{fig:numericalResultsI_physicalNonlinear_simulation1}
    \end{subfigure}

    \vspace{3ex}
    \begin{subfigure}{\columnwidth}
        \centering
        \includegraphics{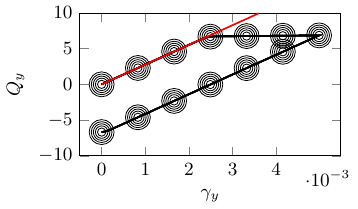}
        \caption{Load case $\gamma$}
        \label{fig:numericalResultsI_physicalNonlinear_simulation2}
    \end{subfigure}

    \vspace{3ex}
    \begin{subfigure}{\columnwidth}
        \centering
        \includegraphics{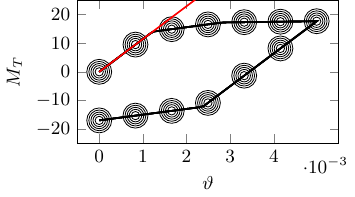}
        \caption{Load case $\vartheta$}
        \label{fig:numericalResultsI_physicalNonlinear_simulation3}
    \end{subfigure}

    \vspace{3ex}
    \begin{subfigure}{\columnwidth}
        \centering
        \includegraphics{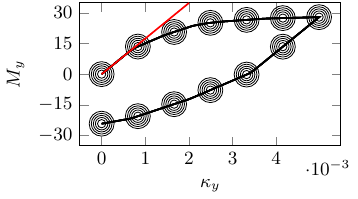}
        \caption{Load case $\kappa$}
        \label{fig:numericalResultsI_physicalNonlinear_simulation4}
    \end{subfigure}

    \caption{Material nonlinearity: Dependence of the homogenized beam stress resultants on the imposed beam strains for elastoplastic material behavior (see Figure \ref{fig:numericalResultsI_geometricNonlinear} for the symbol key)}
    \label{fig:numericalResultsI_physicalNonlinear}
\end{figure}

Figure \ref{fig:numericalResultsI_physicalNonlinear} presents the results as stress resultant--strain resultant curves, together with the materially linear solution for comparison.
Each plot shows the work-conjugate stress resultant as a function of the imposed strain resultant.
All remaining stress resultants are numerically zero and are omitted.
Accordingly, $\matrix{D}\macro$ in \eqref{homogenization_D_structure_01} contains no coupling terms.
The plots also show that $\vectorSym{\sigma}\macro$ is independent of $\ell\rve$ at every load step.

For load case $\varepsilon$, the curve exhibits three sharp changes in slope\footnote{An additional analysis with more load steps was performed to resolve these changes in the plots.}.
The first occurs when all members aligned with the \x-direction yield simultaneously.
The second occurs when all diagonal members in the \ebene{x}{y} and \ebene{x}{z} yield simultaneously.
This causes an abrupt redistribution of axial forces within the RVE, unlike the gradual redistribution in load cases $\vartheta$ and $\kappa$.
For load case $\gamma$, the curve exhibits only one sharp change in slope at the transition from the elastic to the inelastic regime.
At this point, the diagonal members in the \ebene{x}{y} within the RVE yield.
Despite the hardening behavior of these members, their yielding produces an almost horizontal segment of the $Q_y$--$\gamma_y$ curve.

Load cases $\vartheta$ and $\kappa$ exhibit a different response:
the curves transition gradually from the elastic to the inelastic regime.
This occurs because the RVE members experience different loading magnitudes.
Members at the physical boundaries reach the yield stress $\sigma_y$ first, while those in the interior remain elastic.
As loading increases, yielding progresses from the outside toward the interior.
The axial forces are consequently redistributed among the RVE members.
Residual stresses after complete unloading are clearly evident in all load cases.

\subsection{C-shaped cross section} \label{sec:c_shaped_crossSection}
\begin{figure}
    \centering
    \begin{subfigure}{\linewidth}
        \centering\includegraphics{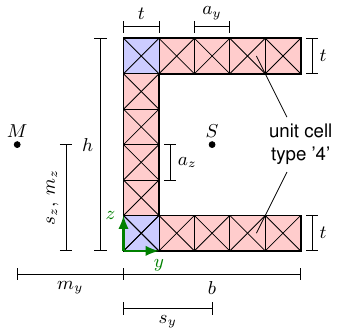}
        \caption{Lattice-like}
        \label{fig:numericalExamples_I_cProfileLattice}
    \end{subfigure}
    \begin{subfigure}{\linewidth}
        \centering\includegraphics{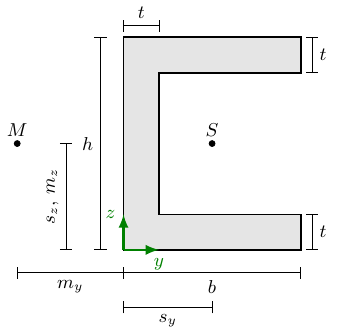}
        \caption{Continuous}
        \label{fig:numericalExamples_I_cProfileConti}
    \end{subfigure}
    \caption{C-shaped cross section: geometry}
    \label{fig:numericalExamples_I_cProfile}
\end{figure}

This example shows that the boundary conditions introduced in Section \ref{sec:constraints_shear} can also be applied in the context of thin-walled open cross-sections.
A C-section composed of type '4' unit cells is considered; see Figure \ref{fig:numericalExamples_I_cProfileLattice}, where only one unit cell through the thickness $t$ is shown for clarity.

Each member is discretized using a geometrically linear finite truss element.
Linear elastic behavior is assumed.
The material parameters are listed in Table \ref{tab:numericalResultsI_materialParameters}.
The region highlighted in red in Figure \ref{fig:numericalExamples_I_cProfileLattice} contains 12 unit cells through its thickness and 48 along its length.
The blue region contains $12\times12$ unit cells.
The resulting section height, width, and flange and web thickness are $h = 72\,a_z$, $b = 60\,a_y$, and $t = 12\,a_z$, respectively, with $a_y$ and $a_z$ specified in Table \ref{tab:numericalExamples_parameters}.
The number of cells in the $x$-direction, $n_x$, and hence the RVE length $\ell\rve$, are varied below.
A continuum C-section with the same dimensions and material parameters is also considered; see Figure \ref{fig:numericalExamples_I_cProfileConti}.

\begin{figure}
    \centering
    \includegraphics{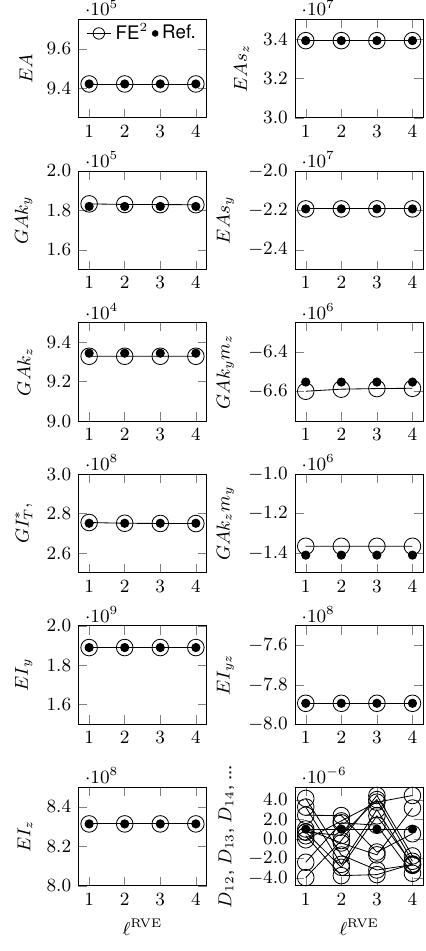}
    \caption{C-shaped cross section: Independence of the homogenized beam stiffness from the length $\ell\rve$ of the selected RVE}
    \label{fig:numericalResults_I_cProfileResults}
\end{figure}

\begin{figure*}
    \centering
    \begin{subfigure}{0.475\textwidth}
        \centering
        \textbf{Continuous}
        \vspace{4ex}
    \end{subfigure}
    \hfill
    \begin{subfigure}{0.475\textwidth}
        \centering
        \textbf{Lattice-like}
        \vspace{4ex}
    \end{subfigure}

    \begin{subfigure}{0.475\textwidth}
        \centering
        \includegraphics[height=5cm]{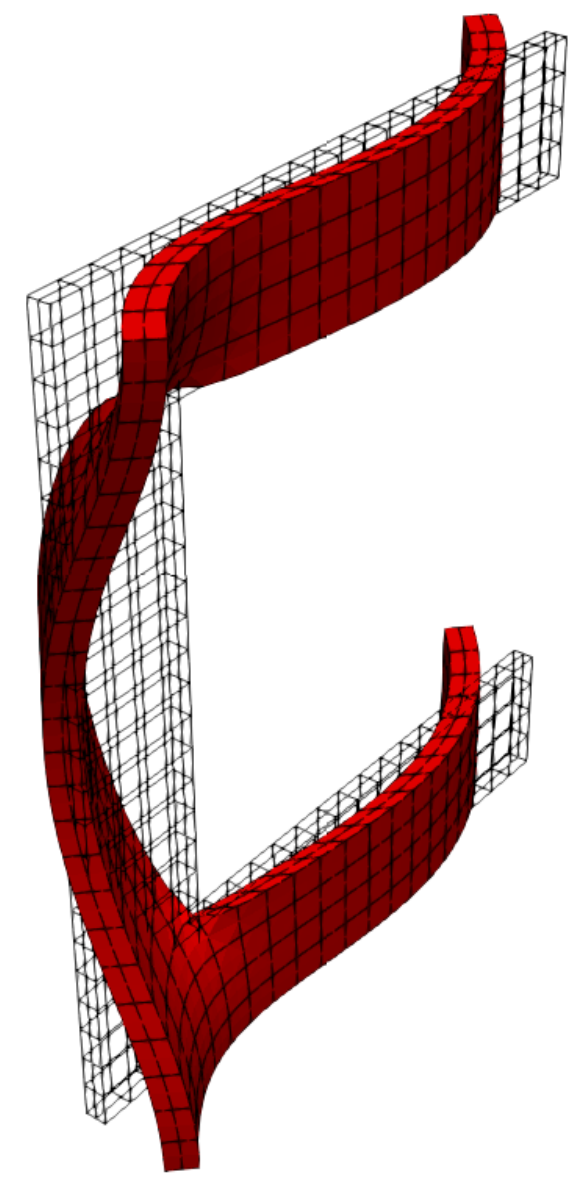}
        \includegraphics[width=4cm]{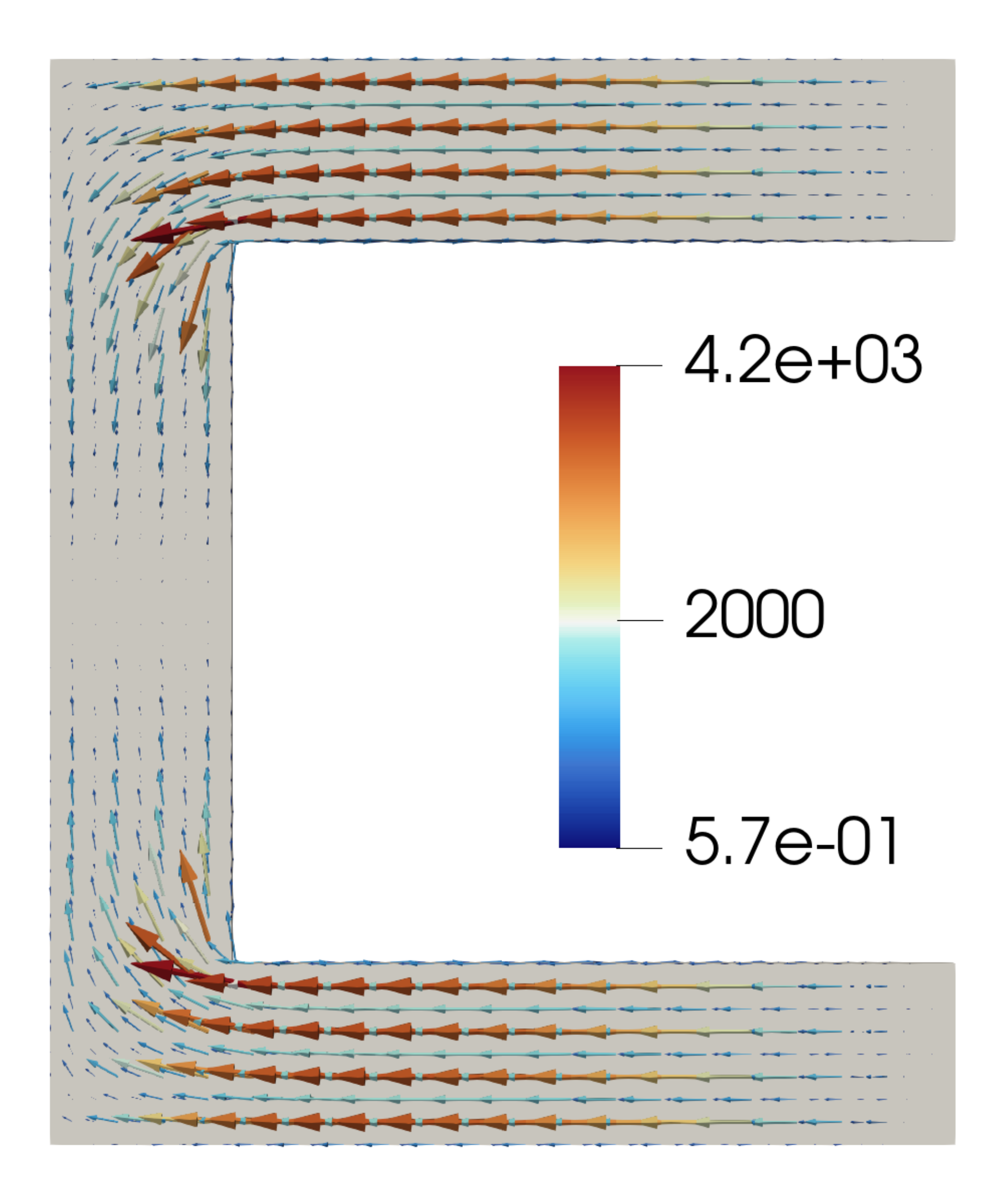}
    \end{subfigure}
    \hfill
    \begin{subfigure}{0.475\textwidth}
        \centering
        \includegraphics[height=5cm]{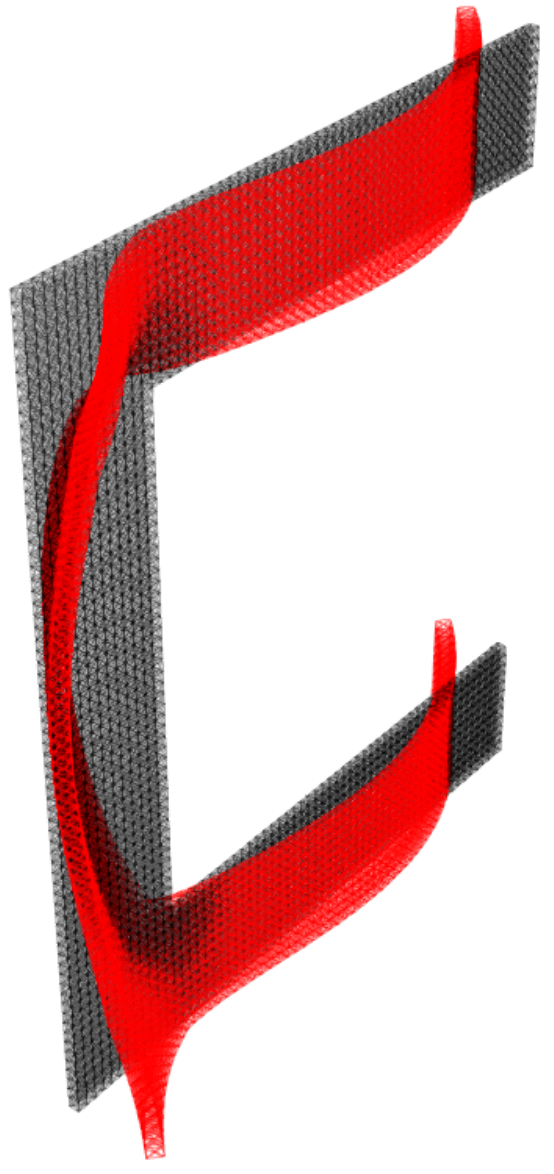}
        \includegraphics[width=4cm]{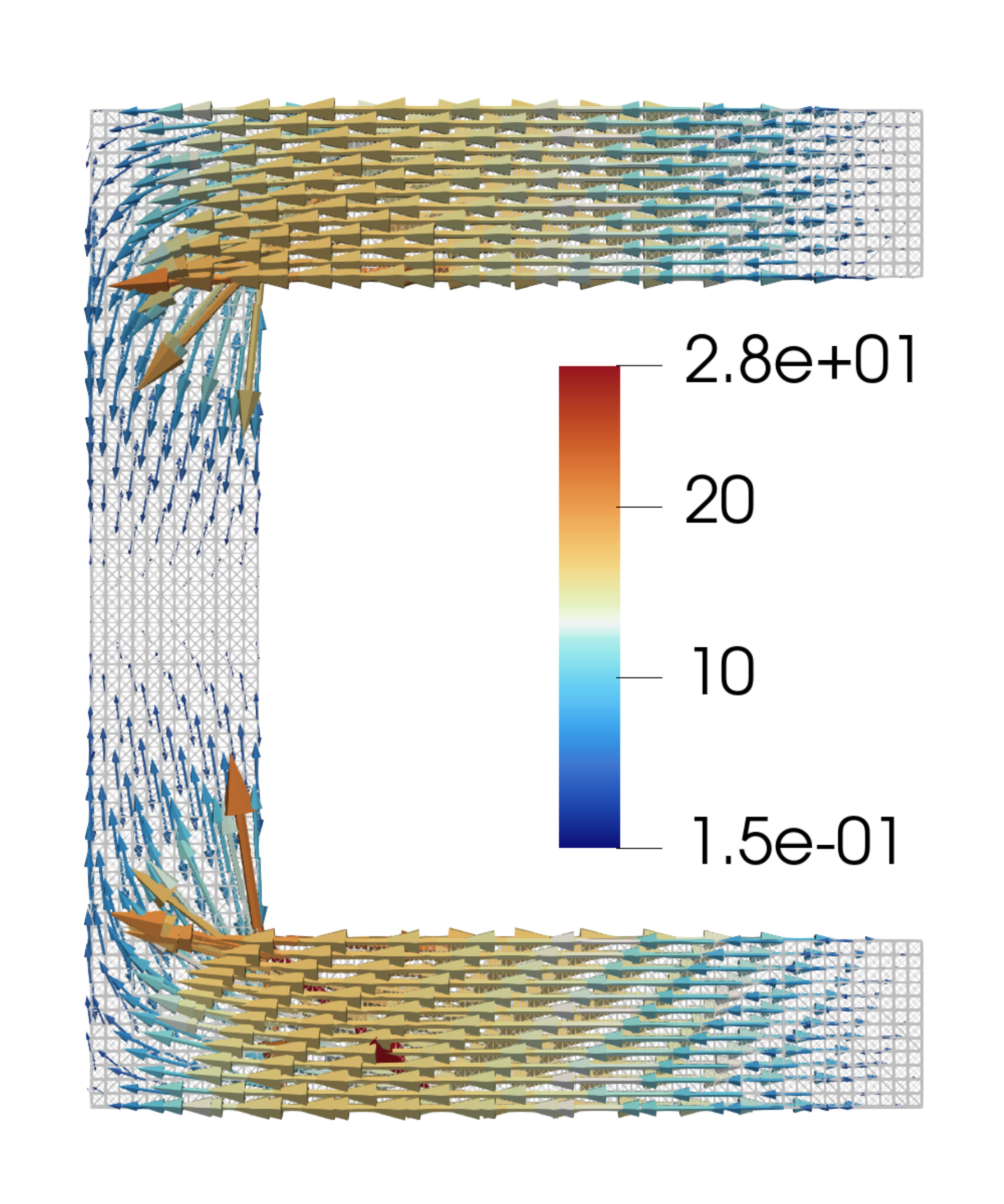}
    \end{subfigure}

    \begin{subfigure}{\textwidth}
        \caption{Load case $\gamma_y$}
    \end{subfigure}

    \begin{subfigure}{0.475\textwidth}
        \centering
        \includegraphics[height=5cm]{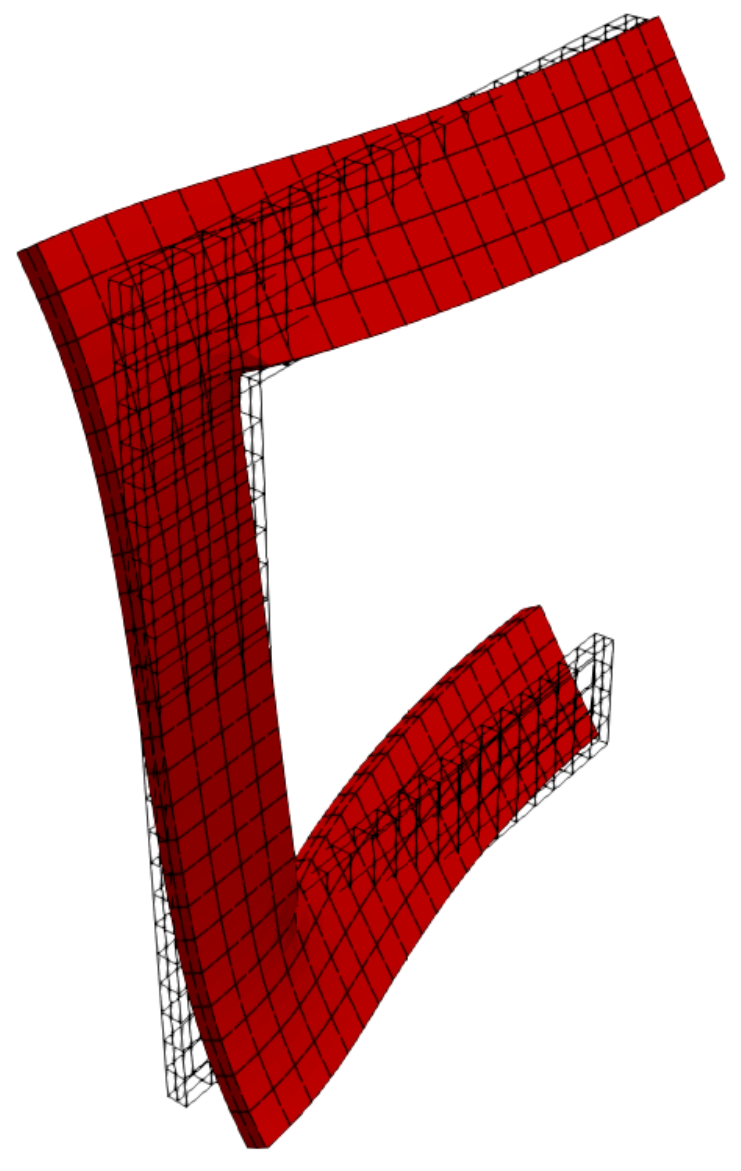}
        \includegraphics[width=4cm]{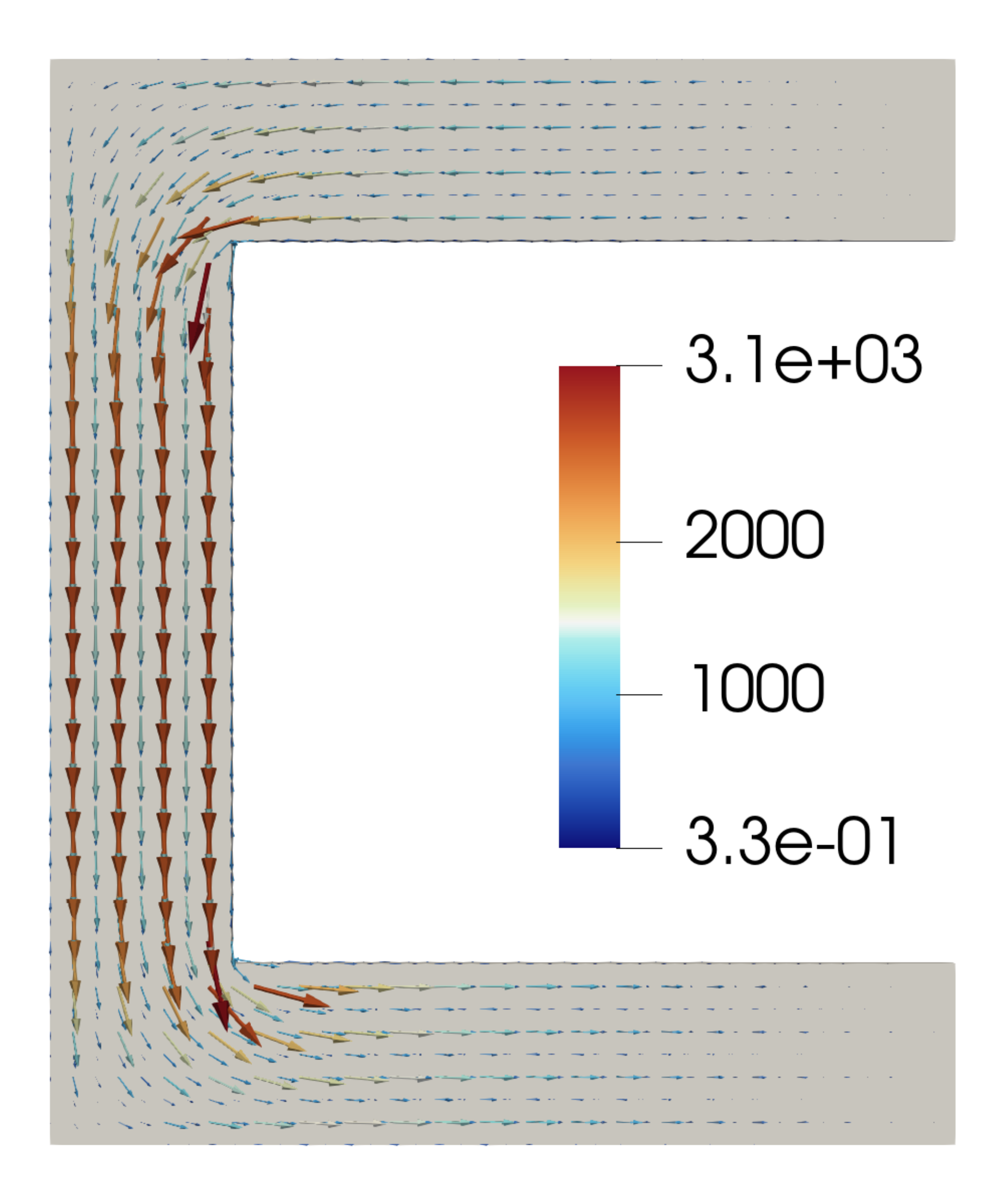}
    \end{subfigure}
    \hfill
    \begin{subfigure}{0.475\textwidth}
        \centering
        \includegraphics[height=5cm]{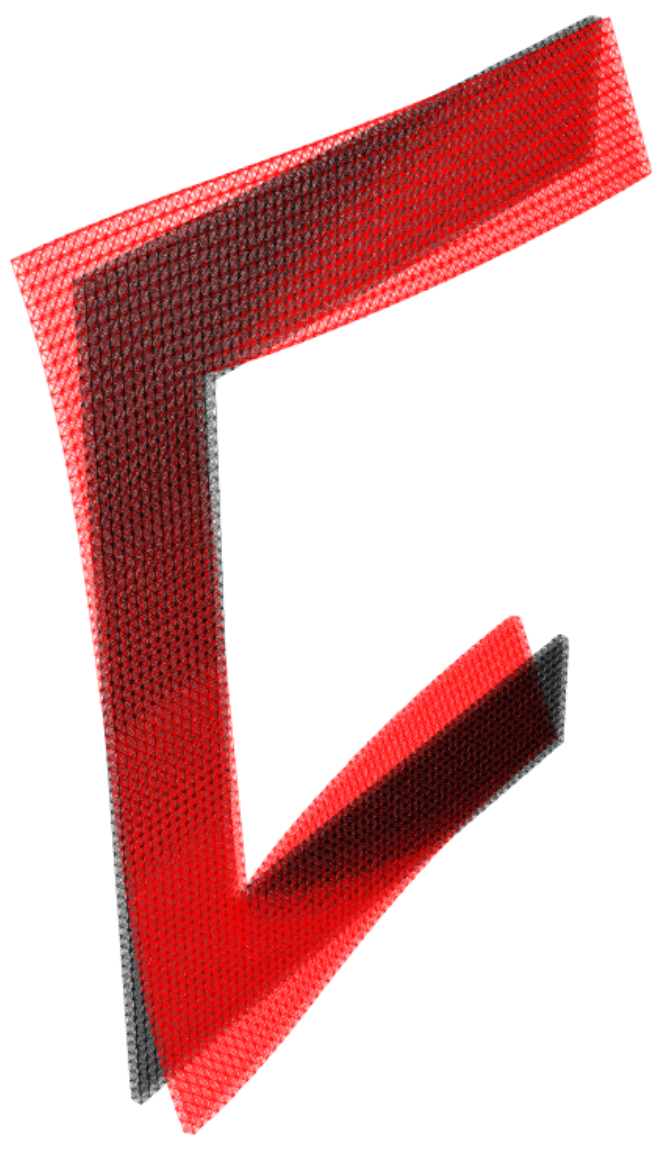}
        \includegraphics[width=4cm]{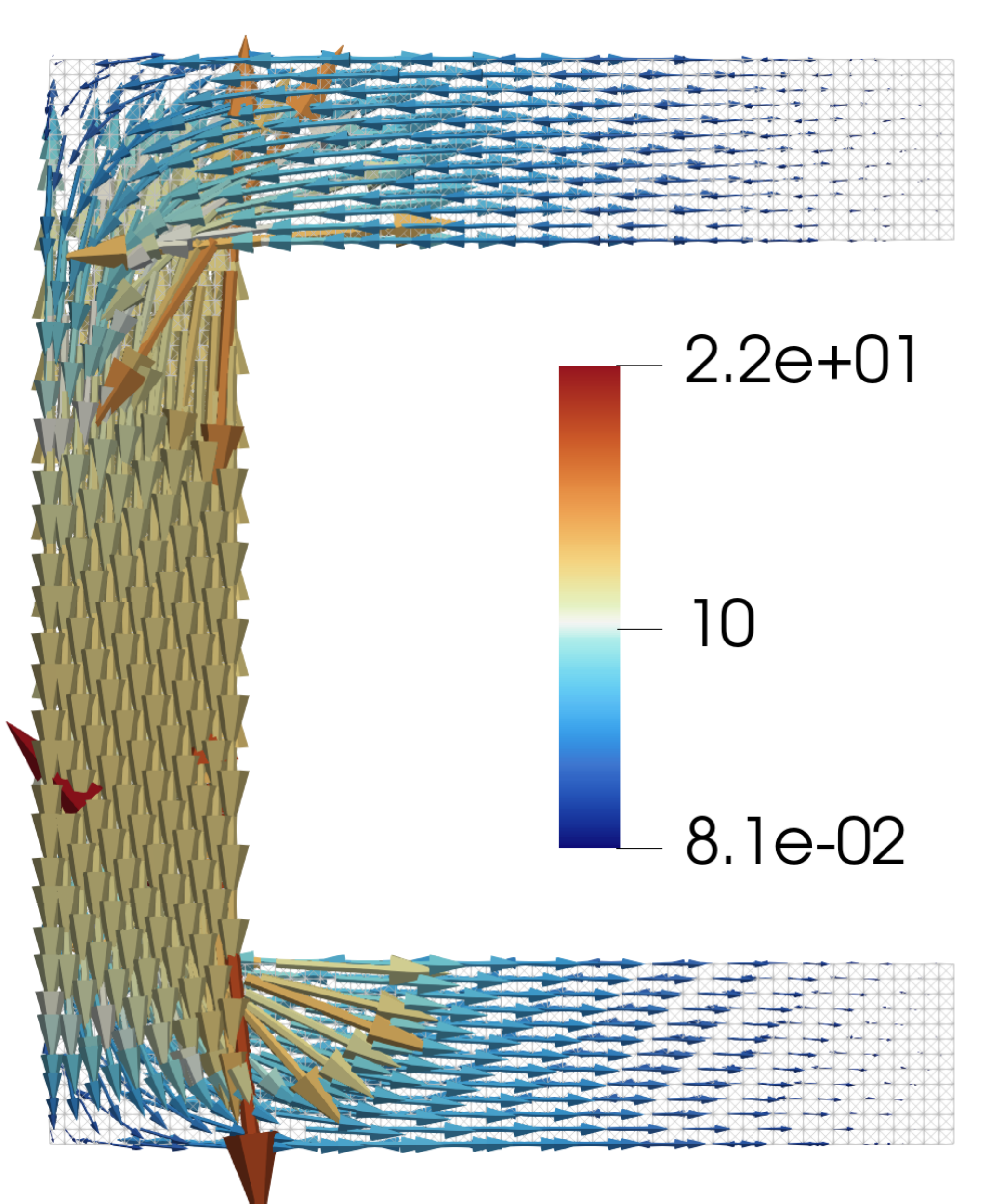}
    \end{subfigure}

    \begin{subfigure}{\textwidth}
        \caption{Load case $\gamma_z$}
    \end{subfigure}

    \begin{subfigure}{0.475\textwidth}
        \centering
        \includegraphics[height=4.8cm]{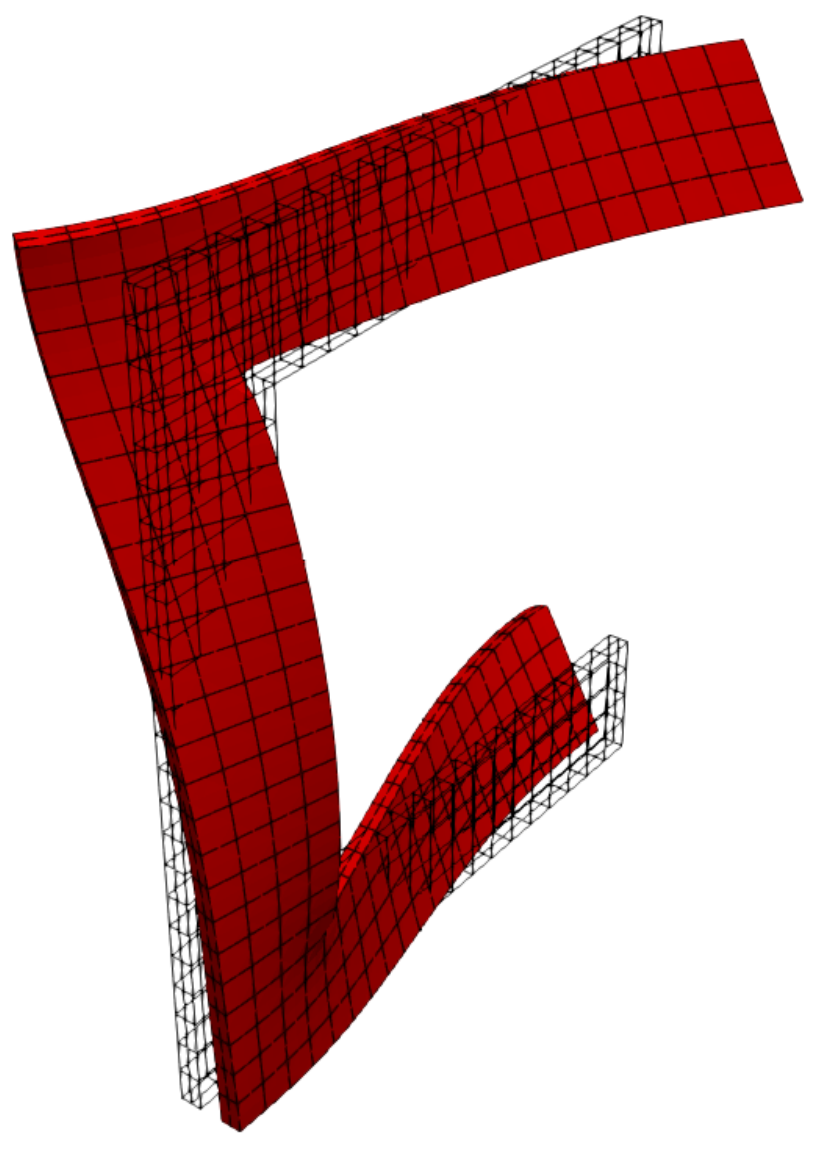}
        \includegraphics[width=4cm]{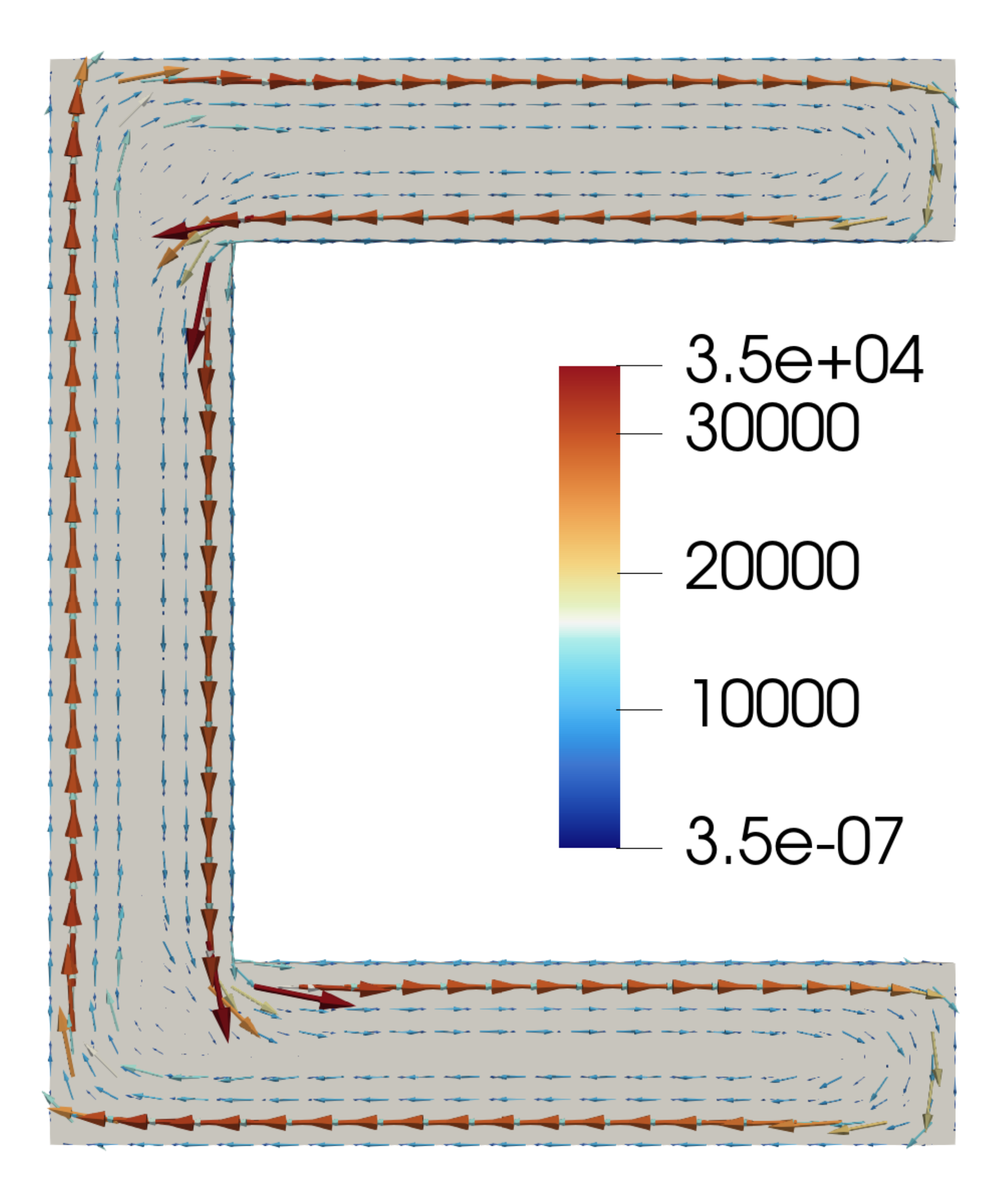}
    \end{subfigure}
    \hfill
    \begin{subfigure}{0.475\textwidth}
        \centering
        \includegraphics[height=4.8cm]{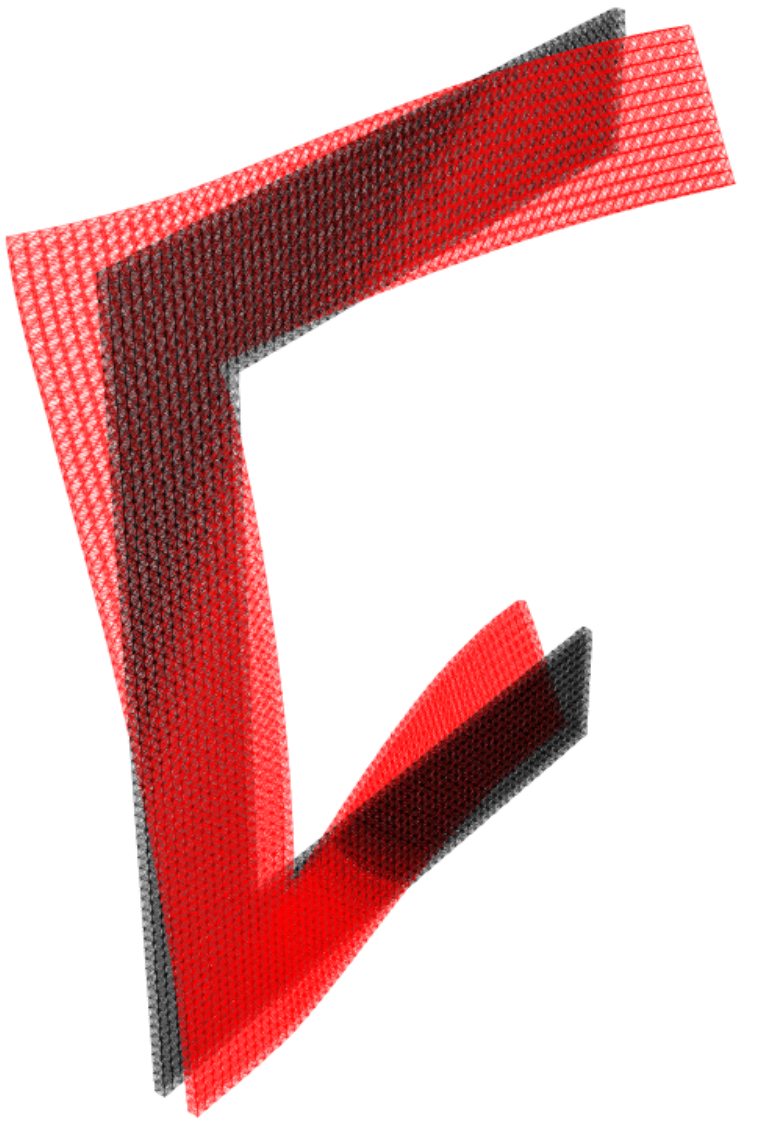}
        \includegraphics[width=4cm]{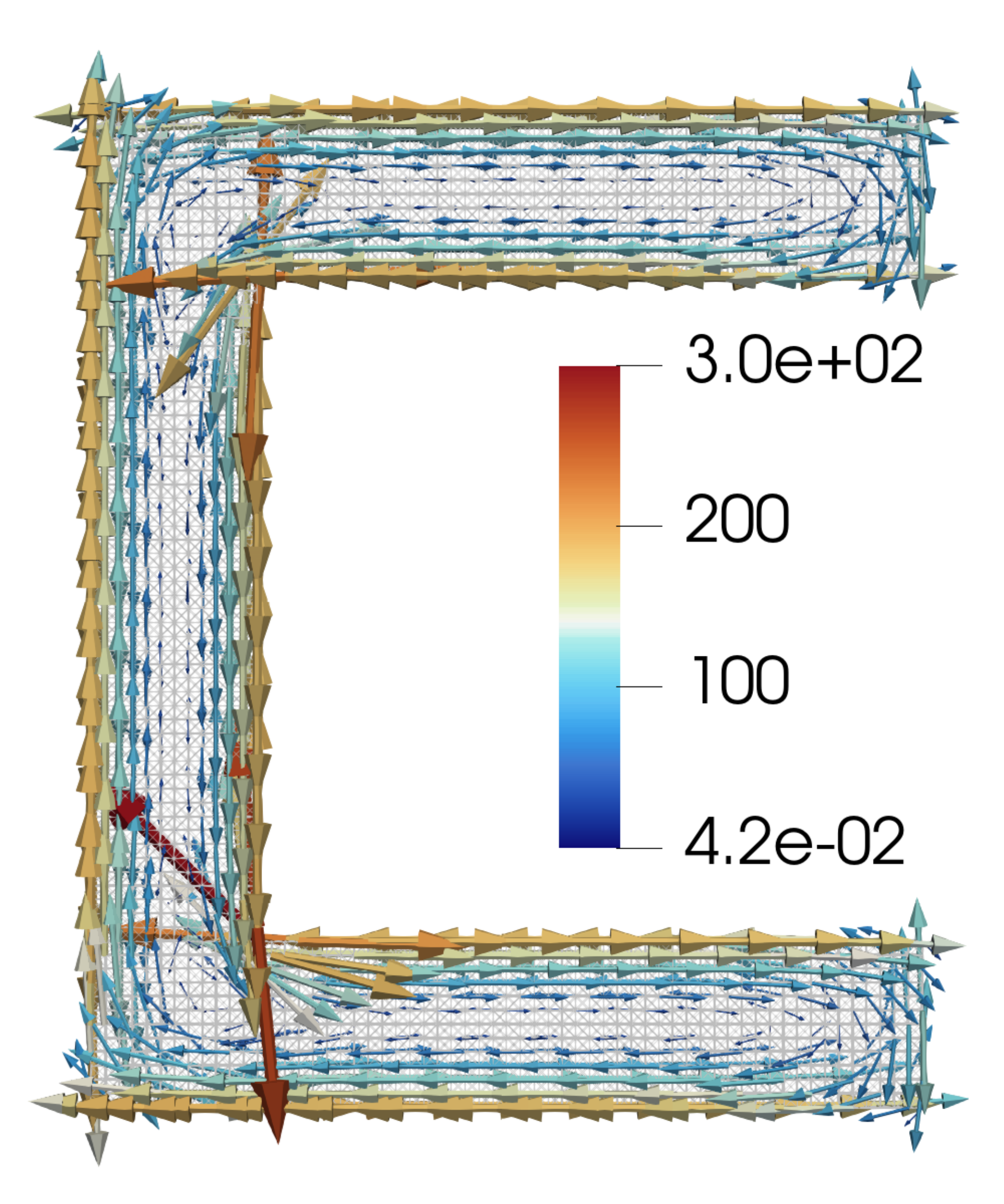}
    \end{subfigure}

    \begin{subfigure}{\textwidth}
        \caption{Load case $\vartheta$}
    \end{subfigure}

    \caption{C-shaped cross section: Deformed RVE and distribution of the calculated constraint forces $\mathbf{T}^+$ (at the model boundary $\Gamma\plus$) for different load cases}
    \label{fig:numericalResults_I_cProfileDeformedMesh}
\end{figure*}

The numerical results are presented in Figures \ref{fig:numericalResults_I_cProfileResults} and \ref{fig:numericalResults_I_cProfileDeformedMesh}, and Table \ref{tab:numericalResults_I_cProfileResults}.
The names and units of the components refer to the quantities listed in Eq.~\eqref{homogenization_D_structure_05}.
Figure \ref{fig:numericalResults_I_cProfileResults} shows all 36 components of $\matrix{D}\macro$ as functions of $\ell\rve$, with numerically zero components grouped in a single plot.
All components are independent of $\ell\rve$ and agree very well with the reference model.
Table \ref{tab:numericalResults_I_cProfileResults} lists the homogenized beam stiffnesses for both the lattice-like and continuum C-sections, together with the reference and analytical solutions, respectively.
The quantitative agreement is also very good.

Figure \ref{fig:numericalResults_I_cProfileDeformedMesh} shows the deformed continuum RVE on the left and the lattice-like RVE on the right.
The loading cases are simple shear in the $y$- and $z$-directions (i.e., $\gamma_y$, $\gamma_z$) and pure torsion (i.e., $\vartheta$).
For comparison, all deformation plots are scaled to show similar displacement magnitudes.
The lattice-like and continuum C-sections exhibit very similar deformation patterns.

Figure \ref{fig:numericalResults_I_cProfileDeformedMesh} also compares the shear force distributions in the lattice-like and continuum C-sections.
The macroscopic strains $\gamma_y$, $\gamma_z$, and $\vartheta$ are imposed on the RVE separately.
The shear force distribution follows from the constraint forces $\mathbf{F}^+_\text{int}$ at the model boundary $\Gamma\plus$; see Eq. \eqref{homogenization_05}$_1$.

For load case $\gamma_z$, the shear flow vanishes at the free flange ends.
It increases approximately linearly from the free ends toward the web.
Along the web, the shear flow is approximately constant.
This behavior occurs in both the continuum and lattice-like mesostructures.
For thin-walled continuum C-sections, the shear stress distribution is known to be linear along the flanges and quadratic along the web.
Since the C-section considered here is no longer thin-walled, this distribution is only approximated.

For load case $\vartheta$, the stress distribution is approximately linear through the section thickness.
The shear forces run parallel to the section boundary, with an almost constant magnitude along the entire boundary.
As before, this behavior is observed for both the continuum and lattice-like mesostructures.

\begin{table*}
        \centering
            \begin{tabular}{p{0.4cm}ccccccc}
                \toprule
                         &                 & \multicolumn{3}{c}{lattice-like (unit cell type '4')} &        \multicolumn{3}{c}{continuous}      \\ \cmidrule(lr){3-5} \cmidrule(lr){6-8}
                         &                 &    \fes    &    Ref.    &  error $e$   &      \fes    &    analytic    & error $e$ \\ \midrule
                $D_{11}$ &      $EA$       & 9.422e+05  & 9.422e+05  &   3.339e-04 \%    &   4.234e+07  & 4.234e+07  &  2.548e-09 \%  \\
                $D_{15}$ &    $EA\,s_z$    & 3.392e+07  & 3.392e+07  &   6.191e-04 \%    &   1.524e+09  & 1.524e+09  & -2.370e-09 \%  \\
                $D_{16}$ &   $-EA\,s_y$    & -2.193e+07 & -2.193e+07 &   5.074e-04 \%    &   -9.798e+08 & 9.797e+08  &  1.673e-09 \%  \\
                $D_{55}$ &    $EI_{y}$     & 1.890e+09  & 1.890e+09  &   5.412e-05 \%    &   8.477e+10  & 8.477e+10  &  1.106e-03 \%  \\
                $D_{66}$ &    $EI_{z}$     & 8.314e+08  & 8.314e+08  &   9.390e-09 \%    &   3.687e+10  & 3.686e+10  & -1.638e-07 \%  \\
                $D_{56}$ &   $-EI_{yz}$    & -7.893e+08 & -7.893e+08 &   1.190e-08 \%    &   -3.527e+10 & -3.527e+10 &  3.638e-09 \%  \\
                $D_{22}$ &   $GA\,k_{y}$   & 1.829e+05  & 1.820e+05  &   5.508e-01 \%    &   8.944e+06  & 8.968e+06  & -2.613e-01 \%  \\
                $D_{33}$ &   $GA\,k_{z}$   & 9.329e+04  & 9.345e+04  &   1.665e-01 \%    &   4.656e+06  & 4.668e+06  & -2.562e-01 \%  \\
                $D_{23}$ &  $GA\,k_{yz}$   & 3.552e-06  & 7.192e-15  &     3.551e-06     &   -7.971e-06 & 0.000e+00  &   -7.971e-06   \\
                $D_{24}$ & $-GA\,k_y\,m_z$ & -6.587e+06 & -6.552e+06 &   5.508e-01 \%    &   -3.220e+08 & -3.228e+08 & -2.613e-01 \%  \\
                $D_{34}$ & $GA\,k_z\,m_y$  & -1.365e+06 & -1.411e+06 &   3.182e-00 \%    &   -6.850e+07 & -6.858e+07 & -1.071e-01 \%  \\
                $D_{44}$ &   $GI_{T}^*$    & 2.750e+08  & 2.751e+08  &   4.851e-04 \%    &   1.337e+10  & 1.340e+10  &  2.463e-01 \%  \\ \midrule
                         &      $s_y$      & 2.327e+01  & 2.327e+01  &         -         &   2.314e+01  & 2.314e+01  &       -        \\
                         &      $s_z$      & 3.600e+01  & 3.600e+01  &         -         &   3.600e+01  & 3.600e+01  &       -        \\
                         &      $m_y$      & -1.464e+01 & -1.510e+01 &         -         &   -1.471e+01 & 1.469e+01  &       -        \\
                         &      $m_z$      & 3.599e+01  & 3.600e+01  &         -         &   3.600e+01  & 3.600e+01  &       -        \\ \bottomrule
            \end{tabular}
    \caption{C-shaped cross section: Verification of homogenized beam stiffnesses by comparison with reference and analytical solutions (in case the reference value is zero, the absolute error is reported instead of the relative error)}
    \label{tab:numericalResults_I_cProfileResults}
\end{table*}

\section{Numerical examples II \label{sec:numericalExamples_II}}
Unlike Section \ref{sec:numericalExamples_I}, the following numerical examples also require solving a macroscopic boundary value problem; see the overview in Figure \eqref{fig:numericalExample_twoStrategies}.
This automatically ensures that the macroscopic strains $\pmb\varepsilon^M$ acting on the RVE are physically meaningful, and that the magnitudes of the individual components are consistent with one another.
The studies focus on the accuracy and robustness of the proposed \fes\ method.

\subsection{Simply supported beam\label{sec:SimplySupportedBeam}}
This example investigates a slender beam with a lattice-like mesostructure under combined loading.
Both elastic and elastoplastic material behavior are considered at the mesoscopic scale.
The influence of geometric nonlinearity is also examined at both scales.
In addition, the study examines the influence of geometrically nonlinear modeling of the RVE on macroscopic behavior.

\begin{figure}[]
    \centering
    \begin{subfigure}{\columnwidth}
        \centering
        \includegraphics{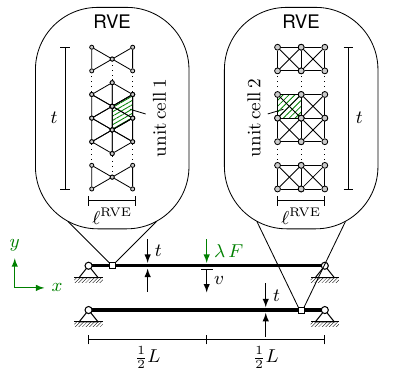}
        \caption{\fes\ model}
        \label{fig:numericalResultsII_beamOnTwoSupports_model_fe2}
    \end{subfigure}
    \begin{subfigure}{\columnwidth}
        \centering
        \includegraphics{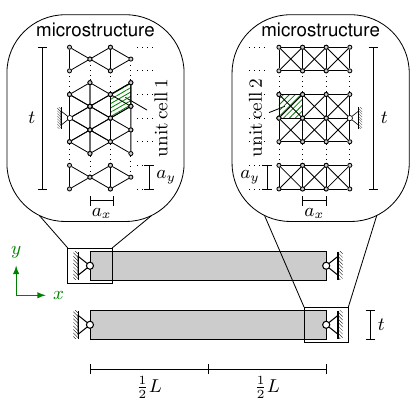}
        \caption{Full scale model}
        \label{fig:numericalResultsII_beamOnTwoSupports_model_fs}
    \end{subfigure}
    \caption{Simply supported beam: geometry, support conditions, loading, and topology of the lattice-like microstructure}
    \label{fig:numericalResultsII_beamOnTwoSupports_model}
\end{figure}

A simply supported beam subjected to a concentrated load is considered, with both supports restrained against translations; see Figure \ref{fig:numericalResultsII_beamOnTwoSupports_model}.
Two centric two-dimensional lattice-like mesostructures, i.e., $s_y=0$, are considered, composed of type '1' or type '2' unit cells with dimensions $a_x$ and $a_y$.
The beam lengths are $L=696\, a_x$ and $600\, a_x$, and the corresponding heights are $t=26\,a_y$ and $16\,a_y$.

The reference model is the fully resolved beam, with each member discretized using one finite truss element.
For unit cell types '1' and '2', this gives 28,898 and 19,800 elements and 10,107 and 10,234 nodes, respectively.
The concentrated load $F$ at midspan is distributed over all nodes at that location.
Each support is modeled as a point support at a node at $y=0$; see Figure \ref{fig:numericalResultsII_beamOnTwoSupports_model_fs}.
An additional constraint is taken into account to prevent the supported node from “pulling out” locally as $F$ increases.

In the proposed FE$^2$ model, the macroscopic beam is discretized using 24 geometrically nonlinear two-node finite beam elements (Timoshenko kinematics, reduced integration of the shear term, and finite rotations).
The selected RVE has length $\ell^\text{RVE}=2\,a_x$ and height $t=26\,a_y$ or $16\,a_y$; see Figure \ref{fig:numericalResultsII_beamOnTwoSupports_model_fe2}.
Each RVE member is modeled using one finite truss element.
For unit cell types '1' and '2', the RVEs contain 97 and 81 elements and 43 and 68 nodes, respectively.
The geometric and material parameters of both RVEs are listed in Table \ref{tab:numericalExamples_parameters}.

\begin{figure}
    \centering

    \begin{minipage}[c][.94\textheight][s]{\columnwidth}
        \centering
    \begin{subfigure}{\columnwidth}
        \centering
        \includegraphics{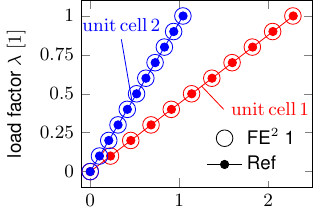}
        \caption{Model approach 1}
        \label{fig:numericalResultsII_beamOnTwoSupports_case1}
    \end{subfigure}
    \vfill
    \begin{subfigure}{\columnwidth}
        \centering
        \includegraphics{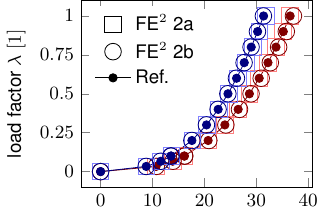}
        \caption{Model approach 2a and 2b}
        \label{fig:numericalResultsII_beamOnTwoSupports_case2}
    \end{subfigure}
    \vfill
    \begin{subfigure}{\columnwidth}
        \centering
        \includegraphics{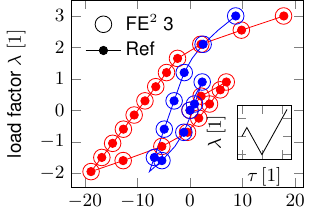}
        \caption{Model approach 3}
        \label{fig:numericalResultsII_beamOnTwoSupports_case3}
    \end{subfigure}
    \vfill
    \begin{subfigure}{\columnwidth}
        \centering
        \includegraphics{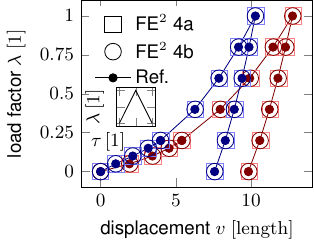}
        \caption{Model approach 4a and 4b}
        \label{fig:numericalResultsII_beamOnTwoSupports_case4}
    \end{subfigure}
    \end{minipage}
    \caption{Simply supported beam: Load-displacement curves for the different modeling approaches}
    \label{fig:numericalResultsII_beamOnTwoSupports}
\end{figure}

Six models are considered in total.
Geometrically linear and nonlinear formulations are distinguished at both scales.
At the mesoscopic scale, linear elastic (E) and elastoplastic (E-P) material behavior are also distinguished.
Their combinations yield the six models listed in Table \ref{tab:numericalResultsII_beamOnTwoSupports_modellingApproach}.
\begin{table*}[!tbp]
    \centering
    \begin{tabular}{c p{0.0cm} cccccc p{0.0cm} cccc  p{0.0cm} cc }
        \toprule
              &  &                         \multicolumn{6}{c}{\fes\ model}                         &  &           \multicolumn{4}{c}{Ref. model}            &  &                     &            \\ \cmidrule(lr){3-8} \cmidrule(lr){10-13}
              &  & \multicolumn{2}{c}{macro} &              \multicolumn{4}{c}{micro}              &  &            &             &            &             &  &                     &            \\ \cmidrule(lr){3-4} \cmidrule(lr){5-8}
              &  & \multicolumn{2}{c}{geo.}  & \multicolumn{2}{c}{geo.} & \multicolumn{2}{c}{mat.} &  & \multicolumn{2}{c}{geo.} & \multicolumn{2}{c}{mat.} &  &                     &            \\ \cmidrule(lr){3-4} \cmidrule(lr){5-6} \cmidrule(lr){7-8} \cmidrule(lr){10-11} \cmidrule(lr){12-13}
        model &  & L          &      NL      &     L      &     NL      &     E      &     E-P     &  & L          &     NL      &     E      &     E-P     &  & $F$                 &    unit    \\ \midrule
         '1'  &  & \checkmark &              & \checkmark &             & \checkmark &             &  & \checkmark &             & \checkmark &             &  & $\tfrac{120}{1000}$ & \unitForce \\ \midrule
        '2a'  &  &            &  \checkmark  & \checkmark &             & \checkmark &             &  &            & \checkmark  & \checkmark &             &  & $120$               & \unitForce \\
        '2b'  &  &            &  \checkmark  &            & \checkmark  & \checkmark &             &  &            & \checkmark  & \checkmark &             &  & $120$               & \unitForce \\ \midrule
         '3'  &  & \checkmark &              & \checkmark &             &            & \checkmark  &  & \checkmark &             &            & \checkmark  &  & $\tfrac{120}{500}$  & \unitForce \\ \midrule
        '4a'  &  &            &  \checkmark  & \checkmark &             &            & \checkmark  &  &            & \checkmark  &            & \checkmark  &  & $\tfrac{120}{50}$   & \unitForce \\
        '4b'  &  &            &  \checkmark  &            & \checkmark  &            & \checkmark  &  &            & \checkmark  &            & \checkmark  &  & $\tfrac{120}{50}$   & \unitForce \\ \bottomrule
    \end{tabular}
    \caption{Simply supported beam: Different modeling approaches for the FE$^2$ model and the reference model (L: linear; NL: nonlinear; E: linear elastic; E-P: elastoplastic)}
    \label{tab:numericalResultsII_beamOnTwoSupports_modellingApproach}
\end{table*}
The concentrated load $F$ is increased monotonically over several steps for models '1' and '2', whereas cyclic loading is applied for models '3' and '4'; see the $\lambda$--$\tau$ plots in Figures \ref{fig:numericalResultsII_beamOnTwoSupports_case3} and \ref{fig:numericalResultsII_beamOnTwoSupports_case4}.
Within each load step, Newton's method converges in two iterations for model '1' and in four to six iterations for the other models, with quadratic convergence.
At every load level, the member axial forces were checked to ensure that no individual RVE member exceeded its Euler buckling load for a pin-ended column (case II).
An accompanying stability analysis for models '2' and '4' confirmed the absence of other global instabilities.

Figure \ref{fig:numericalResultsII_beamOnTwoSupports} presents the load--displacement curves in terms of the midspan deflection.
The FE$^2$ solutions are compared with the corresponding reference solutions.
Model '1', with geometric and material linearity at both scales, gives a linear load--displacement curve. The mesostructure composed of type '2' unit cells has a substantially higher global stiffness than that composed of type '1' cells; see Figure \ref{fig:numericalResultsII_beamOnTwoSupports_case1}.
The results agree very well with the reference model.

Model '2' includes geometric nonlinearity at the macroscopic scale. The mesoscopic model is materially linear and either geometrically linear ('2a') or geometrically nonlinear ('2b').
Both models exhibit the nonlinear load--displacement response expected for this boundary value problem, with characteristic stiffening as the load increases; see Figure \ref{fig:numericalResultsII_beamOnTwoSupports_case2}.
Model '2b' agrees very well with the reference results for both mesostructures.
Model '2a' exhibits small deviations.

In model '3', both scales are geometrically linear.
Elastoplastic behavior with linear isotropic hardening is included at the mesoscopic scale.
This model results in a nonlinear load--displacement curve with the characteristic loading and unloading response; see Figure \ref{fig:numericalResultsII_beamOnTwoSupports_case3}.
Both mesostructures show very good agreement with the reference results.

In model '4', the macroscopic scale is geometrically nonlinear.
The mesoscopic model is elastoplastic and either geometrically linear ('4a') or geometrically nonlinear ('4b').
Both models exhibit nonlinear load--displacement behavior throughout loading and unloading, reflecting the combined effects of geometric and material nonlinearity; see Figure \ref{fig:numericalResultsII_beamOnTwoSupports_case4}.
Model '4b' agrees very well with the reference results for both mesostructures, whereas model '4a' exhibits small deviations.

Overall, FE$^2$ models '1', '2b', '3', and '4b' agree very well with the reference model.
Deviations occur for models '2a' and '4a', which combine a geometrically nonlinear macroscopic model with a geometrically linear mesoscopic model.

\subsection{Spatial frame structure\label{sec:SpatialFrameStructure}}
\begin{figure*}
    \centering
    \begin{subfigure}{\columnwidth}
        \centering
        \includegraphics{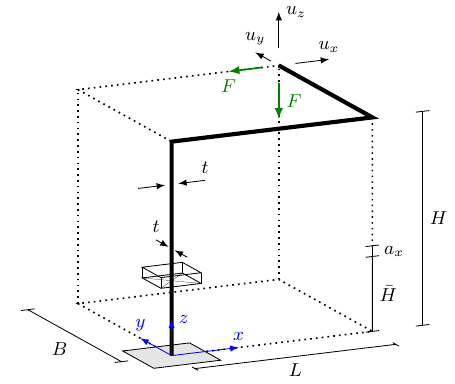}
        \label{fig:numericalResultsII_3memberBeam_model}
    \end{subfigure}
    \begin{subfigure}{\columnwidth}
        \centering
        \includegraphics{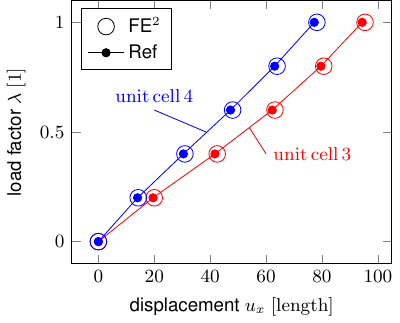}
        \label{fig:numericalResultsII_3memberBeam_results_ux}
    \end{subfigure}

    \begin{subfigure}{\columnwidth}
        \centering
        \includegraphics{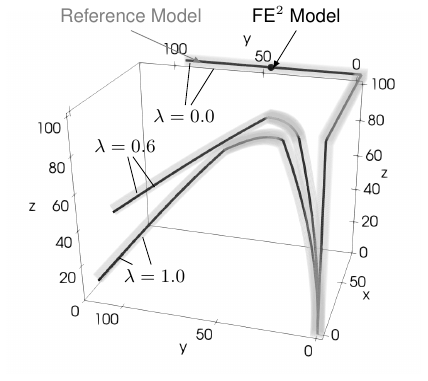}
        \label{fig:numericalResultsII_3memberBeam_bcc_deformedMesh}
    \end{subfigure}
    \begin{subfigure}{\columnwidth}
        \centering
        \includegraphics{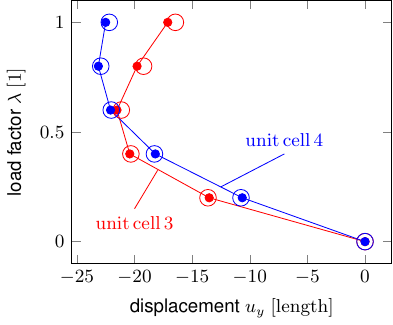}
        \label{fig:numericalResultsII_3memberBeam_results_uy}
    \end{subfigure}

    \begin{subfigure}{\columnwidth}
        \centering
        \includegraphics{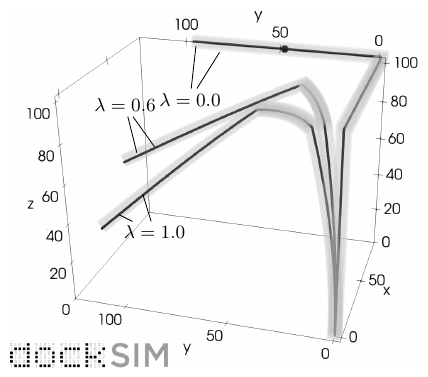}
        \label{fig:numericalResultsII_3memberBeam_fcc_deformedMesh}
    \end{subfigure}
    \begin{subfigure}{\columnwidth}
        \centering
        \includegraphics{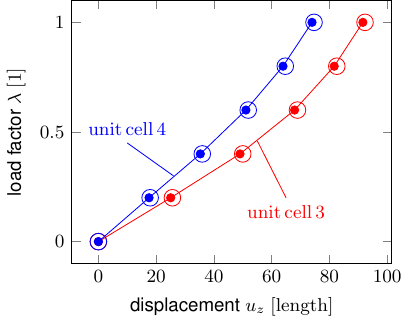}
        \label{fig:numericalResultsII_3memberBeam_results_uz}
    \end{subfigure}

    \caption{Spatial frame structure: System (geometry, support conditions and loading) on the top left and below it the deformed mesh for unit cell '3' and '4' and load-displacement curves (right)}
    \label{fig:numericalResultsII_3memberBeam}
\end{figure*}

This example validates the accuracy and robustness of the proposed FE$^2$ method for a spatial frame structure undergoing large displacements and rotations.
The cantilever shown in the upper left of Figure \ref{fig:numericalResultsII_3memberBeam} consists of three beams connected by rigid joints.

Its dimensions are $L=B=H=100\, a_x$.
The cantilever is clamped at the coordinate origin.
Two concentrated forces $F=12\,\unitForce$ act at its free end.
Each of the three beams is represented by a lattice-like mesostructure composed of pin-jointed truss members, using type '3' or type '4' unit cells.
The mesostructure is prismatic within each beam and has a square cross section with side lengths $t=12 \, a_y$ and $12 \, a_z$.
Each truss member is also assumed to be prismatic, with a square cross section of side lengths $b=h$, and to exhibit linear elastic behavior.
The geometric and material parameters are listed in Table \ref{tab:numericalExamples_parameters}.

The reference model is the fully resolved cantilever, with each member discretized using a geometrically nonlinear finite truss element.
For unit cell types '3' and '4', this gives 981,024 and 837,888 elements and 188,307 and 101,907 nodes, respectively.
The concentrated forces $F$ are distributed over all nodes of the loaded cross section.
The clamped boundary is modeled by restraining all translations of the nodes on that cross section.

In the proposed FE$^2$ model, each of the three beams is represented by one geometrically nonlinear two-node finite beam element (Timoshenko kinematics, reduced integration of the shear term, and finite rotations).
The selected RVE has length $\ell^\text{RVE}=1\,a_x$, height $t=12\,a_y$, and width $12\,a_z$.
Each RVE member is modeled using a geometrically nonlinear finite truss element.
For unit cell types '3' and '4', the RVEs contain 1,945 and 1,993 elements and 482 and 338 nodes, respectively.

The concentrated forces are applied in only five load steps.
Within each load step, Newton's method converges in an average of 11 iterations for the reference model and eight for the FE$^2$ model, with quadratic convergence near the solution.
Table \ref{tab:numericalResultsII_3memberBeam_residual} shows the convergence behavior of the FE$^2$ model in the final load step.
\begin{table}
    \centering
    \begin{tabular}{ccc}
        \toprule
        &       \multicolumn{2}{c}{$\|\vector{R}\|_2$}       \\ \cmidrule(lr){2-3}
        Iter. &       unit cell type '3'        &       unit cell type '4'        \\ \midrule
        0   & \expnumber{3.394}{-01} & \expnumber{3.394}{-01}  \\
        1   & \expnumber{9.364}{+02} & \expnumber{7.919}{+02}  \\
        2   & \expnumber{8.643}{+00} & \expnumber{8.543}{+00}  \\
        3   & \expnumber{2.595}{+00} & \expnumber{7.685}{-01}  \\
        4   & \expnumber{1.561}{+00} & \expnumber{1.114}{+00}  \\
        5   & \expnumber{2.590}{-02} & \expnumber{5.607}{-03}  \\
        6   & \expnumber{1.345}{-03} & \expnumber{1.736}{-04}  \\
        7   & \expnumber{1.049}{-08} & \expnumber{5.060}{-10} \\ \bottomrule
    \end{tabular}
    \caption{Spatial frame structure: Convergence behavior in the final load step}
    \label{tab:numericalResultsII_3memberBeam_residual}
\end{table}
At every load level, the member axial forces were checked to ensure that no individual RVE member exceeded its Euler buckling load for a pin-ended column (case II).
An accompanying stability analysis (models '2' and '4') confirmed the absence of other global instabilities.

The numerical results are presented in various ways.
Figure \ref{fig:numericalResultsII_3memberBeam} shows the deformed finite element meshes.
Here, the FE$^2$ results are represented by the macroscopic beam model, with a black solid line marking the beam reference axes.
And the results of the reference model (fully resolved lattice structure) appear as a gray area.
The deformed meshes show very good qualitative agreement for both mesostructures.

For a quantitative comparison, the displacement at the load application point is considered, where the largest absolute displacement occurs.
The corresponding load-displacement curves are shown on the right side of Figure \ref{fig:numericalResultsII_3memberBeam}.
The response is nonlinear, most prominently in the displacement component $u_y$.
All displacement components agree very well with the reference results
for both mesostructures.
The small deviations arise partly from the more restrictive kinematic assumptions of the macroscopic beam model compared with the fully resolved model; these assumptions are not valid near supports, frame corners, and concentrated loads.
The cantilever composed of type '4' unit cells is stiffer than that composed of type '3' cells.

\begin{figure}
    \centering
    \begin{subfigure}{\columnwidth}
        \centering
        \includegraphics{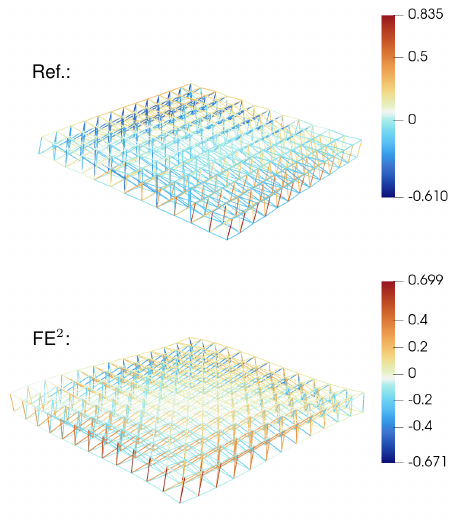}
        \caption{Unit cell type '3'}
    \end{subfigure}

    \begin{subfigure}{\columnwidth}
        \centering
        \includegraphics{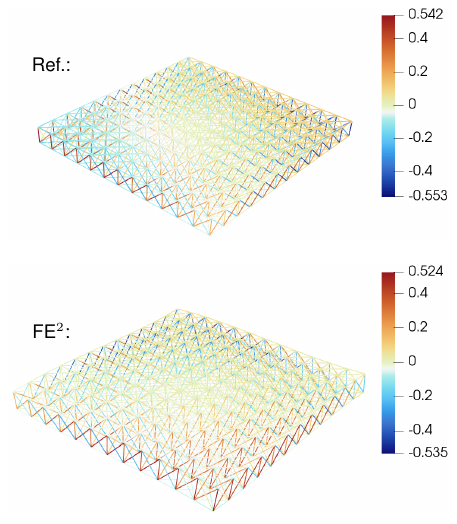}
        \caption{Unit cell type '4'}
    \end{subfigure}
    \caption{Spatial frame structure: Deformed configuration of the section of thickness $a_x$ shown in Figure \ref{fig:numericalResultsII_3memberBeam} at a height of $\bar{H}$ on the vertical beam for the 5th load step}
    \label{fig:numericalResultsII_3memberBeam_deformedMesh_ref_vs_fe2}
\end{figure}

Finally, the local response is examined in detail at a highly loaded location of the cantilever.
A section of length $a_x$ is cut out from the beam that runs in the z-direction at a height of $\bar{H}=50\,a_x$,
i.e., the region $(-t,-t,\bar{H})\le (x,y,z) \le (t,t,\bar{H}+a_x)$.
This location is marked as a section in Figure \ref{fig:numericalResultsII_3memberBeam}.
Figure \ref{fig:numericalResultsII_3memberBeam_deformedMesh_ref_vs_fe2} compares the deformed meshes of the FE$^2$ and reference models.
For the FE$^2$ model, this is the RVE associated with the beam element integration point closest to the selected section.
The extracted cross section exhibits warping caused by the shear and torsional loading in the beam considered.
The axial force in each truss member is also shown in color.
Very good agreement is obtained for the type '4' mesostructure.
The type '3' mesostructure exhibits larger deviations of approximately 15\%.
These deviations again arise from the overly restrictive kinematic assumptions of the macroscopic beam model, whose influence propagates through the structure, particularly from frame corners and clamped boundaries.
This influence is clearly more significant in the type '3' microstructure than in the type '4' microstructure.

\section{Conclusions and outlook \label{sec:conclusion}}
This work developed an \fes\ framework for shear-deformable beams with periodic lattice-like mesostructures.
The mesostructure is represented by truss members.
The \fes\ formulation is based on the Hill--Mandel condition.
Since a structural model is used at the macroscopic scale, the RVE cross-sectional dimensions match those of the lattice structure. Its extent along the beam reference axis must instead be selected subject to scale separation.
Periodic boundary conditions are formulated for the RVE, linked to the beam kinematics, and supplemented by additional constraints.
These suppress rigid-body translations and rotations and allow the homogenized quantities to be determined independently of the RVE length.
The axial stiffnesses of truss members on the model boundaries are also adjusted to preserve periodicity and eliminate parasitic boundary effects.
Homogenization yields beam stress resultants and stiffnesses in a form consistent with a continuum beam model.

The proposed \fes\ model is first verified and validated using isolated RVE analyses, with only a single material point at the macroscopic scale.
More complex examples are then considered, requiring the solution of a macroscopic boundary value problem as well.
These studies include two- and three-dimensional beam structures with different mesostructures.
Under geometric and material linearity, the results demonstrate that the proposed constraints ensure independence of the homogenized quantities from the RVE length for centric and eccentric RVEs with compact and open cross sections.
The homogenized material matrix also has the same basic structure as that of a continuum Timoshenko beam.
Very good agreement with fully resolved reference models is obtained.

The studies are further extended to material and geometric nonlinearity.
For materially nonlinear behavior ($J_2$ plasticity), the homogenized quantities also remain independent of the selected RVE length.
Geometrically nonlinear mesoscopic modeling introduces parasitic axial forces and additional coupling terms in the homogenized material matrix.
For the mesostructures investigated, however, these effects have only a minor influence on the results.

The \fes\ simulations of more complex two- and three-dimensional beam structures also agree very well with fully resolved reference models.
This holds for both geometrically and materially linear and nonlinear analyses.
In the nonlinear analyses, the macroscopic Newton iteration exhibits the expected quadratic convergence near the solution.

The proposed \fes\ model therefore provides very good results, particularly under geometric linearity.
Further research is required for geometrically nonlinear behavior at the mesoscopic scale, since the influence of the observed parasitic effects cannot readily be generalized to arbitrary mesostructures.
An appropriate extension would also enable detailed investigations of complex mesoscopic stability phenomena.
Future work could also extend the framework to lattice structures whose boundary nodes at either end do not lie in the same plane, which are excluded from the present study as discussed in Section \ref{sec:latticeStructure}.

The present \fes\ method represents the mesoscopic lattice using truss members.
Extending it to beam-based mesostructural models is therefore a natural next step.
The authors have already published initial results in this direction \cite{Ochs.2026}.
A more detailed analysis of heterogeneity along the beam axis, particularly for three-dimensional mesostructures, is another promising direction.

\backmatter
\section*{Statements and Declarations}
\bmhead{Funding} The authors did not receive support from any organization for the submitted work.

\bmhead{Competing interests} The authors have no competing interests to declare that are relevant to the content of this article.

\bmhead{Data availability} Data will be made available upon reasonable request.

\bmhead{Author contributions}
JO: conceptualization, methodology, software, validation, formal analysis, investigation, visualization, writing—original draft, and writing—review and editing. JW: conceptualization, methodology, formal analysis, supervision, writing—original draft (substantial rewriting and reformulation of the mathematical derivations and the scientific presentation throughout the manuscript), and writing—review and editing. All authors read and approved the final manuscript.

\bmhead{ORCID iDs}
\begingroup
\raggedright
\noindent Julian Ochs: {\small\href{https://orcid.org/0009-0009-4871-3951}{0009-0009-4871-3951}}\par
\noindent Jens Wackerfu{\ss}: {\small\href{https://orcid.org/0000-0002-1377-1678}{0000-0002-1377-1678}}\par
\endgroup
\begin{appendices}
\renewcommand{\theHequation}{appendix.\theequation}

\section{Continuous mesostructures\label{app:constraints_shear_conti}}
The constraint derived for lattice-like mesostructures in Section \ref{sec:constraints_shear} builds on the '\textit{constant moment constraint}' introduced for beams with continuum mesostructures \cite{klarmann2020} and extended to shells in \cite{gruttmann2024}.
The following derivation shows that this constraint can be simplified for isotropic linear elastic material behavior.
The simplified form then provides the basis for its extension to lattice-like mesostructures.

Consider the bending moment $M_y(x)$ at a position $x$ in a continuum RVE.
It is obtained from the normal stress distribution $\sigma_{xx}(x,y,z)$ over the cross section\footnote{The same reasoning applies to the bending moment $M_z(x)$.}:
\begin{equation} \label{eq:appendix_cmc_01}
    M_y(x) = \int_{y_-}^{y_+} \int_{z_-}^{z_+} z \, \sigma_{xx}(x,y,z) \, \text{d}y\text{d}z \; .
\end{equation}
Using the constitutive relation $\pmb\sigma=\mathbf{C}\,\pmb\varepsilon$, the stress component $\sigma_{xx}$ is expressed as
\begin{equation} \label{eq:appendix_cmc_02}
    \begin{split}
        \sigma_{xx}(x,y,z)
        &= C_{11}(x,y,z) \, \varepsilon_{xx}(x,y,z) \\
        &+ C_{12}(x,y,z) \, \varepsilon_{yy}(x,y,z) \\
        &+ C_{13}(x,y,z) \, \varepsilon_{zz}(x,y,z) \; ,
    \end{split}
\end{equation}
where $\varepsilon_{ij}=\frac{1}{2}(u_{i,j}+u_{j,i})$ are the components of the linearized Green--Lagrange strain tensor.
The RVE is assumed to have a homogeneous composition in the $x$-direction (cf. types 'AA', 'AB1', and 'AB2' in Figure \ref{fig:RVE_classification} for the two-dimensional case), so the material matrix components satisfy
\begin{equation} \label{eq:appendix_cmc_03}
    C_{ij}(x,y,z) = \text{const. in $x$} \;.
\end{equation}

The 'constant moment constraint' requires the bending moment distribution $M_y(x)$ to be symmetric in the $x$-direction, i.e.,
\begin{equation} \label{eq:appendix_cmc_05}
    \int_{-\tfrac{1}{2}\ell\rve}^{\tfrac{1}{2}\ell\rve} M_y(x)\,x\,\d x \overset{!}{=} 0 \, .
\end{equation}
However, the constraint \eqref{eq:appendix_cmc_05} does not prevent a pure rigid-body rotation about the $y$-axis when a shear strain is imposed.
To prevent this rotation, the first term in \eqref{eq:appendix_cmc_01}, i.e., '$C_{11}\varepsilon_{xx} \,x\,z$', is integrated by parts with respect to $x$, and the resulting boundary term is omitted; see \cite{gruttmann2024}.
This yields
\begin{equation} \label{eq:appendix_cmc_06}
    -\int_V u_xC_{11}z \,\d V + \int_V (C_{12}\varepsilon_{yy}+C_{13}\varepsilon_{zz})xz \,\d V \overset{!}{=} 0 \,.
\end{equation}
The conditions under which the second integral in \eqref{eq:appendix_cmc_06} vanishes are examined below, first considering only its first term:
\begin{equation} \label{eq:appendix_cmc_07}
    \int_V C_{12}(x,y,z)\,\varepsilon_{yy}(x,y,z)\,x\,z \overset{!}{=} 0 \; .
\end{equation}
Using \eqref{eq:appendix_cmc_03}, condition \eqref{eq:appendix_cmc_07} is satisfied if the strain $\varepsilon_{yy}(x,y,z)$ is symmetric in the $x$-direction within the RVE.
Applying the same reasoning to the second term of the second integral gives the corresponding conditions on $\varepsilon_{zz}(x,y,z)$.
Together, these conditions are
\begin{equation} \label{eq:appendix_cmc_08}
    \left .
    \begin{split}
        \varepsilon_{yy}(x,y,z) \\
        \varepsilon_{zz}(x,y,z)
    \end{split}
    \right \} \; = \text{sym. in $x$} \; .
\end{equation}
Numerical RVE studies under geometric linearity were used to determine the strain distributions for each of the six macroscopic strains separately.
They confirmed the symmetries required by \eqref{eq:appendix_cmc_08} for each macroscopic strain, and therefore for arbitrary linear combinations, satisfying \eqref{eq:appendix_cmc_07}.
The constraint \eqref{eq:appendix_cmc_05} therefore simplifies to
\begin{equation} \label{eq:appendix_cmc_09}
    \int_V u_x(x,y,z) \,C_{11}(x,y,z) \,z \,\d V \overset{!}{=} 0 \; .
\end{equation}

For the special case of an RVE with a homogeneous composition in all three spatial directions, i.e.,
\begin{equation} \label{eq:appendix_cmc_10}
    C_{ij}(x,y,z) = \text{const. in $x$, $y$ and $z$} \; ,
\end{equation}
eq. \eqref{eq:appendix_cmc_09} reduces directly to the constraint introduced in \cite{geers2007}:
\begin{equation} \label{eq:appendix_cmc_11}
    \int_V u_x(x,y,z) \, z \,\d V \overset{!}{=} 0 \;.
\end{equation}

\paragraph{Eccentric RVE.}
An additional constraint is required for eccentric RVEs \cite{klarmann2020}.
Otherwise, the homogenized shear stiffnesses $GA_{sy}$ and $GA_{sz}$ depend on the selected RVE length $\ell^{RVE}$.
This 'constant normal force constraint' is given by
\begin{equation} \label{eq:appendix_cnc_01}
    \int_{-\tfrac{1}{2}\ell\rve}^{\tfrac{1}{2}\ell\rve} N(x)\,x\,\d x \overset{!}{=} 0
\end{equation}
with
\begin{equation} \label{eq:appendix_cnc_02}
    N(x) = \int_{y_-}^{y_+} \int_{z_-}^{z_+} \sigma_{xx}(x,y,z) \, \text{d}y\text{d}z
\end{equation}
The integral \eqref{eq:appendix_cnc_02} can be simplified using the same arguments as above, yielding
\begin{equation} \label{eq:appendix_cnc_03}
    \int_V C_{11}(x,y,z) \, u_x(x,y,z) \, \d V \overset{!}{=} 0
\end{equation}

\end{appendices}
\bibliography{lit}

\end{document}